\documentclass[conference]{IEEEtran}

\IEEEoverridecommandlockouts
\usepackage{cite}
\usepackage{amsmath,amssymb,amsfonts}
\usepackage{amsmath}
\usepackage{graphicx}
\usepackage{subfigure}
\usepackage{textcomp}
\usepackage{xcolor}
\usepackage{booktabs}
\usepackage{multirow}
\usepackage{bigstrut}
\usepackage{hyperref}
\usepackage{tikz}
\usepackage{calc}

\usepackage{makecell, longtable}                % 用于划分表格

\usepackage{stfloats}
    
\usepackage[skip=1ex]{caption}
\usepackage{lipsum}

\def\BibTeX{{\rm B\kern-.05em{\sc i\kern-.025em b}\kern-.08emT\kern-.1667em\lower.7ex\hbox{E}\kern-.125emX}}

\begin{document}

\title{A survey of sketches in traffic measurement: Design, Optimization, Application and Implementation\\
{}
\thanks{}
}

\author{\IEEEauthorblockN{\textsuperscript{} }
Shangsen Li, Lailong Luo, Deke Guo, Qianzhen Zhang, Pengtao Fu
\IEEEauthorblockA{\textit{Science and Technology on Information System Engineering Laboratory,} \\
\textit{National University of Defense Technology,} Changsha Hunan 410073, P.R. China.}
}
\maketitle

\begin{abstract}

Network measurement probes the underlying network to support upper-level decisions such as network management, network update, network maintenance, network defense and beyond.
Due to the massive, speedy, unpredictable features of network flows, sketches are widely implemented in measurement nodes to approximately record the frequency or estimate the cardinality of flows.
At their cores, sketches usually maintain one or multiple counter array(s), and rely on hash functions to select the counter(s) for each flow.
Then the space-efficient sketches from the distributed measurement nodes are aggregated to provide statistics of the undergoing flows.
Currently, tremendous redesigns and optimizations have been proposed to improve the sketches for better network measurement performance.
However, existing reviews or surveys mainly focus on one particular aspect of measurement tasks.
Researchers and engineers in the network measurement community desire an all-in-one survey that covers the entire processing pipeline of sketch-based network measurement.
To this end, we present the first comprehensive survey of this area.
We first introduce the preparation of flows for measurement, then detail the most recent investigations of design, aggregation, decoding, application and implementation of sketches for network measurement.
To summarize the existing efforts, we carry out an in-depth study of the existing literature, covering more than 90 sketch designs and optimization strategies.
Furthermore, we conduct a comprehensive analysis and qualitative/quantitative comparison of the sketch designs.
Finally, we highlight the open issues for future sketch-based network measurement research.

\end{abstract}


\section{Introduction}

Measurement-based performance evaluation of network traffic is a fundamental prerequisite for network management, such as load balance\cite{DBLP:conf/nsdi/Al-FaresRRHV10}, routing, fairness\cite{DBLP:conf/hoti/KabbaniAYPP10}, instruction detection\cite{DBLP:journals/compsec/Garcia-TeodoroDMV09}, caching\cite{DBLP:journals/tos/EinzigerFM17}, traffic
engineering\cite{DBLP:conf/sigcomm/VasudevanPSKAGGM09}, performance diagnostics\cite{DBLP:conf/sigcomm/CurtisMTYSB11} and policy enforcement\cite{DBLP:conf/sigcomm/SommersBDR07}.
With the exploration of network users and network scale, the underlying network traffic becomes massive and unpredictable, which challenges and complicates all the network functions.
Network measurement probes the underlying flows and provides the metadata to support network-related decisions and threat diagnosis.
Efficient network measurement requires various levels of traffic statistics, both as an aggregate and on a per-flow basis\cite{DBLP:conf/infocom/Ben-BasatEFK17a}.
These statistics include the number of distinct flows (flow cardinality), the number of packets for a flow (flow size), the traffic volume contributed by a specific flow (flow volume) and the time information.
These statistics together profile the network and provide essential information for network management and protection/defense.

% Table generated by Excel2LaTeX from sheet 'Sheet1'
\begin{table}[htbp]
  \centering
  \caption{Comparison with existing surveys}
    \begin{tabular}{llll}
    \toprule
    First author & \# Variants & \#Measurement tasks & Year \\
    \midrule
    Cormode\cite{DBLP:journals/vldb/CormodeH10} & $\le 15$ & 3     & 2010 \\
    Cormode\cite{Cormode11sketchtechniques} & $\le 20$ & 3     & 2011 \\
    Zhou\cite{High-Speednetwork} & $\le 10$ & 4     & 2014 \\
    Yassine\cite{DBLP:journals/imm/YassineRS15} &---& --- & 2015 \\
    Gibbons\cite{DBLP:books/sp/16/Gibbons16} & $\le 10$ & 1     & 2016 \\
    Yan\cite{DBLP:journals/comsur/YanYGL16} & ---& 1 & 2016 \\
    This survey & $\ge 90$ & 13    & 2021 \\
    \bottomrule
    \end{tabular}%
  \label{comparison}%
\end{table}%

The term \emph{sketch} refers to a kind of synopsis data structure to record the frequency or estimate the cardinality of items in a multiset or stream approximately.
Basically, sketches hash each packet according to the interested flow key into a union of cells or counters.
Sketches offer an efficient method to record the existence of the active flow and its statistical information for later passive traffic measurement tasks.
As a compact summary, sketches are typically much smaller than the size of input and adaptive to solve network measurement tasks.
Moreover, sketches guarantee constant time overhead of element insertion and information query operations with guaranteed accuracy.

%Sketch offers an efficient method to record the existence of active flow and its volume information in passive traffic measurement. It acts as a synopsis to efficiently store and retrieve the interested information. The core idea of the sketch includes two aspects: one is the counter and updating design, which is to record statistical information; another is the counter organization or sketch architecture design.

%However, software servers have plenty of memory but incur new challenges of achieving both high performance and high accuracy.

The primary motivation of conducting this survey is two-fold.
Firstly, a lot of sketch designs and optimizations have been proposed to perform measurement tasks in recent years.
Existing surveys, however, are somehow out-of-date and do not cover these new proposals.
The latest survey\cite{DBLP:books/sp/16/Gibbons16} was published four years ago and only covers a small partition of work.
Secondly, the existing surveys\cite{DBLP:journals/imm/YassineRS15}\cite{DBLP:journals/comsur/YanYGL16} solely focus on a particular or several aspects of measurement tasks and lack of a comprehensive  view on sketch design and optimization.
%Yassine et al.\cite{DBLP:journals/imm/YassineRS15} makes effort on passive traffic measurement in SDN mostly, while Yan\cite{DBLP:journals/comsur/YanYGL16} provides a comprehensive survey of SDN-based DDoS attack detection and mitigation solutions.
The earlier work\cite{DBLP:journals/vldb/CormodeH10}\cite{Cormode11sketchtechniques} only covered less than 20 sketch variants and were published nearly ten years ago.
These existing surveys did not provide comprehensive reviews on sketch-based network measurement.
As shown in Table. \ref{comparison}, compared with the existing surveys, our survey covers more sketch variants and considers more measurement tasks.
In particular, our survey concentrates on the design and optimization for the sketch from a perspective of dataflow in sketch algorithm.

\begin{figure*}
  \centering
  % Requires \usepackage{graphicx}
  \includegraphics[width=7in]{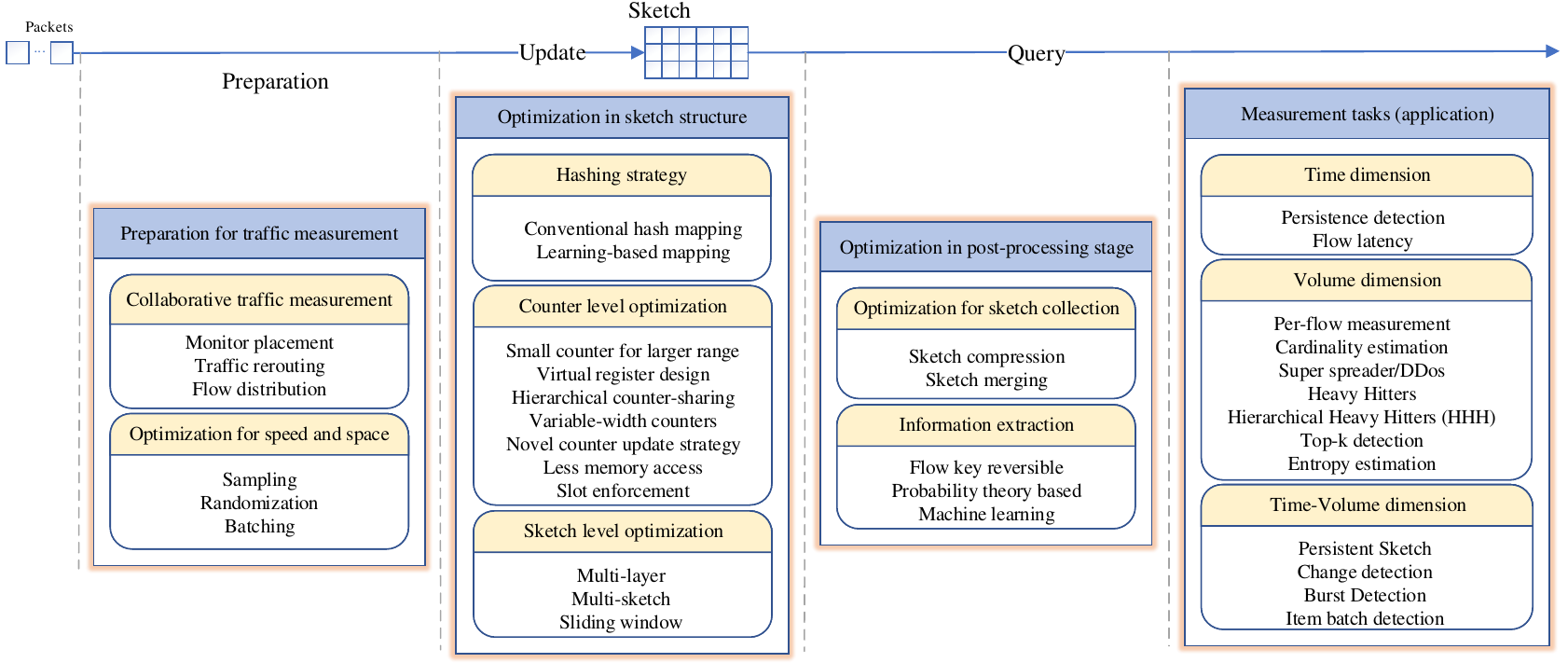}\\
  \caption{Framework of this survey}\label{work}
\end{figure*}

In this survey, we first classify the sketch design and optimization from the perspective of data process pipeline of the sketch algorithm.
Specifically, from the data recording stage to the information extraction stage, the core components for a sketch data structure are update strategy, design of data structure, query strategy, and the enabled functionality.
Thus, we survey the existing works from three-dimension, i.e., preparation for sketch-based network measurement, optimization in sketch data structure, and optimization in the post-processing stage.
Besides, we further provide a comprehensive survey of sketch-based measurement tasks and implementation.
The comprehensive framework is illustrated in Fig. \ref{work}.
For each measurement task, we compare the corresponding sketch design quantitatively and qualitatively.

Following the above directions, the remainder of this paper is organized as follows. Section \ref{based} gives a basic description of sketch-based network measurement.
Section \ref{prepare} details the related work about preparation for traffic measurement, including collaborative traffic measurement and packet preprocessing.
Section \ref{sketch structure} focuses on the design and optimization in the sketch data structure, including hashing strategy, counter level optimization, and sketch level optimization.
Section \ref{post} focuses on the process when extracting information from the sketch and gives a detailed summary about technology for sketch compression, sketch merging, data supplement, and information extraction.
In Section \ref{app}, we compare the sketch-based methods from the perspective of measurement tasks in terms of both the time dimension and volume dimension.
Further, we conclude the related hardware and software implementation of sketch-based measurement in that section.
Finally, we give the related open issues in Section \ref{summary}.

\section{Sketch based network measurement}\label{based}

Network measurement is indispensable for understanding the performance of a network infrastructure and managing networks efficiently.
There are two ways to measure network performance: active or passive techniques.

In active measurement, network flows are continuously measured by sending a series of probe packets over the network paths.
With the help of such probes, network operators are able to measure one-way delay, round-trip-time (RTT), the average packet loss, connection bandwidth and adjust the forwarding policies in case of load changes.
It estimates the network performance by tracking how the probe packets are treated in the network\cite{DBLP:conf/globecom/SuWXH14}.
The accuracy is closely associated with the probe frequency in general.
Active measurement offers different granularity of end-to-end QoS measurement and is flexible since you can measure what you want.
However, sending probes frequently may impose significant measurement overhead that may disturb critical traffic.

By contrast, passive measurement provides an efficient tool for charging, engineering, managing and securing the communication tools\cite{DBLP:journals/ton/HuLZ0CCW14}.
Without probing packets, passive measurement monitors the traffic traversing the measurement points to collect statistical information.
In particular, a passive measurement system is able to summarize flow-level or packet-level traffic, estimate flow/packet-size distribution, evaluate the fine-grained volume of network traffic with different attributes for usage-based charging, and more.
Besides, passing measurement can help identify abnormal traffic patterns to ease network management tasks, such as traffic engineering, load balancing, and intrusion detection.
Especially, a flowkey is often defined as the five fields in the packet headers, i.e., for IPv4 protocol, a flow identifier includes Source IP address (32 bits), Destination IP address (32 bits), Source Port (16 bits), Destination Port (16 bits), and Protocol (8 bits), i.e., $ f =\left<SrcIP,SrcPort,DstIP,DstPort,Protocol \right>$.
A flow $f$ is defined as a set of packets that share the same flow key, which can be flexibly assigned according to high-level need.
Therefore, the number of flow identifiers can be as large as $2^{104}$ \cite{Forouzanwithsophiachungfegan2003TCP}.

\begin{figure}
  \centering
  \includegraphics[width=3in]{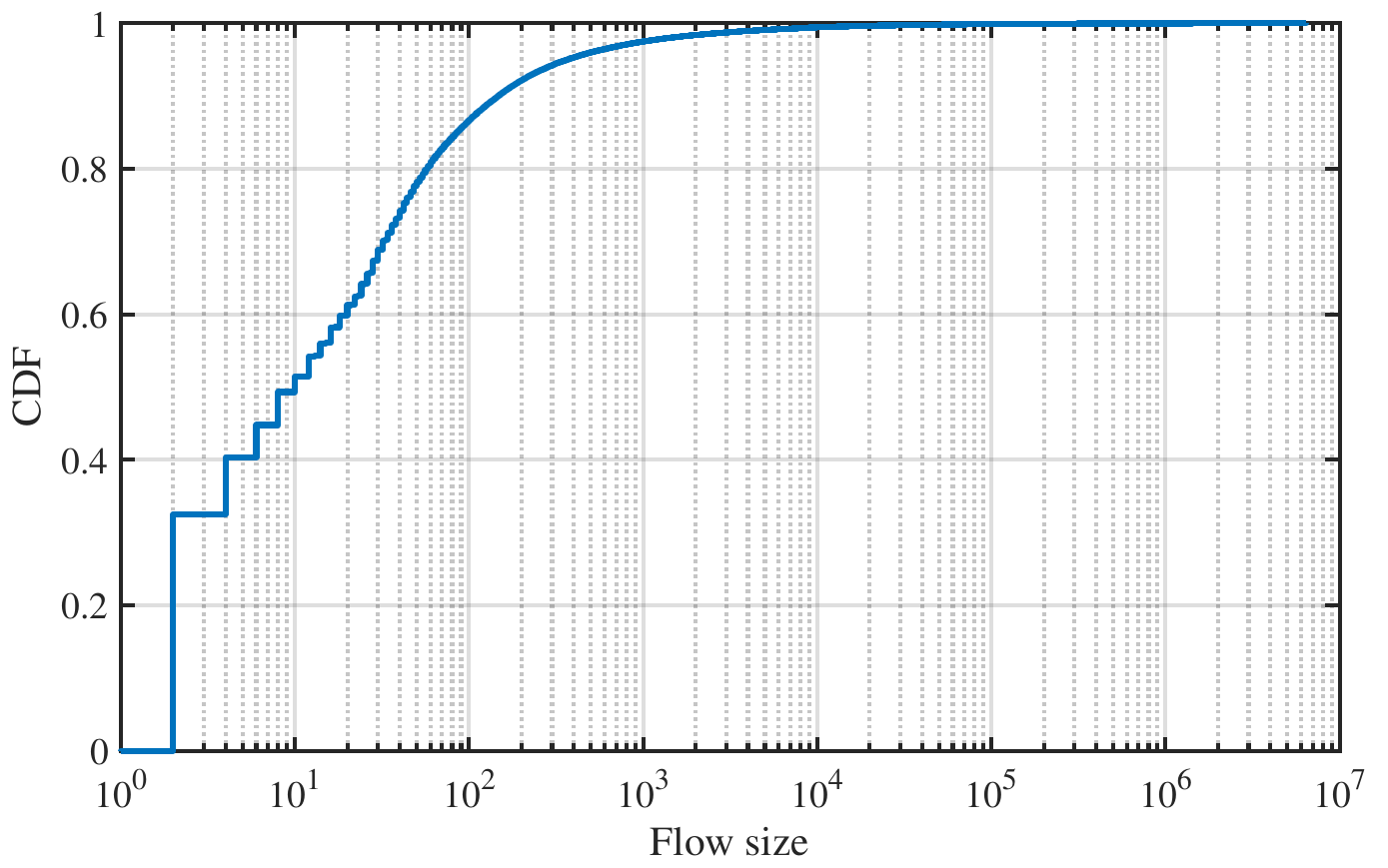}\\
  \caption{The CDF of flow size distribution. There are 424.6 million IPv4 packets in total from WIDE traffic trace\cite{MAWIWork21:online}.}\label{distribution}
\end{figure}

Active measurement obtains the network state, such as bandwidth, by injecting probe packets into the network.
However, the measurement packets will disrupt the network, especially when sending measurement traffic with high frequency.
In contrast, passive measurement measures network without creating or modifying any traffic.
And we concentrate on sketch-based passive network measurement in this survey.

\subsection{Challenges of passive measurement}

The challenge of passive measurement mainly lies in two aspects, i.e., resources limitations and unpredictable traffic characteristics.
Therefore, most researchers envision an algorithm that can obtain interesting statistical information of traffic with limited resources.
%The main research focuses on network measurement is to design an algorithm, which can efficiently obtain the interested statistical information within limited resources.

\textbf{Resource Limitation.} When implementing a measurement task, a key challenge is how to perform it on a small high-speed memory efficiently\cite{DBLP:conf/icnp/XiaoQMC14}.
Nowadays, the core routers forward most packets with the fast-forwarding path between network interface cards directly, without going through the CPU nor main memory.
To keep up with such high speed, measurement tasks are expected to perform in SRAM.
However, the SRAM capacity is limited and is shared by all online network functions of routing, management, performance and security.
Hence, the available space for measurement tasks is limited.
To address this challenge, as stated in Section \ref{SRAM},\ref{DRAM},\ref{SRAM-DRAM}, lots of proposals have been investigated to fit the measurement algorithms into SRAM, hybrid SRAM-DRAM, and DRAM.
Moreover, researchers have tried to perform measurement tasks with FPGA and TCAM as summarized in Section \ref{FPGA}\ref{TCAM} as the development of hardware.
Furthermore, with the emergence of software-defined networking (SDN), efforts have been made to perform measurement tasks in P4Switch and Open vSwitch, as shown in Section \ref{P4}\ref{OVS}\ref{sdn}.

\textbf{Flow Unpredication.} Real network traffic is always high-speed and non-uniform distributed.
Such characteristics further bring challenges to measurement algorithm design.
On the one hand, the high-speed feature make it difficult to record the size of traffic flows accurately.
On the other hand, the skewness of flow in the network further aggregates the measurement task.
Usually, the flow size/volume follows Zipf\cite{DBLP:conf/conll/Powers98} or Power-law\cite{5288290} distribution, as illustrated in Fig. \ref{distribution} wherein traffic trace was collected by MAWI Working Group\cite{MAWIWork21:online} from its samplepoint-G at 2:00 pm to 2:15 pm on 29/02/2020.
Specifically, most flows are small in size, often known as mouse flows; while a few of them are extremely large, commonly known as elephant flows.
For most measurement tasks, elephant flows are more important than the small ones.
Therefore, the size of each counter should be capable of counting the largest flows accurately.
In effect, the network operators may not know the size of elephant flows in advance, which makes it tricky to set the length of each counter.

The optimal measurement strategy for a specific measurement objective typically assumes a priori knowledge about the traffic characteristics.
However, both traffic characteristics and measurement objectives can dynamically change over time, potentially rendering a previously optimal placement of monitors suboptimal.
For instance, a flow of interest can avoid detection by not traversing the deployed monitors.
The optimal monitor deployment for one measurement task might become suboptimal once the objective changes.
We survey the work which targeted at handling the skewness of flows in Section. \ref{sketch structure}.
Further, packet batching and sampling strategies, which are proposed to match up with the high line rate, are detailed in Subsection. \ref{process}.
Note that, as illustrated in Fig. \ref{NWWM}, the packets of a flow may go through multiple paths from source to destination.
To maximize the flow coverage and minimize the redundancy, the switch in the middle should take part in measurement tasks.
The related work are summarized in Subsection. \ref{coll}.
%In the traces we analyzed, long duration flows from chat applications can be as high as 48${\%}$. Second, more than half of all hosts participating in long duration flows get blacklisted by real-time black lists.\cite{DBLP:conf/infocom/ChenJC10}

\subsection{Performance requirements for network measurement}

Due to the limitations and challenges of passive network measurement, we further elaborate the performance requirement of measurement tasks.

\textbf{High Accuracy.} Most of the existing measurement methods provide approximate results.
They provide theoretical guarantees that the estimated result has a relative error $\epsilon$ with a confidence probability $1\mathrm{-}\delta$, where $\epsilon$ and $\delta$ are configurable parameters between 0 and 1\cite{DBLP:conf/sigcomm/HuangLB18}.
Novel updating strategies, which are designed to improve measurement accuracy, can be found in Subsection. \ref{update}.
And several multi-layer sketch designs to improve the measurement accuracy are detailed in Subsection. \ref{layer}.

\textbf{Memory Efficiency.} Take a compact data structure to perform traffic measurement is a core fundamental requirement for both hardware and software implementations.
Because of the vast number of flows passing through a network device, memory space becomes a significant concern.
For a hardware implementation, high memory usage aggravates chip footprints and heat consumptions, thereby increasing
manufacturing costs\cite{DBLP:conf/sigcomm/HuangLB18}.
For software switches, although severs have plenty of memory\cite{DBLP:conf/sosr/AlipourfardMZ0Y18}, high memory usage will deplete the memory of co-located applications (e.g., VMs or containers).
The related work for improving space efficiency are detailed in Section \ref{count}.
Furthermore, sampling strategies are also deployed to reduce the space overhead, as specified in Section \ref{sample}.

\textbf{Real Time Response.} Most measurement tasks should respond to traffic anomalies in time to avoid potentially catastrophic events.
However, some sketch designs can support queries only after the entire measurement epoch.
For example, before querying a certain flow, Counter Braids\cite{DBLP:conf/sigmetrics/LuMPDK08} must obtain the information of all flows in advance, and decoding should be executed offline in a batched way.
As a result, the query speed is significantly slowed down.
Additionally, several work that focus on the most recent information are detailed in Section \ref{sliding}.

\textbf{Fast Per-packet Processing.}Processing packets at high speed by limiting the memory access overhead of per-packet updates is essential for network measurement.
For example, a reduction in latency for partition/aggregate workloads can be achieved if we can identify traffic bursts in near-real time\cite{DBLP:conf/sigcomm/CurtisMTYSB11}.
The main challenges of high processing speed stem from high line rates, the large scale of modern networks, and frequent memory accesses.
A hardware implementation usually prefers running algorithms in TCAM or SRAM, rather than DRAM, for better high-speed processing.
For a software implementation, as can be envisioned in upcoming SDN and NFV realizations, the ability to perform computations on time greatly depends on whether the data structures fit in the hardware cache and whether the relevant memory pages can be pinned to avoid swapping\cite{DBLP:conf/infocom/Ben-BasatEFK17a}. The related works can be found in Section \ref{process}\ref{SRAM-DRAM}\ref{SRAM}. The per-flow measurement methods are discussed in Section \ref{sketch structure}. The optimization strategies for less memory access are detailed in Section \ref{access}.

\textbf{Distributed Scalability.} Due to the space limitation of a single switch, it is advisable to deploy a collaborative traffic measurement system wherein each switch is only responsible for a part of the underlying flows in the network.
For more details of collaborative traffic measurement, the related works are introduced in Section \ref{coll}.

\textbf{Generality.} There are various network measurement tasks.
Designing a specific algorithm for each different type of task is costly and inefficient.
Moreover, it requires sophisticated resource allocation across different measurement tasks to provide accuracy guarantees\cite{DBLP:conf/sigcomm/MoshrefYGV14}\cite{DBLP:conf/conext/MoshrefYGV15}.
A general sketch-based measurement method means to support a variety of network-measurement tasks simultaneously.
Typical methods, which are general or partially general to traffic measurement tasks, are listed in Table. \ref{general}.

\textbf{Reversibility.} Although traditional sketch designs can answer the point query, they fail to return the corresponding flow identifier from the counters.
Thus, many efforts have been made to make sketches invertible and capable of identifying the  interested flows from the sketches directly. The details are described in Section \ref{reversible}.

\subsection{Sketch as a perfect choice}

\begin{figure}
  \centering
  % Requires \usepackage{graphicx}
  \includegraphics[width=3in]{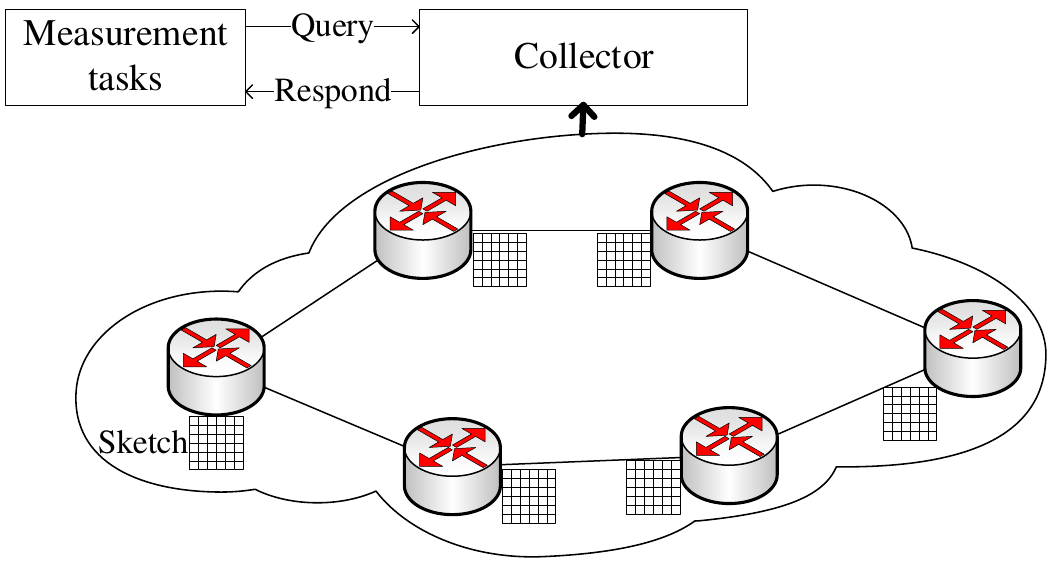}\\
  \caption{Sketch-based network traffic measurement}\label{Sketch}
\end{figure}

Sketch-based network measurement belongs to the passive network measurement, which does not cause any overhead in the network since it does not send any probe packets.
It allows the processing of local traffic states and the global behavior of traffic flows passing through specific network partitions.
%Passive measurement provides detailed information about the nodes being measured.
For passive network measurement, monitors require full access to the network devices such as routers and switches, as shown in Fig. \ref{Sketch}.
Generally, the sketch-based traffic measurement can be divided into the following steps: 1) the distributed measurement points acknowledge their tasks and collect the corresponding traffic statistical information passively, and 2) the collectors merge the measurement results from the subordinate monitors and respond to the measurement query.

As illustrated in Fig. \ref{work}, sketch-based network measurement experiences the same pipeline as stream processing.
It is concerned with processing a long stream of traffic packets in one pass using a small working memory to estimate specific statistics of flows/packets approximately.
To enable numerous per-flow counters in a scalable way, replacing exact numbers with estimated counters is an intrinsic feature of sketch-based measurement.
These estimates reach an acceptable trade-off memory occupation and estimation accuracy.
Sketch offers an efficient method to record the existence of active flows and their information in passive network measurement with accuracy guarantees.
As a compact summary, sketches are typically much smaller than the size of the input.

A sketch organizes its working memory as a synopsis data structure, specializing in capturing as much information about the target statistics as possible.
For the vast flow key space, we are interested in the identity of active flows (during the measurement epoch).
To record the massive active flows, a brute-force method is to count and record them directly.
However, such a plain design will incur high memory access and computing overhead.
By contrast, a sketch-based method usually maps the flow keys into a table with multiple buckets/cells within constant time.
The corresponding buckets/cells will store the interested information about the flows for later queries within constant time.
Various sketches are proposed for different statistics about the flows. %They have a common online operation called hash and increment. In these algorithms, an incoming item is fed to a hash function, and the hash result is treated as the index into an counters array. The corresponding counter is then incremented\cite{DBLP:conf/sigmetrics/ZhaoXL06}. Then by specific information extraction strategies, we can get desired statistics information by performing a query on the sketch.
More generally, there are two primary operations in a sketch-based algorithm: \textbf{\emph{update}} and \textbf{\emph{query}}.
The flows in a measurement epoch are represented by the sketch with the update operations and thereafter queried for upper-level network analysis.
Note that, different sketches may provide distinct update and query strategies.
Moreover, most kinds of sketches enable function distributed measurement scalability by merging operations.
Some sketches embed additional information into cells to enable reversibility.

%The first operation updates the sketch using the key and value of each item in the data stream; the latter is used to obtain the statistics of the data stream (e.g., the most frequent items or the number of distinct elements in the data stream) from the sketch following the query operation.
%Each sketch-based algorithm uses different methods to implement its update and query operations.

%\begin{figure}
%  \centering
%  % Requires \usepackage{graphicx}
%  \includegraphics[width=3in]{Two_D}\\
%  \caption{The information needed for traffic measurement}
%  \label{TwoD}
%\end{figure}

\subsection{Count-Min sketch as an example}

As the most widely applied sketch, Count-Min (CM) sketch\cite{DBLP:conf/latin/CormodeM04} has been used to perform per-flow measurement, heavy hitter detection, entropy estimate\cite{DBLP:conf/esa/BhuvanagiriG06}, inner-product estimation\cite{DBLP:conf/sigmod/RusuD07}, privacy preserve\cite{DBLP:journals/ccr/RoughanZ06}, clustering over a high-dimensional space\cite{DBLP:conf/sigmod/RusuD07} and personalized page rank\cite{DBLP:conf/www/SarlosBCFR06}.

The CM sketch can represent high dimensional data (like the 5-tuple flow key space) and answer point query with strong accuracy guarantees within constant time.
With such design, CM sketch supports various queries, such as heavy hitters, flow size distribution, top-$k$ detection, and beyond.
Since the data structure can efficiently process updates in the form of insertion or deletion, it is capable of working under the traffic stream at high rates.
The data structure maintains the linear projection of the data with several other random vectors, which are defined implicitly by simple hash functions.
Increasing the range of hash functions or the number of hash functions improves the accuracy of the summary\cite{DBLP:journals/ftdb/CormodeGHJ12}.
Because of this linearity, CM sketches can be scaled, added, and subtracted, to produce summaries of the corresponding scaled network traffic.
However, due to the  irreversible process of hash, CM sketch can not support flow key reversibility.

\begin{figure}
  \centering
  \includegraphics[width=3in]{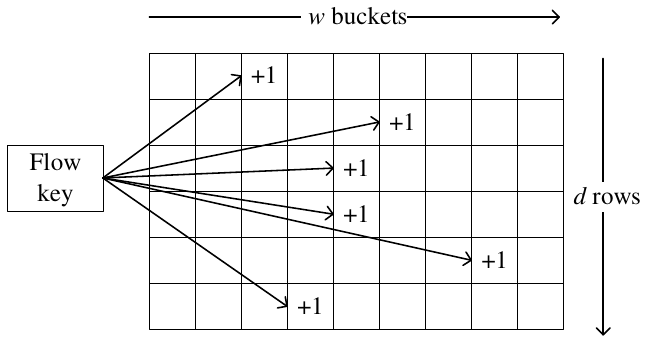}\\
  \caption{The CM sketch structure}\label{Countmin}
\end{figure}

Specifically, it hashes an arbitrary flow into each row of the hash table and leverages the corresponding counters to record the multiplicity or size information.
It consists of \emph{d} arrays, ${A_1...A_d}$,  each of which contains \emph{w} counters, as shown in Fig. \ref{Countmin}.
When measuring the network traffic, each flow key is extracted from the packet header and hashed by the \emph{d} independent hash functions, ${h_1...h_d}$.
Thereafter, the $d$ corresponding counters, i.e., ${\forall \ 1\le k\le d,A_k\left[ h_k\left( f \right)\%w \right]}$ are added up by the flow sizes.
Because computing each hash function takes $O(1)$ (constant) time, the total time to perform an update is $O(d)$, independent from $w$.
To query the size of a flow $f$, CM returns the minimum among the $d$ counters as the estimate. Moreover, CM only incurs over-estimated errors following accuracy guarantee.

\noindent \textbf{Theorem 1\cite{DBLP:conf/latin/CormodeM04}}: If ${w\mathrm{=}\lceil \frac{e}{\varepsilon} \rceil}$ and ${d\mathrm{=}\lceil \ln \frac{1}{\delta} \rceil}$, the estimate ${\hat{s}_i}$ has the following guarantees: ${s_i<\hat{s}_i}$; and with probability at least ${1-\delta}$.
\begin{equation}\label{Theorem1}
 \hat{s}_i\le s_i+\varepsilon \lVert f \rVert _1
\end{equation}
where $s_i$ is the real size of the $i^{th}$ flow in the flow set $f$; $\hat{s}_i$ represents the point query result of the $i^{th}$ flow from the CM sketch; and $e$ is the base of the natural logarithm and it is a constant chosen to optimize the space overhead for a given accuracy requirement.

\section{Preparation for network measurement}\label{prepare}

\begin{figure}
  \centering
  % Requires \usepackage{graphicx}
  \includegraphics[width=3.5in]{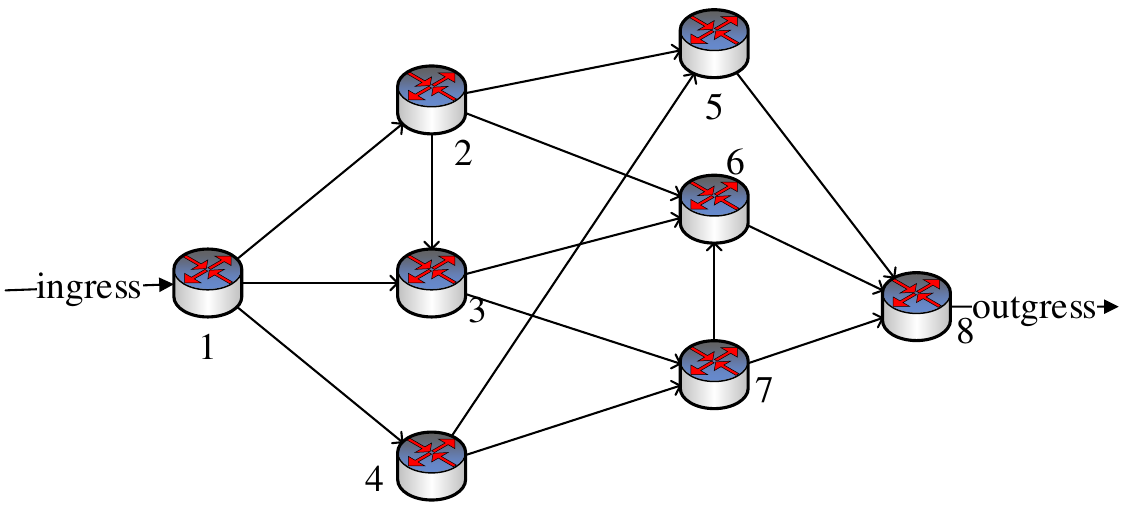}\\
  \caption{Example topology for multiple routing path}\label{NWWM}
\end{figure}

In practice, the measurement module is integrated as software inside a switch or deployed as standalone measurement hardware that taps into a communication link.
As shown in Fig. \ref{Sketch}, once a measurement module is placed or deployed on a link, it can capture packets carried by the link according to its configuration.
As illustrated in Fig. \ref{NWWM}, each flow may traverse multiple switches on its routing path, and any of them can perform measurement tasks.
Moreover, introducing a measurement module on a link needs additional deployment cost, including hardware/software investment, space cost and maintenance cost\cite{DBLP:conf/infocom/SuhGKT05}.
There is no need for one switch to measure all passing traffic.
Therefore, the network-wide measurement functions should be installed as efficiently as possible.
Thus, performing measurement environment preparation is necessary before measuring the flows.
Besides, the actual measurement operation performed by monitors also factors into its operating cost.
Due to the per-packet operation cost of each monitor depends mainly on the line speed, it is nontrivial to perform network measurement for high-speed traffic within limited resources.
It is necessary to pre-process traffic packets whenever possible to reduce the computation and storage overhead.

%There are two method to achieve this target in the pre-processing stage, which is \textbf{sampling} and \textbf{batching}.
%With the observation that each flow may traverse multiple switches on its routing path and its measurement can be performed by any one of them. In order to observe a large fraction of a network's traffic, multiple links should be measured concurrently since only a relatively small fraction of the traffic can be seen at any single measurement point in a large  network. In fact, there is no need to require each switch to measure all its flows because otherwise flows with multiple hops will be unnecessarily measured multiple times redundantly. By assigning each flow to a single switch for measurement, every switch only measures a subset of passing flows, which reduces not only memory requirement but also processing overhead. This promising approach to relieve the space and processing overhead constraints in a single monitor point is called \textbf{\emph{collaborative traffic measurement}}\cite{DBLP:conf/infocom/XuCMH19} or \textbf{\emph{distributed traffic measurement}}\cite{DBLP:conf/infocom/SuhGKT05}.

%In this subsection, we will give a survey of \textbf{monitor placement}, \textbf{traffic rerouting} and \textbf{flow distribution} to conduct the collaborative traffic measurement.

\subsection{Preparation for measurement  environment}\label{coll}

%In passive traffic measurement, there is no more probe packet need to be injected into the network. But the tapped hardware or installed software that monitor the traffic may be very expensive due to the requirements for processing and storing collected data. It is thus very important to minimize the number of such hardware or software to install and maintain in the network.

% Table generated by Excel2LaTeX from sheet 'Sheet1'
\begin{table*}[htbp]
  \centering
  \caption{Collaborative traffic measurement}
    \begin{tabular}{llll}
    \toprule
    \multicolumn{1}{c}{\multirow{14}[8]{*}{Collaborative measurement}} &       & Methods & Core idea \\
\cmidrule{2-4}          & \multirow{4}[2]{*}{Monitoring Placement} & Suh et al. \cite{DBLP:conf/infocom/SuhGKT05} & Minimize cost and maximize coverage \\
          &       & Sharma and Byers\cite{DBLP:conf/pam/SharmaB05} & {Minimize duplication of effort} \\
          &       & Cantieni et al. \cite{DBLP:conf/conext/CantieniIBDT06} & Decide which switch should be actived \\
          &       & OpenTM\cite{DBLP:conf/pam/TootoonchianGG10} &Choose which switches to measurement \\
\cmidrule{2-4}          & \multirow{2}[2]{*}{Traffic Rerouting} & MMPR\cite{DBLP:journals/tnsm/HuangCCL12} & \multirow{2}[2]{*}{Make traffic adaptive to netowrk and requirement} \\
          &       & MeasuRouting\cite{DBLP:journals/ton/RazaHCSS12} &  \\
\cmidrule{2-4}          & \multirow{7}[2]{*}{Flow Distribution} & OpenWatch\cite{DBLP:conf/conext/Zhang13} & {Compute the optimal balanced task assignment} \\
          &       & LEISURE\cite{DBLP:conf/ancs/ChangHLC11} &Consider load-balancing under different objectives \\
          &       & cSamp\cite{DBLP:conf/nsdi/SekarRWZKA08} & {Specify hash ranges per OD-pair per router} \\
          &       & cSamp-T\cite{DBLP:conf/comsnets/SekarGRZ10} & {Without relying on OD-identifiers} \\
          &       & DECOR\cite{DBLP:conf/iwqos/ShenA12} & {From local to global optimization} \\
          &       & DCM\cite{DBLP:conf/sigcomm/YuQL14} &Leverage the BF\cite{DBLP:journals/cacm/Bloom70} to allocate responsibility \\
          &       & Xu et al.\cite{DBLP:conf/infocom/XuCMH19} &Leverage hash range to allocate responsibility \\
    \bottomrule
    \end{tabular}%
  \label{collaborative}%
\end{table*}%

\subsubsection{Monitor Placement}

\

%The focus of this subsection is the related works to minimize the overhead in terms of hardware, software, maintenance cost in order to get the expected measurement performance in the collaborative traffic measurement scenario.
In collaborative traffic measurement scenarios, it is urgent to minimize the overhead of hardware, software, and maintenance, with the guarantee of measurement performance.
To ensure network-wide coverage, previous work focus on the optimal deployment of monitors across the network with the target to maximize the measurement utility, such as the coverage of measured traffic and the efficiency of measurement functions, and minimize the cost of the measurement.
The optimal strategies are in the form of traffic routing path or the optimal measurement point allocation in the whole measurement-enable network.

Under the measurement functionality un-powered situation in switch, Suh et al. \cite{DBLP:conf/infocom/SuhGKT05} consider the minimum cost and maximum traffic coverage problems under various budget constraints.
They derive out a monitor placement strategy and specify the sampling rates of each monitor in order to maximize the fraction of sampled IP flows.
They first find the links should be measured and then determines the sampling rates with an optimization algorithm.
Besides, Minimum Edge Cost Flow (MECF) \cite{DBLP:conf/conext/ChaudetFLRV05} studies the problem of assigning tap devices for passive measurement.
They consider three main problems, the first one maximizes the volume of captured traffic under cost constraints.
The second problem minimizes the deployment cost to achieve a measurement objective and the last one minimizes both installation and operational cost under the same objective.
The authors present a combinatorial view of the problem and derive complexity and approximately results and efficient and versatile Mixed Integer Programming (MIP) formulation.
For passive measurement, they study the problem of sampling packets and then present an efficient way of placing monitor devices and how to control their sampling rates.
They show that all these problems are NP-complete and they present heuristics to approximate the optimal solution for each one. They evaluate the performance of the proposed algorithms with simulation on topologies discovered by the Rocketfuel utility and with generated traffic matrices. In addition, this work only considers the static placement of monitors.

In practice, current routers deployed in operational networks are already equipped with measurement modules (e.g., Netflow\cite{DBLP:journals/rfc/rfc3954}, Openflow\cite{mckeown2008openflow}).
There is no need to turn on all these measurement functionalities because of their expensive operational cost and measurement redundancy.
Because there are potentially hundreds of candidate measurement points to choose for network-wide measurements.
Cantieni et al. \cite{DBLP:conf/conext/CantieniIBDT06} focus on which monitors should be activated and what sampling rate should be set on these monitors in order to achieve a given measurement task with high accuracy and low resource occupation.
Sharma and Byers\cite{DBLP:conf/pam/SharmaB05} propose space-efficient data structures for gossip-based protocols to approximately summarize sets of measured flows.
With some fine-tuning of the methods, they ensure that all flows are observed by at least one monitor, and only a tiny fraction of flows are measured redundantly.
OpenTM\cite{DBLP:conf/pam/TootoonchianGG10} explores several algorithms to choosing switches for query.
They further, reveal that there is a trade-off between the accuracy of measurements and the worst-case maximum load on individual switches.
The work shows a non-uniform distributed query strategy that tends to query switches closer to the destination with a higher probability has a better performance compared to the uniform schemes. %OpenTM reads byte and packet counters kept by OpenFlow switches for active flows and therefore incurs a minimal overhead on network elements. At the same time, the highest level of accuracy is preserved, because the TM is derived directly without making any simplifying mathematical and statistical assumptions. OpenTM shows the possibility of direct measurement of TM (traffic matrix) with the least overhead as long as the infrastructure (OpenFlow) provides the appropriate feature set for measurements.

Measurement point  placement problem focuses on the optimal placement strategies under the measurement functionality un-powered situation\cite{DBLP:conf/infocom/SuhGKT05}\cite{DBLP:conf/conext/ChaudetFLRV05} or the measurement activation problem in measurement powered situation\cite{DBLP:conf/conext/CantieniIBDT06}\cite{DBLP:conf/pam/TootoonchianGG10}.
As the core of collaborative measurement, these works are essential for setting up the environment for sketch-based measurement.
\subsubsection{Traffic Rerouting}

\

The above works focus on the optimal measurement point placement across the network.
However, both traffic characteristics and measurement tasks can dynamically change over time so that a previous optimal placement of measurement points may be suboptimal for the current network state.
When it is not feasible to adaptively redeploy/reconfigure measurement infrastructures to cater to such changing requirements\cite{DBLP:journals/ton/RazaHCSS12}, a possible strategy is to reroute the traffic across these monitoring points to get better measurement performance.

MeasuRouting\cite{DBLP:journals/ton/RazaHCSS12} addresses the problem where different measurement points have different measurement abilities for certain kinds of traffic by rerouting interested traffic across measurement points.
%strategically re-routing traffic sub-populations over fixed monitors.
Building upon MeasuRouting, Measurement-aware Monitor
Placement and Routing (MMPR)\cite{DBLP:journals/tnsm/HuangCCL12} considers that not only the number of deployed monitors is limited, but the traffic characteristics and measurement objectives change both continually and simultaneously.
The MMPR framework jointly optimizes monitors placement, dynamical turning on and off the monitors, and dynamic routing strategy to achieve maximum measurement utility, quantifying how well each flow is monitored.
MMPR formulates the problem as a MILP (Mixed Integer Linear Programming) problem and proposes several heuristic algorithms to approximate the optimal solution and reduce the computation complexity.
Considering dynamics issues to estimate the flow importance and configure the routing table entries of both MeasuRouting and MMPR, Dynamic Measurement aware Routing (DMR)\cite{DBLP:journals/network/HuangCRS11} summarizes these challenges and proposes solutions.

\subsubsection{Flow Distribution/Allocation}

\

A natural problem of collaborative measurement is how to assign the flows to the measurement nodes/switches to ensure both load balance and measurement completeness.
Such a problem is formulated as a flow distribution/allocation problem.
By allocating each flow to a sequence of switches, every switch only measures a subset of flows to save memory and computation resources.
Hitherto, there are a dozen works that try to tackle this question.
DCM\cite{DBLP:conf/sigcomm/YuQL14}, OpenWatch\cite{DBLP:conf/conext/Zhang13} and LEISURE\cite{DBLP:conf/ancs/ChangHLC11} focus on the measurement load balance strategies making.
While the hash-based works\cite{DBLP:conf/nsdi/SekarRWZKA08}\cite{DBLP:conf/comsnets/SekarGRZ10} \cite{DBLP:conf/infocom/XuCMH19} make efforts on lightweight implementation of strategies.

Distributed and Collaborative Monitoring system (DCM)\cite{DBLP:conf/sigcomm/YuQL14} leverages Bloom filters to record which flows should be measured by the switch.
It requires each switch to store two Bloom filters, with the first one encoding the set of flows to be measured locally and the second one helping remove false positives from the first filter.
However, the memory overhead is significant because it takes a Bloom filter 10 bits per flow on average to ensure a false-positive ratio of less than 1{\%}.
Moreover, the lookup of each Bloom filter takes $k$ hash operations and $k$ memory accesses, where $k$ is 7 for an optimal Bloom filter with 1{\%} false positive ratio.
OpenWatch\cite{DBLP:conf/conext/Zhang13} designs an efficient method to offload the measurement loads from the ingress switches to other switches along with the routing paths.
It leverages the global view of network topology and routing path for each flow to compute the optimal task assignment for all the switches in the network. %OpenWatch tests and therefore selects the switches for the measurement target to achieve good performance.%On the one head, for each aggregation flow, the involved switches which are responsible for measurement task is as less as possible; on the other head, the switch select the smallest counter to cover the flow, which aims to cover more aggregation flow in later task assignment. With these two rules, OpenWatch tests and therefore selects the switches for the measurement target, this strategy can achieve good performance.
By contrast, Load-EqualIzed meaSUREment (LEISURE)\cite{DBLP:conf/ancs/ChangHLC11} balances the network measurement across distributed monitors.
Specifically, it considers various load-balancing problems under different objectives and studies their extensions to support different deployment scenarios. %It takes routing matrix, the topology, measurement infrastructure deployment and measurement requirements of tasks as inputs, and decides which available monitors should participate in each specific measurement task and how much they need to measure to optimize the load-balancing objectives. Ideally the load balancing objective is to have identical workload for all monitors where workload denotes the normalized traffic amount that each monitor measures.
%In this work, the load-balancing objective is mainly defined as two terms: 1) minimizing the variance of workloads across all monitors or 2) minimizing the maximum workload among them.

Several works leverage hash functions to tradeoff computation and space overhead when allocating flows across route paths.
cSamp\cite{DBLP:conf/nsdi/SekarRWZKA08} is a centralized hash-based packet selection system, which allows distributed monitors to measure disjoint sets of traffic without explicit communications.
%thus eliminating an possibly ambiguous measurements across the network.
cSAMP specifies the set of flows identified by OD-pairs (origin-destination) that each measurement point is required to record by considering a hybrid measurement objective that maximizes the total flow-coverage subject.
 %to ensuring that the optimal minimum fractional coverage of the task can be achieved.
An improved method cSamp-T\cite{DBLP:conf/comsnets/SekarGRZ10} provides comparable measurement capabilities to cSamp without relying on OD-pair identifiers.
Each router only uses only local information, such as packet headers and local routing tables, rather than the global OD-pair identifiers.
%Leveraging results from the theory of maximizing submodular set functions, cSamp-T provides near-ideal performance in maximizing the total flow coverage in the network. Further, with a small amount of targeted upgrades to a few routers, cSamp-T nearly optimally maximizes the minimum fractional coverage across all OD-pairs.
DECOR\cite{DBLP:conf/iwqos/ShenA12} provides a solution to coordinate network resources and avoids controller bottleneck, message delay, and a single point of failure.
DECOR divides a network into smaller pieces called optimization units.
Therefore, it tries to achieve local optimization in each optimization unit and then extends to global optimization.
%DECOR can be divided into two parts: (a) hash range arrangement: assign hash range to nodes along a path to achieve local optimization, and (b) quota distribution: nodes assign resource quota to paths going through them to avoid the conflict of local optimizations.
The DECOR framework can be applied to cSAMP, as cSAMP is also a resource assignment strategy.
The comprehensive experiments conclude that the DECOR-based cSAMP is superior to others.
Xu et al.\cite{DBLP:conf/infocom/XuCMH19} propose a new lightweight solution to flow distribution problems for collaborative traffic measurement in SDN.
The proposed framework minimizes the memory/space overhead on each switch with a single sampling probability value $p$.
The processing overhead is at most one hash operation per packet to implement sampling if the packet is not recorded by one of the other switches on the routing path;
if the packet is recorded earlier, the hash operation will not be performed.
%The method maintains a state bit on each packet head, initially as 0, which indicates that this packet has not been measured.
Thus, the controller can decide the optimal Probability Assignment for Ingress Switches (PAIS) and Network-wide Switch Probability Assignment (NSPA) collaborative measurement scenarios.

Compared with \cite{DBLP:conf/conext/Zhang13,DBLP:conf/ancs/ChangHLC11,DBLP:conf/nsdi/SekarRWZKA08,DBLP:conf/comsnets/SekarGRZ10,DBLP:conf/iwqos/ShenA12,DBLP:conf/sigcomm/YuQL14},
up to now,
the method proposed in \cite{DBLP:conf/infocom/XuCMH19} is the most lightweight strategy to complete the flow distribution problem. By allocating different hash range to a series of switch, NSPA elegantly allocates the responsibility to the switch across the link.
In this way, the additional processing overhead is at most one hash operation per packet to implement sampling, and the additional space overhead on each switch is only a single sampling probability value $p$.

\subsection{Preparation for measurement data}\label{process}

Intuitively, after deploying a measurement environment, measurement points can directly perform measurement tasks.
However, in the high-speed network, performing measurement on each packet is a heavy overhead task.
Sampling, batch processing, and randomization are powerful tools to reduce the processing
overhead in various systems\cite{DBLP:conf/infocom/JangMMMN20}.

\subsubsection{Traffic Sampling}\label{sample}

\

Production network measurement tools, such as NetFlow\cite{DBLP:journals/rfc/rfc3954} and sFlow\cite{sFloworg84:online}, always sample the network flows for resource saving and fast processing.
However, they incur low estimation accuracy since many flows are missed.
Given a sampling probability of $p$,
%there are two ways to sample the flow, i.e., flow sampling and packet sampling.
flow sampling means that every monitor processes one packet per $1/p$ packets in each flow.
while packet sampling indicates that every monitor processes one packet per $1/p$ packets of all traffic.
The linear sampling strategies can both miss flows and sample the same packets for multiple times along the routing path.
Some non-linear sampling sketches have been proposed to overcome the drawback of uniform random sampling strategy.

To realize the constant relative error, small values of flow size are incremented with high probability while large ones with low probability.
The authors in\cite{DBLP:conf/infocom/KumarX06} propose to set the packet sampling rate as a decreasing function of the flow sizes.
This strategy can significantly increase the packet sampling rate of small and medium flows by sampling less large flows.
By doing so, more accurate estimations of various network statistics can be guaranteed.
However, the exact sizes of all flows are available only if we keep per-flow information for every flow, which is prohibitively expensive for highspeed links.
Sketch guided sampling (SGS)\cite{DBLP:conf/infocom/KumarX06} solves the estimation problem by using a synopsis data structure called counting sketch to estimate the approximate sizes of all flows.
Adaptive Random Sampling\cite{DBLP:conf/icc/ChoiPZ03} bounds the sampling error (relative estimation error) within a pre-specified tolerance level and proposes an adaptive random sampling technique to adjust the sampling probability and minimize the number of samples.

%Sample and hold\cite{DBLP:journals/tocs/EstanV03} keeps near exact counts of heavy hitters, flows with high packet counts. For each incoming packet, the router checks if it is tracking this packet's flowkey, defined over one or more fields of the 5-tuple. If yes, the router updates that counter. If not, the flowkey for this packet is selected with probability \emph{p}, and the router keeps an exact count for this selected flowkey subsequently.
%MRSCBF\cite{DBLP:journals/jsac/KumarXW06} consider the same sampling probability is not friendly in terms of accuracy and storage efficiency. It employs multiple SCBFs, but operating at different resolutions (sampling probability). An incoming of packet will result in an insertion into each SCBF \emph{i} with a sampling probability ${p_i}$. The higher ${p_i}$ value corresponds to higher resolution. The inspiration is that elements with low multiplicities will be estimated by filter(s) of higher resolutions, while elements with high multiplicities will be estimated by filters of lower resolutions.
Adaptive Non-Linear Sampling (ANLS)\cite{DBLP:conf/infocom/HuWTLCC08} updates the counters of sketch with sampling probability function $p(c)$, where $c$ is the counter value and $p(c)$ is a pre-defined function.
In this work, the authors provide the general principles to guide the selection of sampling function $p(c)$ for sampling probability adjustment.
The intuition of ANLS is to implement a large sampling rate for small flows while a small sampling rate for large flows.
The authors also derive the unbiased flow size estimation, the bound of the relative error, and the bound of the required counter size for ANLS.
It updates a counter from $c$ to ${c\mathrm{+}1}$ with probability $p(c)$.
%, where $c$ is the counter value and ${p\left( c\right)\mathrm{=}1/\left[ f\left( c\mathrm{+}1 \right)\mathrm{-}f\left( c \right) \right]}$.
Self-tuning ANLS\cite{DBLP:conf/cse/HuL09} selects one specific sampling function according to the principles
ANLS\cite{DBLP:conf/infocom/HuWTLCC08} proposed. Set the $p(c)$ as follow:
\begin{equation}\label{ANLS}
p\left( c \right)\mathrm{=}\frac{1}{\left( 1\mathrm{+}a \right) ^c\mathrm{-}1}
\end{equation}
where \emph{a} is a pre-defined parameter ${0<a<1}$.
This paper focus on the method which adjusts the parameter \emph{a} during the statistic process in order to provide the best and ideal tradeoff between accuracy and counting range. %To keep the ANLS in a stable state, which achieves balance between accuracy metric and counting range metric. Self-tuning ANLS defines the accuracy utility is one and the counting range utility is zero initially. With the increase of parameter, the counting range utility increases and the accuracy utility decreases. When the two utility curves meets, the cross point is the equilibrium tuning point where the parameter should be set.
%Then to keep the inverse estimations before and after the parameter tuning as the same, renormalizing the counter when the counter overflows according the parameter changing is implemented.
DISCO\cite{DBLP:journals/ton/HuLZ0CCW14} further extends ANLS by regulating the counter value to be a real increasing concave function of the actual flow length to support the counting of flow size, flow volume counting, and flow byte count simultaneously.

However, in order to accurately estimate the largest counter, the above methods compromise the accuracy of the smaller flows\cite{8347005}, where the sampling probability function can be scaled to achieve a higher counting range at the cost of a larger estimation error.
Independent Counter Estimation
Buckets (ICE-Buckets)\cite{8347005} first presents a closed-form explicit representation of an optimal estimation function, which is used to determine these probabilities and estimate the real value of a counter.
%According to the extensive study of this function using rigorous mathematical analysis, including the relation between its relative error, memory complexity, estimation symbol range, and even bound the probability of the actual error exceeding a certain value, ICE Buckets is proposed to improve the accuracy for all counters.
ICE Buckets separate the flows into buckets and configure the optimal estimation function according to each bucket's counting range, thereby significantly reducing the overall error by efficiently utilizing multiple counter scales.

Both linear and nonlinear sampling may incur the problem that some flows are over-estimated (sampled more than the sampling probability), while some are under-estimated (sampled less than the sampling probability).
Starting with a simple idea that ``independent per-flow packet sampling provides the most accurate estimation of each flows'', SketchFlow\cite{DBLP:conf/infocom/JangMMMN20} performs an approximated systematic sampling for flows to provide a better tradeoff between accuracy and overhead for a given sampling rate $1/p$.
This strategy is achieved by recognizing a sketch saturation event for a flow and only sample the saturated packets.
However, SketchFlow is only a general sampler strategy and can not be deployed to measure the traffic individually.

Generally speaking, sampling is the optimal strategy to reduce the per-flow measurement.
From the uniform sampling, adaptive sampling\cite{DBLP:conf/infocom/KumarX06}\cite{DBLP:conf/icc/ChoiPZ03}\cite{DBLP:conf/cse/HuL09}\cite{DBLP:conf/infocom/HuWTLCC08}\cite{DBLP:journals/ton/HuLZ0CCW14} to separated estimated-bucket strategy\cite{8347005}, the estimation accuracy keep increasing.
Although sampling still suffers from minor estimation errors, these strategies have played an essential role in sketch-based network measurement.

\subsubsection{Randomization Processing}\label{random}

\

%Sekar et al.\cite{DBLP:conf/imc/SekarRZ10} revisit the case for a "minimalist" approach in which a small number of simple yet generic router primitives collect flow-level data from which different traffic metrics can be estimated. This paper demonstrates the feasibility and promise of such a minimalist approach using flow sampling and sample-and-hold\cite{DBLP:journals/tocs/EstanV03} as sampling primitives and configuring these in a network-wide coordinated fashion using cSamp\cite{DBLP:conf/nsdi/SekarRWZKA08}. The paper shows that this proposal yields better accuracy across a collection of application-level metrics than dividing the same memory resources across metric-specific algorithms. Moreover, because a minimalist approach enables late binding to what application level metrics are important, it better insulates router implementations and deployments from changing measurement needs.

Different from packet sampling, randomization processing is an alternative approach to reduce the packet processing overhead in which only the sampled counters are updated.

%Univmon\cite{DBLP:conf/sigcomm/LiuMVSB16} proposes a framework composed of an online stage and an offline stage to maintain an overall sketch for various tasks.
%At the online stage, Univmon maintains $\log \left( n \right)$ copies of a "${L_2}$ heavy-hitter(L2-HH)" sketch.
%Each sketch is fed with the substream of a half smaller size.
%Each substream is defined recursively by the substream before it and is created by sampling the previous one with the probability 1/2.
%Repeating this procedure \emph{k} times reduces the dimensionality of the stream by a factor of ${2^k}$.
%Then, by summing across heavy hitters of all these recursively defined vectors, Univmon creates a single "recursive sketch", which gives a proper estimate of \emph{G-sum} for multiple measurement tasks.

Randomized HHH (Hierarchical Heavy Hitters)\cite{DBLP:conf/sigcomm/Ben-BasatEFLW17} proposes a randomized constant-time algorithm for HHH.
Unlike the deterministic algorithms whose update complexity is proportional to the hierarchy's size \emph{H}, RHHH randomly samples only a single prefix to update using its respective instance of heavy-hitters rather than updating all prefixes for each incoming packet.
This randomization strategy decreases the update time from $O(H)$ to $O(1)$, but requires a specific number of packets to provide desired accuracy guarantees.

%BUS\cite{DBLP:conf/icccn/EinzigerLW17} is a sampling method for estimating flow volumes. BUS samples packets with probability proportional to their size. That is, larger packets are sampled with higher probability than smaller ones. The number of sampled packets in BUS is proportional to the total byte traffic. In this way, BUS enables the underlying algorithm simply count the number of sampled packets and still provide reliable estimations for per-flow byte volume.

NitroSketch\cite{DBLP:conf/sigcomm/LiuBEKBFS19} is a careful synthesis of rigorous yet practical sketch design to reduce the number of per-packet CPU and memory operations.
Like the mechanism used in Randomized HHH\cite{DBLP:conf/sigcomm/Ben-BasatEFLW17}, NitroSketch only samples a small portion of packets by geometric sampling, and the sampled packets need to go through one hash computation, update to one row of counters, and occasionally to a top-\emph{k} structure.
The experiments demonstrate that the accuracy is comparable to unmodified sketches while attaining up to two orders of magnitude speedup and a 45\% reduction in CPU usage.

To catch up with the line rate and reduce the processing overhead, both RHHH\cite{DBLP:conf/sigcomm/Ben-BasatEFLW17} and NitroSketch\cite{DBLP:conf/sigcomm/LiuBEKBFS19} takes the randomization strategy to update the sampled counters in their sketches.
Although they all reduce the updating overhead, the corresponding query is complicated to cope with the incomplete recorded information, and a minimal volume of packets is required to achieve the desired accuracy guarantee.

%SA\cite{DBLP:conf/infocom/YangXLLWBL19} proposes a self-adaptive counters mechanism which is some like the floating-point number representation in computer science to enlarge counting range. The key idea behind the SA counter is that when a counter is going to overflow, for each insertion, instead of always increasing it, we increase it with a predefined probability. SA counters divides the bits into two parts, sign part and counter part. For different sign part, the counter part will increase with a pre-defined probability which is related to the number represented by the sign part. In this way, SA counter can make a small counter able to represent both small and large value.

\subsubsection{Batch Processing}

\

\begin{figure}
  \centering
  % Requires \usepackage{graphicx}
  \includegraphics[width=3.5in]{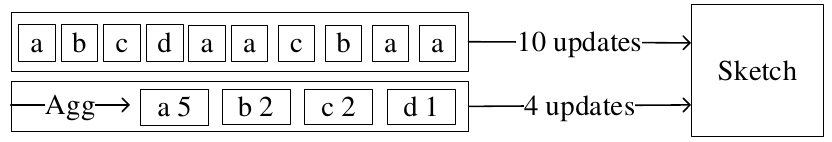}\\
  \caption{Agg-Evict strategy}\label{Aggregation}
\end{figure}

Due to the high speed of traffic packets and the per-flow processing requirement, the operational overhead is rather heavy.
In contrast to the sampling strategy, the inspiration of batching is to aggregate the packets for the same flow and then update them together as a whole to reduce the overall computation overhead, as illustrated in Fig. \ref{Aggregation}.

Randomized DRAM-based counter scheme\cite{DBLP:conf/ancs/ZhaoWLX09} takes a new counter update by looking up the cache to see if there is already a waiting update request to the same counter.
The scheme modifies that request in the cache rather than creating a new request (e.g., change the request from $+1$ to $+n$).
%%Since these two updates will result in only one (instead of two) eventual DRAM access (read and write), there is no incentive (toward degrading the performance of system) for an adversary to access the same counter repeatedly within a sliding window of $C$ cycles.
%iSTAMP\cite{DBLP:conf/infocom/MalboubiWCS14} is an intelligent Traffic (de)Aggregation and Measurement Paradigm, which partitions TCAM entries of switches/routers into two parts to: 1) optimally aggregate part of incoming flows for aggregate measurements, and 2) de-aggregate and directly measure the most informative flows for per-flow measurements. iSTAMP then processes these aggregate and per-flow measurements to effectively estimate network flows using a variety of optimization techniques.
%AMON\cite{DBLP:journals/jsac/KallitsisSBM16} leverages the high-performance packet measurement PF-RING and is readily deployable on commodity hardware. AMON examines all packets, partitions traffic into sub-streams by using rapid hashing and computes certain real-time data products. The resulting data structures provide views of the intensity and connectivity structure of network traffic at the speed of line rate. In the context of network measurement software system,
Agg-Evict\cite{DBLP:journals/ccr/ZhouAYY18}, a typical network measurement tool, also employs the same idea to leverage the benefits of caching.
It constructs a data structure, which improves the efficiency of network traffic measurement in software and proposes a low-level mechanism to improve the efficiency of various schemes.
Skimmed Sketch\cite{DBLP:conf/edbt/GangulyGR04} and Augmented Sketch (ASketch)\cite{DBLP:conf/sigmod/RoyKA16} focus on aggregating the most frequent items in the filter and evict the cold items into the second stage, while Cold filter\cite{DBLP:conf/sigmod/Zhou0J0YLU18} captures the cold items in the first filter and evicts the hot items to the second stage.
ASketch and Skimmed Sketch need two-direction communication between two stages to exchange elements because of the difficulty to capture hot items accurately.
However, Cold filter only needs one-direction communication and less memory access.

The core idea in batch processing \cite{DBLP:conf/ancs/ZhaoWLX09}\cite{DBLP:journals/ccr/ZhouAYY18}\cite{DBLP:conf/sigmod/RoyKA16}\cite{DBLP:conf/sigmod/Zhou0J0YLU18}
 is to take the incoming packets for a specific flow into a batch and then update them as a whole into the synopses data structure.
In this way, the memory access to the sketch is significantly reduced.
However, this strategy needs to augment an additional cache to maintain the information of the previous packets' flow.

\subsection{Summary and lessons learned}

In this section, we cover the aspect of preparation before measurement, including measurement environment and measurement data preparation.
By preparing the environment, each monitor is optimally placed or activated in the corresponding measurement point to get the best measurement performance considering measurement cost.
And all of them are allocated the responsibility of measurement tasks.
By data preparation, each measurement point performs the strategies on the raw traffic stream to reduce the measurement overhead.
Note that both sampling and randomization have to balance the tradeoff between accuracy and processing overhead.
After the preparation phase, the sketch algorithms can be installed and performed in each monitor point.

\section{Optimization in sketch structure}\label{sketch structure}

\begin{figure*}
  \centering
  % Requires \usepackage{graphicx}
  \includegraphics[width=6.5in]{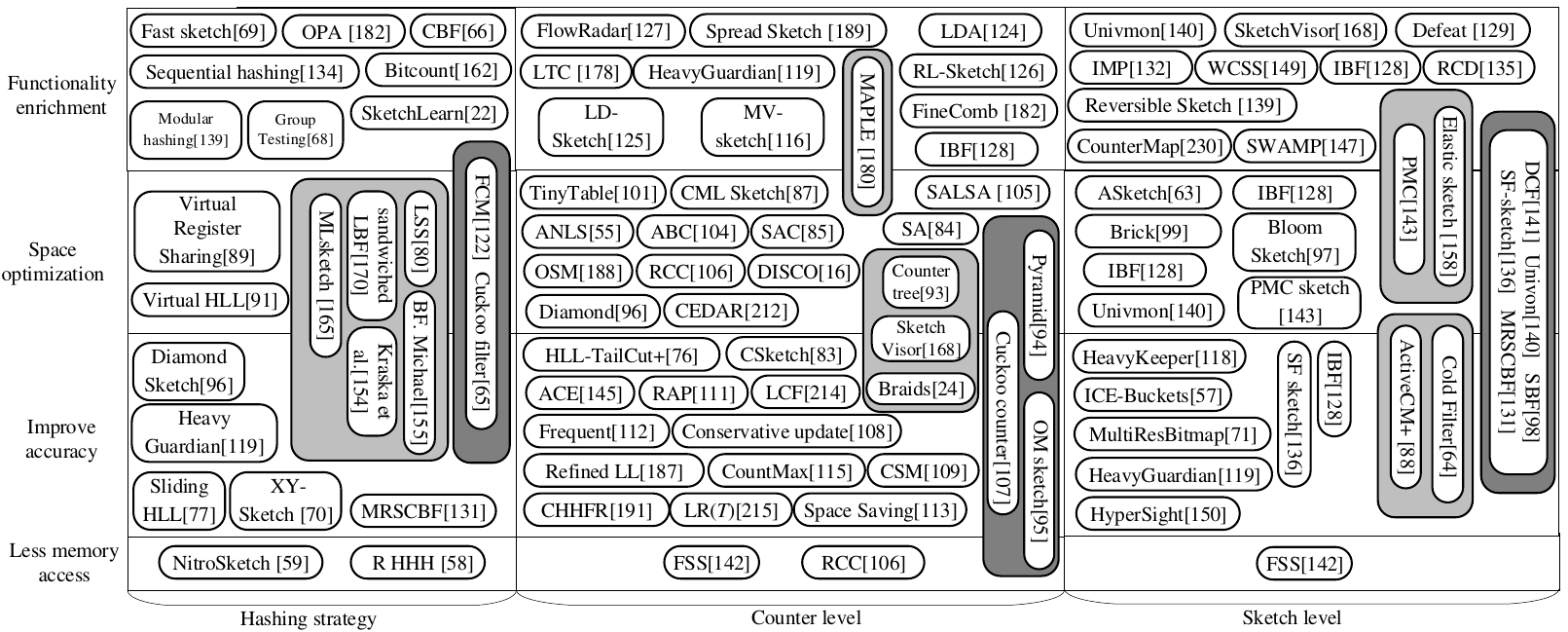}\\
  \caption{The taxonomy of the existing sketch methods. On the one head, the sketches are classified in three main aspects from the structure optimization perspective, i.e., hashing strategy, counter level optimization and sketch level optimization. On the other head, to improve the performance, dozens of variants devote themselves to improve the space efficiency, increase measurement accuracy, incur less memory access and  enrich the functionalities.
  }\label{combine}
\end{figure*}

The sketches are responsible for recording the active flows and the corresponding size or volume information.
The flow information is recorded in the sketch for later information retrieval and extraction in the updating stage.
As shown in Fig. \ref{combine}, a lot of strategies have optimized sketches in terms of \textbf{hashing strategy}, \textbf{counter level optimization} and \textbf{sketch level optimization}.
In this section, we will review these work in detail with the consideration of performance optimization.

\subsection{Hashing strategy}

A sketch algorithm needs a feasible amount of space to store the counters and a flow-to-counter association rule so that arriving packets can update corresponding counters at line speed.
This subsection gives a detailed summary of hashing strategies, including traditional and learn-based hash strategies.
Almost all existing sketch-based network measurements follow these different flow-to-counter rule designs as the basis.

\subsubsection{Conventional hash-based data structures}

\

The existing sketches are mainly based on traditional hash functions to maintain the mapping from flow keys to cells/buckets in the data structures.

\textbf{\emph{Hash-based frequency estimation.}} We regard CM Sketch\cite{DBLP:conf/latin/CormodeM04}, BF\cite{DBLP:journals/cacm/Bloom70} and Cuckoo filter\cite{DBLP:conf/conext/FanAKM14} as the hash base for frequency estimation.
 CM sketch\cite{DBLP:conf/latin/CormodeM04} consists of \emph{d} arrays, ${A_1,...,A_d}$,  each of which contains \emph{w} counters.
 It hashes incoming packets into each row of the hash table and locates one counter in each row to update the multiplicity or size information.
Bloom filter\cite{DBLP:journals/cacm/Bloom70} leverages the $k$ bits in a vector to represent the membership of an element within set.
Despite the constant-time complexity and the space-efficiency features, Bloom filter cannot support the deletion and record the multiplicity of flows.
 %The reason is that resetting the $k$ corresponding bits from 1s to 0s may cause the mis-deletion of other elements, which are also mapped to these bits.
To this end, Counting Bloom Filter (CBF)\cite{DBLP:journals/ton/FanCAB00} replaces each bit in the vector with a counter of multiple bits.
Whenever a flow is mapped into cells, the $k$ counters will be incremented by the flow size.
Then the deletion of elements is straightforwardly realized by decreasing the $k$ corresponding counters. %Consequently, the deletion of element $x$ will not affect the membership information of other elements.
Cuckoo filter (CF) is a hash table with $b$ buckets, and each bucket has $w$ slots to accommodate at most $w$ elements.
%When inserting a key, the cuckoo hashing table applies two perfect random hash functions to determine the two possible buckets where the key can be inserted\cite{richa2001power}.
Fan \emph{et al}.\cite{DBLP:conf/conext/FanAKM14} further represents the elements by recording their fingerprints in the cuckoo filter.
They further apply the partial-key strategy to determine the candidate buckets during the insertion phase and locate the candidate bucket\cite{richa2001power} during the reallocation phase.
However, CF can not support constant-time insertion because of the unbalanced probe during its reallocation phase.

Instead of maintaining mapping from flow key to flow counters directly, a novel strategy chooses to maintain mapping from flow key to several bit/bits counters according to bit-representation of flow key.
Group Testing\cite{DBLP:journals/ton/CormodeM05} arranges a number of items into a group together to find out active items.
It firstly decides which group the item belongs to and then updates these group counters as follows.
The first counter records the total number of items and the following counters are increased according to bits representation of the item. And the majority item can be identified by traversing the total counters.
Fast sketch\cite{DBLP:conf/infocom/LiuCG12} takes the same strategy as Group Testing\cite{DBLP:journals/ton/CormodeM05}.
It hashes each incoming packet into several predefined numbers of rows and adds its size to the first counter in each row.
After that, the corresponding counters also adds the flow size according to the bit representation of the quotient of the flow.
By collecting and adding sketches from multiple switches, the first counter in each row is firstly checked to recover the heavy-change flows.
SketchLearn\cite{DBLP:conf/sigcomm/HuangLB18} also takes the bit-mapping strategy to extract large flows from the multi-level sketch and leaves the residual counters for small flows to form Gaussian distributions.
XY-Sketch\cite{DBLP:conf/www/LiuX21} investigates the decomposition-and-recomposition framework to transform the problem of item frequency estimation into the problem of probability estimation.
Specifically, it decomposes each item into a sequence of sub-items and hashes them into corresponding rows of the sketch.
When responding to a point query, XY-Sketch recomposes the element and returns the product of all the sub-items probabilities appearing in the data stream and the total number of items.

%Modular hashing\cite{DBLP:journals/ton/SchwellerLCGGZDKM07} partitions flow keys into several segments which are therefore hashed separately with independent hash functions. The final hash result of flow key is composed of these independent hash result. Although modular hashing has more hash operations compared with hashing the entire key directly, it is an efficient strategy to perform reverse hashing by cross product of multiple modular reverse mapping to represent candidate flow set.

\textbf{\emph{Hash-based cardinality estimation.}} To count the number of distinct elements in a flow stream, different hash strategy was proposed.
Bitmap\cite{DBLP:journals/ton/EstanVF06} maps all the stream uniformly among a bit array.
By encoding each element as an index of the bit array, replicated elements can be filtered automatically.
However, the space of this strategy is linear to the cardinality\cite{DBLP:conf/edbt/MetwallyAA08}.
Another strategy is to prepare an array of sample buckets with a reducing sampling probability exponentially.
Each bucket is allocated a bit to record whether the bucket receives the stream elements.
By extracting information for this bit array, PCSA\cite{DBLP:journals/jcss/FlajoletM85}, LogLog\cite{DBLP:conf/esa/DurandF03},
HyperLogLog\cite{DBLP:conf/edbt/HeuleNH13}, HLL-TailCut+\cite{DBLP:conf/infocom/XiaoZC17}, Sliding HyperLogLog\cite{DBLP:conf/icdm/ChabchoubH10} all take this underlying data structure as the base to estimate the cardinality.

Recently, Zhou et al.\cite{DBLP:journals/pomacs/ZhouZMCO19} proposed a generalized sketch framework, including bSketch (based on the structure underlying the counting Bloom filter), cSketch (based on the structure of CountMin) and vSketch (based on the memory sharing mechanism), which aims to incorporate the sketch design under a general implementation structure, with the flexibility of plug-n-play and many options for the tradeoff.
Besides, to reverse flow key from sketches, a lot of work separately hash separated bit or a partition of flow key, instead of the whole flow key, to record the statistical information of bit or bits.
The details can be referred to in Section \ref{reversible}.
As the basis of the sketch algorithm, flow-to-counter rule is the first step to obtain the index of corresponding counters inside the sketch.

\subsubsection{Learning-based mapping}

\

Apart from the traditional hash strategies, a new trend of combining learning-based methods, such as $k$-means clustering and neural network, with the hash table is emerging to reduce the space overhead and improve estimation accuracy.

To compute the optimal mapping scheme, Bertsimas et al.\cite{DBLP:journals/corr/abs-2007-09261} propose a mixed-integer linear optimization
formulation, and an efficient block coordinate descent algorithm to compute the near-optimal hashing scheme for the seen flows.
A multi-class classifier is responsible for mapping elements to counters based on their features for the unseen flows.
Fu et al.\cite{DBLP:conf/infocom/FuLS0C20} put forward a class of locality-sensitive sketch (LSS) by formulating a theoretical equivalence relationship between the sketching error and the approximation error of the $k$-means clustering. Composed with a cuckoo table as a per-filter to test if it is a seen flow and a cluster model to collect the flow with the closest cluster center for this flow, LSS can efficiently mitigate the error variance optimize the estimation.
Hsu et al.\cite{DBLP:conf/iclr/HsuIKV19} combines a neural network to predict the log of the packet counts for each flow.
The heavy flows are separately recorded from non-heavy ones by this learned oracle to reduce the interruption between them.
Theoretical analysis proves that the error of the learned CM sketch is up to a logarithmic factor smaller than that of its non-learning counterpart. Based on\cite{DBLP:conf/iclr/HsuIKV19}, Aamand\cite{DBLP:journals/corr/abs-1908-05198} further provide a simple tight analysis of the expected error incurred by CM Sketch and the first error bounds for both standard and learned version of CSketch\cite{DBLP:journals/pvldb/CormodeH08}.

By such designs, the learning-based sketches show a new trend for measurement.
We believe there will be a body of such work in the future.
Although learning-based mapping incurs less space overhead, the implementation of such designs requires substantial computation resources, which may not be advisable for computation-scarce situations.
Besides, the updating of the out-of-data learning model is also no-trivial.

\subsection{Counter level optimization}\label{count}

Given the limited on-chip memory, if the counters in the sketch use many bits, the number of counters will be small, leading to poor estimation accuracy.
In this case, most counters only record mouse flows; hence, those untapped bits are wasted.
By contrast, if the counters occupy only a few bits, the sketch fails to represent elephant flows.
Thus, many efforts have been made to address this dilemma.

\subsubsection{Small counter for larger range}\label{compress}

\

\begin{figure}
\centering
\subfigure[SA] {\includegraphics[width=2in]{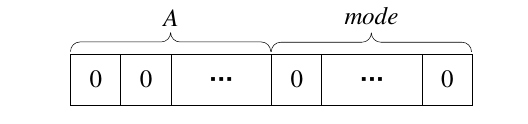}}
\subfigure[When insert an item $(e,1)$ into SA]{\includegraphics[width=3in]{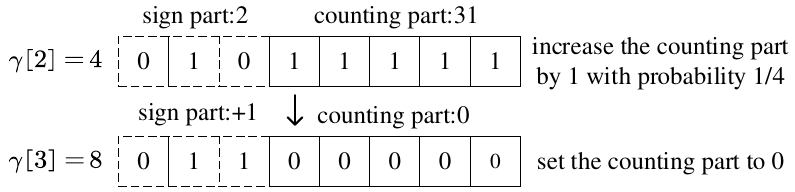}}
\subfigure[Dynamic SA] {\includegraphics[width=2in]{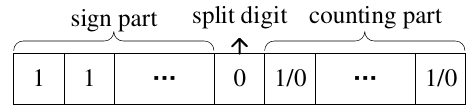}}
\caption{Self-adaptive counter\cite{DBLP:conf/infocom/YangXLLWBL19} and small active counter\cite{DBLP:conf/infocom/Stanojevic07}}
\label{SA}
\end{figure}

It is space-efficient to use a small counter to keep approximate counts of large numbers, as the resulting expected error can be rather precisely controlled.
Firstly proposed in Approximate Counting\cite{DBLP:journals/cacm/Morris78a}, Count-Min-Log Sketch\cite{DBLP:journals/corr/PitelF15}, SAC\cite{DBLP:conf/infocom/Stanojevic07} and SA\cite{DBLP:conf/infocom/YangXLLWBL19} have led this idea to network measurement.

Traditional sketches use $binary$-based counters to record the frequency of flows.
The counter is increased by 1 once incoming a corresponding packet. %when recording flow size.
{CML Sketch\cite{DBLP:journals/corr/PitelF15} replaces the classical $binary$-based counters by $log$-based counting cells.
The $log$-based counters increase with a predefined probability ${x^{-c}}$, where \emph{x} is the $log$ base and \emph{c} is the current estimation.
Consequently, the $log$-based counter can use fewer bits and increase the number of counters for the same storage space to improve the accuracy.
However, CML Sketch sacrifices the ability to support deletions for a more extensive counting range.

Small active counters (SAC)\cite{DBLP:conf/infocom/Stanojevic07} divides a cell into two part: a $k$-bit estimation part $A$ and a $l$-bit exponent part \emph{mode}, as shown in Fig. \ref{SA}(a).
Like floating-point number representation, the estimation of SAC is ${\hat{n}\mathrm{=}A\mathrm{\cdot} 2^{r\mathrm{\cdot} mode}}$, and \emph{r} is a global variable for all counters.
An incoming packet with size \emph{c} will cause the counter $A$ update as follows.
If $c\mathrm{\ge}2^{r\cdot mode}$, SAC first increases $A$ by $\lfloor \frac{c}{2^{r\cdot mode}} \rfloor$ and then increases $A$ by 1 with probability given by normalized value of the residue.
If $c\mathrm{<}2^{r\cdot mode}$, SAC increases $A$ by 1 with probability $\frac{c}{2^{r\mathrm{\cdot}mode}}$.
When $A$ overflows, the $mode$ part will be increased.
However, if $mode$ overflows, renormalization step will move the counting range to a higher scale, which is only supported by increasing the global variable $r$ and renormalize all counters at once to avoid estimation collision. This up-scaling strategy is inflexible.
%However, SAC allocates a fixed space for $A$ and $model$, the up-scale can only be supported by increasing the global variable $r$. When overflow occurs, SAC need to renormalize all the counter at once to avoid the estimation collision.
ActiveCM\cite{DBLP:conf/infocom/XiaoTC20} uses a variant of compressed counters in \cite{DBLP:conf/infocom/Stanojevic07} and enlarges a 32-bit counter the maximum counting range, which is far greater than $2^{32}$.

Compared with SAC\cite{DBLP:conf/infocom/Stanojevic07}, Self-Adaptive counters (SA)\cite{DBLP:conf/infocom/YangXLLWBL19} counter is a more generic technique to adapt to different counting ranges. %nd takes a similar strategy as SAC.
Let $n$ denote the number of bits in a counter.
For each counter, it has $s$-bits for sign and ($n\mathrm{-}s$)-bits for counting.
Moreover, SA augments an expansion array, $\gamma \left[ 0 \right] ,\gamma \left[ 1 \right] ,\cdots ,\gamma \left[ k\mathrm{-}1 \right]$, where $k\mathrm{=}2^s$.
When updating a SA counter with value 1, SA first gets the sign part $s_0$, and then adds 1 to the counting part $c_0$ with probability $\frac{1}{\gamma \left[ s_0 \right]}$, as Fig. \ref{SA}(b) illustrated.
If the counting part reaches $2^{n-s}$, the sign part will be increased by one, and the counting part will be set to zero. Finally, the estimation can be calculated as follows in Equ. \ref{SA1} and Equ. \ref{SA2}:

\begin{equation}\label{SA1}
\left\{ \begin{array}{l}
	stage\left[ 0 \right] \mathrm{=}0\\
	stage\left[ i \right] \mathrm{=}2^{n-s}\times \sum_{j\mathrm{=}0}^{i-1}{\gamma \left[ j \right] ,\ i>0}\\
\end{array} \right.
\end{equation}

\begin{equation}\label{SA2}
estimate\mathrm{=}c_0\times \gamma \left[ s_0 \right] +stage\left[ s_0 \right]
\end{equation}

\noindent To address the inflexibility of the fixed-length sign part, Dynamic SA (DSA) augments a split bit and dynamically adjusts the sign bits, as shown in Fig. \ref{SA}(c). The length of the sign part is initialized to 0.
Except for the split digit, all other bits are used to count. With the value represented by the counter becomes larger, SA moves the place of split bit and increases the bits of sign part. The length of the sign part is increased to $i$,
%the split digit is moved right by $i$ bits, and
 the counting part is shortened by $i$ bit. The $stages$ change to:

\begin{equation}\label{SA3}
\left\{ \begin{array}{l}
	stage\left[ 0 \right]\mathrm{=}0\\
	stage\left[ i \right]\mathrm{=}\sum_{j=0}^{i-1}{\left( \gamma \left[ j \right] \times 2^{n-j-1} \right) ,\ i>0}\\
\end{array} \right.
\end{equation}

\noindent In this way, DSA can accurately record mouse flows while being able to deal with elephant flows. %The estimate can be calculated using Equ. \ref{SA2}.

The above algorithms all focus on a single counter optimization for a more extensive counting range.
CML Sketch leverages a fixed $log$ base and updates counters according to the current count.
In comparison, SAC and DSA take more elastic strategies to enlarge the count range of little bits.
All these three estimators can represent large values with small symbols at the price of a minor error because they all increase the hashed counters with probabilities.

\begin{figure}
\centering
\subfigure[Register sharing\cite{DBLP:journals/ton/XiaoCZCLLL17}] {\includegraphics[width=3.5in]{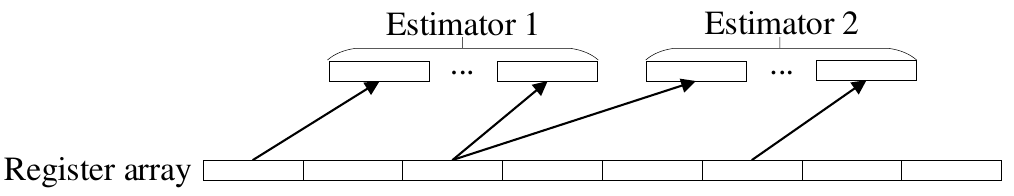}}
\subfigure[Bits sharing\cite{DBLP:journals/ton/YoonLCP11}]{\includegraphics[width=3.5in]{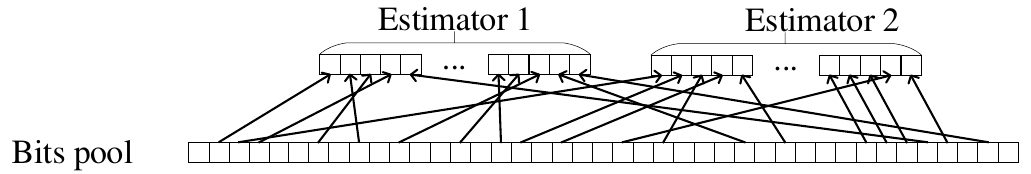}}
\caption{Virtual register design}
\label{register}
\end{figure}

\subsubsection{Virtual register for spread estimation}\label{virtual}

\

In this subsection, we focus on spread estimation, which aims to record the distinct connections with hosts and takes cardinality estimation register to solve DDos, port scan, and super spreader detection.
More works about cardinality estimation tasks are detailed in Section \ref{card}.
And the spread estimation can be divided into register sharing and bits sharing, as illustrated in Fig. \ref{register}.

\textbf{\emph{For register-sharing.}}
Virtual HyperLogLog Counter (VHC)\cite{DBLP:conf/globecom/ZhouZCXC16} shares HLL registers in a shared pool to measure flow size.
VHC contains an online encoding module to record the packets in real-time, while the offline estimation module measures the size of all flows based on the counters recorded by the online module.
%To encode a packet, VLC only needs to calculate one hash function, generate two pseudo-random numbers, and perform at most two memory accesses, i.e., reading counter and writing counter back if its value changes. At the same time, the probability of updating the counter is relatively small if the total number of packets mapped to the counter is large.
Furthermore, Virtual Register Sharing\cite{DBLP:journals/ton/XiaoCZCLLL17} dynamically creates an estimator for each flow by randomly selecting several registers from a multi-bits registers pool to estimate the cardinality of elephant flows.
The register is the basic cardinality estimation unit, such as PCSA\cite{DBLP:journals/jcss/FlajoletM85}, LogLog\cite{DBLP:conf/esa/DurandF03} and
HyperLogLog\cite{DBLP:conf/edbt/HeuleNH13}.
By sharing registers among many flows, as shown in Fig. \ref{register}(a), the space of register is fully utilized.
The array of registers are physical, while the estimators are logical and created on the fly without additional memory allocation.

\textbf{\emph{For bits-sharing.}} Compact spread estimator (CSE)\cite{DBLP:journals/ton/YoonLCP11} creates a virtual bit vector for each source by taking bits from a shared pool of available bits.
It counts the number of distinct \emph{DstIP} for each \emph{SrcIP} and can be abstracted as multiple-set version of linear counting\cite{DBLP:journals/tods/WhangVT90}.
Every individual \emph{SrcIP} has its own virtual bits vector for linear counting, as illustrated in Fig. \ref{register}(b), a portion of which is randomly shared with another \emph{SrcIP}'s virtual vector.

Both register-sharing and bits-sharing construct virtual estimators with common memory space.
Using a compact memory space can estimate the cardinality with a wide range and reasonable accuracy guarantees.
These space-efficient strategies enable on-chip implementation of spread estimation for network measurement with line speed.

\subsubsection{Hierarchical counter-sharing for skewed traffic}\label{hierar}

\

%To solve this problem, a lot of hierarchical counter-sharing based schemes have been designed to record the flow sizes, such as Counter Braids\cite{DBLP:conf/sigmetrics/LuMPDK08}, Counter Tree\cite{DBLP:journals/ton/ChenCC17}, Pyramid Sketch\cite{DBLP:journals/pvldb/YangZJCL17}, OM sketch\cite{DBLP:conf/globecom/ZhouLJ0DL17} and Diamond sketch\cite{DBLP:conf/infocom/0003GSWSL18}.
%As shown in Fig.10, the hierarchical structure of the counters may contains multiple layers. In detail, the lower layer counters mainly record the information of mouse flows, while the high layer counters record the number of overflows from the lower layers and can be shared by multiple flows to reduce space overhead.

In conventional sketches, all the counters are allocated the same size and tailored in order to accommodate the maximum flow size.
However, the elephant flows only occupy a small partition of the traffic. Therefore, the high-order bits in most counters of conventional sketches are wasted. This waste leads to space inefficiency.
To solve this problem, a lot of works, such as Counter Braids\cite{DBLP:conf/sigmetrics/LuMPDK08}, Counter Tree\cite{DBLP:journals/ton/ChenCC17}, Pyramid Sketch\cite{DBLP:journals/pvldb/YangZJCL17}, One memory
access sketch (OM sketch)\cite{DBLP:conf/globecom/ZhouLJ0DL17} and Diamond sketch\cite{DBLP:conf/infocom/0003GSWSL18} have designed hierarchical counter-sharing scheme to record the flow size, as shown in Fig. \ref{hierarchical}.
They organize the counters as a hierarchical structure in which the higher layers possess fewer memory.
The lower layer counters mainly record the information of mouse flows, while the high layers record the number of overflows at the low layers (the significant bits of the elephant flows size).
Moreover, higher layer counters can be shared by multiple flows in order to reduce space overhead.

\begin{figure}
  \centering
  % Requires \usepackage{graphicx}
  \includegraphics[width=3in]{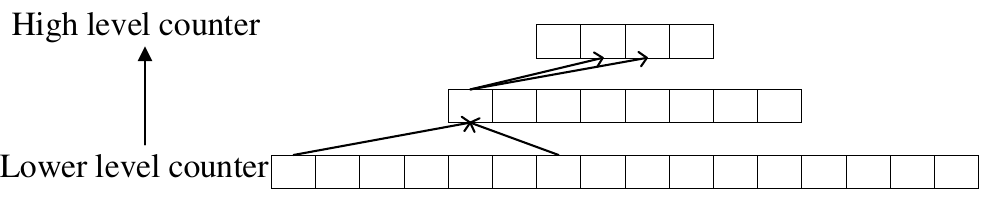}\\
  \caption{Hierarchical counter-sharing}\label{hierarchical}
\end{figure}

Counter Braids\cite{DBLP:conf/sigmetrics/LuMPDK08} compress counters into multi-layers by ``braiding" a hierarchy of counters with random graphs.
Braiding results in drastic space reduction by sharing counters among flows.
%and using random graphs generated on-the-fly with hash functions avoids the storage of flow-to-counter association.
A random mapping strategy is implemented to construct the relation between flows and the first-layer counters or two consecutive layers of counters.
Multiple flows share the common bits by sharing the high-layer counters to act as the overflow counters.
The low-layer and high-layer counters function as an biger counters to record the size of flows.
However, before querying a specific flow, the information of all flows must be obtained in advance, and the decoding should be executed offline in batching way.
As a result, the query speed is significantly slowed down.

%And it fails to count flows with very limited memory.

Counter tree\cite{DBLP:journals/ton/ChenCC17} is a two-dimensional counter sharing scheme, including horizontal counter sharing and vertical counter sharing.
Physical counters are logically organized in a tree structure with multiple layers to compose a new big virtual counter.
Different child counters may share the same parent counters in vertical counter sharing. Virtual counters are shared among different flows in horizontal counter sharing.
The virtual counter array for a particular flow consists of multiple counters pseudo-randomly chosen from the virtual counters.
However, as pointed in \cite{DBLP:conf/globecom/ZhouZCXC16}, the experiments based on real network trace data show that Counter tree cannot work well under a tight memory space.
Pyramid Sketch\cite{DBLP:journals/pvldb/YangZJCL17} further devises a framework that not only prevents counters from overflowing without the need to know the frequency of the hottest item in advance but also can achieve high accuracy, update speed, and query speed simultaneously.
It consists of multiple counter layers where each layer has half counters of the last layer.
The first layer only has pure counters to record multiplicity, while other layers are composed of hybrid counters.
Different from the Bloom sketch\cite{DBLP:conf/infocom/ZhouJLZ0L18}, Pyramid sketch uses two flag bit in hybrid counter for one child counter to record whether the child counter has overflowed or not.

Unfortunately, as pointed out in SA\cite{DBLP:conf/infocom/YangXLLWBL19}, for an elephant flow, the hierarchical counter-sharing strategy needs to access all layers, requiring many memory accesses for each insertion and query.
OM sketch\cite{DBLP:conf/globecom/ZhouLJ0DL17} has multiple layers with a decreasing number of counters.
%The high layer acts as the overflow counters of the low layer counters.
When an overflow occurs the first time, the flag in the low-layer counter will be set.
To alleviate the speed decline in the worst case, OM sketch only takes a two-layer counter array to record multiplicity information.
Moreover, OM Sketch leverages \emph{word constraint} to constrain the corresponding hashed counters within one or several machine words.
To improve the accuracy of OM sketch, OM Sketch
%uses the fingerprint check technique, which
records the fingerprints of the overflowed flows in their corresponding machine words at the lower layer to distinguish them from non-overflowed flows during queries.
Cold Filter\cite{DBLP:conf/sigmod/Zhou0J0YLU18} also consists of two counter layers.
%And for hot items, if the corresponding counters in the first overflow concurrently, Cold Filter will resort to the second layer to record the additional multiplicity.
Bloom sketch\cite{DBLP:conf/infocom/ZhouJLZ0L18} and Pyramid Sketch\cite{DBLP:journals/pvldb/YangZJCL17} take flag bit to record if the low layer has overflowed.
By contrast, Cold Filter will not change the low layer counters when the lower counter overflows and take no flag bit.
For hot items, the size is the addition of two-layer counters.
Furthermore, cold filter takes \emph{one-memory-access} strategy to handle the memory-access bottleneck, and the details can be referred to in Section \ref{access}.

\subsubsection{Variable-width counters for space efficiency}\label{variable}

\

Unlike the fixed-width counter design, many efforts have designed variable-width counters to improve space efficiency and estimation accuracy under skewed traffic.

\textbf{\emph{Counter resizability.}}
Spectral bloom filters (SBF)\cite{DBLP:conf/sigmod/CohenM03} extends bits vector of BF\cite{DBLP:journals/cacm/Bloom70} as counters vector.
The counters, packed one after another, are located by a hierarchical indexing structure for fast memory access.
%And a hierarchical indexing structure was used to locate counters that are packed next to each other in order to fast random accesses (reads).
However, an update causing the width of counter $i$ to grow will cause a shift of counters $i\mathrm{+}1, i\mathrm{+}2, \cdots$, which can have a global cascading effect, making it prohibitively expensive when there can be millions of counters\cite{DBLP:journals/ton/HuaXLZ11}.
Although the expected amortized cost per update remains constant and the global cascading effect of counter shift is small in the average case, the worst case cannot be tightly bounded.
Therefore, SBF with variable-width encoding is not an effective counter solution as it cannot ensure quick per-packet updates when a packet arrives.
%, forcing it to become a primarily read-only data structure in the sense that updates should be orders of magnitude less frequent than queries.
Bucketized Rank Indexed Counters
(BRICK)\cite{DBLP:journals/ton/HuaXLZ11} adopts a sophisticated variable-width encoding of counters, which uses fixed-size buckets and \emph{Rank Indexing Technique}\cite{DBLP:conf/icnp/HuaZLX08}
to support variable-length counters with a restricted counter sum.
 Specifically, it bundles a fixed number of counters randomly selected from the array into buckets.
And Brick only allocates enough bits to each counters, i.e., if its current value is $c$, Brick allocates $\lceil \lg \left( c \right) + 1\rceil $ bits to it.
In each bucket, Brick reorganizes counters into sub-counter arrays with a decreasing number of counters. Moreover, multiple counters are composed of a new variable-width counter. %Brick reduces the total memory size by utilizing the features among all the counters.
%A multi-level partition scheme is also designed to address the problem that the baseline bucketing scheme does not allow for efficient read and write accesses.
Brick can hold more counters as the average counter is shorter than the largest one\cite{8347005}.
Unfortunately, with the average counter value increasing, the encoding becomes less efficient.

\textbf{\emph{Bucket Expandability.}}
Inspired by \emph{Rank Indexing Technique}\cite{DBLP:conf/icnp/HuaZLX08}, TinyTable\cite{DBLP:conf/icdcn/EinzigerF16} reorganizes space into chained fingerprints and counters.
Using the same block for both fingerprints and counters, Tinytable chains associate counter and fingerprints where each chain always starts with a fingerprint and that counterparts are always associated with the fingerprint to the left if necessary.
In order to handle bucket overflows, TinyTable dynamically modifies the bucket size according to the actual load in that specific bucket.
If there is not enough space in the bucket, Tinytable will borrow space from the following buckets by the \emph{Bucket Expand} operation, which takes its inspiration from a linear probing hash table.
This strategy may cause the bucket to overflow again, in that case, TinyTable repeats the process.
However, the complexity of additions and removals increases with the table load.
Compared with SBF\cite{DBLP:conf/sigmod/CohenM03}, TinyTable only requires a single hash function and accesses memory in add/remove operations in a serial manner.
We also note that, Counting Quotient Filter (CQF)\cite{DBLP:conf/sigmod/PandeyBJP17} augments variable-sized counters in quotient filter\cite{DBLP:journals/pvldb/BenderFJKKMMSSZ12} just like TinyTable\cite{DBLP:conf/icdcn/EinzigerF16} does.
%The strategies, i.e., resizability and variable-sized counters, enable the CQF to use substantially less space for Zipfian and other skewed input distributions. In the CQF, if a particular element occurs multiple times, then the slots immediately following that element's remainder will hold an encoding of the number of times that element occurs.
To make CQF work, an ``\emph{escape sequence}" is proposed to determine whether a slot holds a remainder or part of a counter and how many slots are used by that counter.

\textbf{\emph{Bits borrow or combination.}} Adjacent Borrow and Carry (ABC)\cite{DBLP:conf/bigdataconf/Gong0ZYC0L17} takes a novel bit reorganization strategy, which is designed for non-uniform multiset.
The experiments on real-world datasets show that when an elephant flow is recorded, the neighboring counters are often empty or occupied by mouse flows.
As a result, the unused bits can be borrowed by the adjacent elephant flows to enlarge their count ranges.
By borrowing bits from the adjacent counter or combining the mapped counter with its adjacent counter into a big counter, ABC can improve the memory efficiency at the cost of three bits to mark combined counters.
However, the encoding of ABC is cumbersome and slows down sketch significantly.
% of the sketch. %With complicated operations of $bits$-$borrowing$ and $combination$, ABC can make full use of space.
Self-Adjusting Lean Streaming Analytics (SALSA)\cite{DBLP:journals/corr/abs-2102-12531} dynamically re-sizes counters by merging neighbors to represent larger numbers.
Compared with ABC, SALSA only needs one bit to mark combined counters and encodes more efficiently.

The variable-width counter provides more flexible and elastic space efficiency for sketch design.
With an overhead of complicated operations, bit-shifting expansion\cite{DBLP:conf/sigmod/CohenM03}, rank-index-technique\cite{DBLP:journals/ton/HuaXLZ11}\cite{DBLP:conf/icdcn/EinzigerF16}, bit-borrowing/combination\cite{DBLP:conf/bigdataconf/Gong0ZYC0L17} and fingerprint-counter chain\cite{DBLP:conf/icdcn/EinzigerF16}\cite{DBLP:conf/sigmod/PandeyBJP17} all provide variable-sized counters, which enables the counters to use substantially less space for \emph{Zipfian} and other skewed flow distributions.
Such variable-width counters make a proper tradeoff between space efficiency and the encoding operation overhead.

\subsubsection{Optimization for less memory access}\label{access}

 \

While sketches can provide an estimation from multiple shared counters, they can not achieve both high accuracy and line speed at the same time.
Accessing multiple counters requires multiple memory accesses and hash computations.
Frequent memory access can become the bottleneck of sketch-based measurement.
Several techniques have been developed to reduce memory access and hash computation when updating or querying sketches to handle this dilemma.

OM sketch\cite{DBLP:conf/globecom/ZhouLJ0DL17} leverages the $word\ acceleration$ strategy to constrain the corresponding counters within one or several machine words and achieves nearly one memory access for each insertion. Specifically, for the lower layer $L_l$, $d_l\mathrm{+}1$ hash functions are associated with it.
The first hash functions are used to locate one machine word at the low layer, and the other $d_l$ hash functions are used to locate the $d_l$ counters in this machine word.
For the higher layer $L_h$, $d_h\mathrm{+}2$ hash functions are associated with it. The first two hash function is used to locate two machine words at the high layer, and the other $d_h$ hash functions are used to locate the $d_h$ counters in this machine word.
Cold filter\cite{DBLP:conf/sigmod/Zhou0J0YLU18} also proposes its $one$-$memory$-$access$ strategy to access the low layer $L_1$.
Specifically, it confines the $d_1$ counters within a machine word of $W$ bits to reduce the memory access.
Cold filter further splits the value produced by a hash function into multiple segments, and each segment is used to locate a machine word or a counter.
In this way, OM Sketch and Cold filter can locate multiple counters within one machine word.
To reduce the time consumption of memory access, Recyclable Counter with Confinement (RCC)\cite{DBLP:journals/ton/NyangS16} also limits a virtual vector to one memory block so that it requires only one memory block access to read and write a virtual vector.
To maintain only one memory access for each bucket operation, Cuckoo counter\cite{DBLP:conf/ancs/QiLYLL19} configures each bucket as 64 bits.
Furthermore, to efficiently handle skewed data streams, Cuckoo counter implements entries with unequal lengths to insulate mice flows from elephant flows.

By confining the counters for a specific flow within one or several words, sketches can limit the memory access and speed up insertion and query overhead.
The above sketches achieve less memory access overhead at the cost of lower accuracy.

%Recyclable Counter (RCC)\cite{DBLP:journals/ton/NyangS16} has both a deterministic counting segment, which is a simple accumulator and probabilistic counting segments depending on Linear counting\cite{DBLP:journals/tods/WhangVT90} and CSE\cite{DBLP:journals/ton/YoonLCP11}. A virtual memory block allocated to a flow in RCC is recyclable to prevent saturation and is confined to one memory block for speed constraint. Whenever a memory block for a flow reaches its maximum capacity, its estimation result is accumulated in the deterministic counting segment. Moreover, there is the built-in data structure that holds flow labels and their sizes, so it is easy to retrieve the list of large flows directly from the table. In terms of encoding speed, RCC uses about one memory access and one hash computation. Unlike other previously proposed schemes, for decoding, RCC demands about three memory accesses and two hash calculations.

\subsubsection{Counter update strategies for higher accuracy}\label{update}

\

Novel counter updating strategies are proposed to improve the estimation accuracy of sketches.
In this subsection, we review such strategies by classifying them as CM-Sketch variants, top-$k$ detection and various-hash design.

\textbf{\emph{CM-Sketch variants.}} Conservative update\cite{DBLP:journals/tocs/EstanV03} increases counters as less as possible.
%An incoming packet will only increase the smallest of the counters, while other counters are set to the maximum of their old values and the updated smallest counter.
The intuition is that, since the point query returns the minimum of all the $d$ values, CM should update a counter only when it is necessary.
This design avoids unnecessary update of counter values to lessen the over-estimation problem.
CM-CU sketch also applies a conservative update method.
Specifically, CM-CU updates a flow $f$ with size $c$ as ${\max \left\{ sk\left[ k,h_k\left( f \right) \right] ,\ \widehat{c}\left( f \right) \mathrm{+}c \right\}}$, where $ sk\left[ k,h_k\left( f \right) \right]$ represents the value of counters before updating, and $\widehat{c}\left( f \right)$ is the query result of flow $f$ before updating, i.e., $\forall \ 1\le k\le d, \widehat{c}\left( f \right)\mathrm{=} min\{sk\left[ k,h_k(f) \right]\}$.
SBF\cite{DBLP:conf/sigmod/CohenM03} also optimizes with the ``\emph{minimum increase}" scheme.
The minimum increase scheme prefers conservative insertion, i.e., only increase the smallest counter(s).
 %As for the recurring minimum scheme, it further initializes a secondary SBF to store the elements with a single minimum. Therefore, the error rate of query can be further decreased.
Count sketch (CSketch)\cite{DBLP:journals/pvldb/CormodeH08} introduces one more hash function $g_k()$ for the $k^{th}$ array to map flows onto $\left\{ +1,-1 \right\}$.
When updating a flow $f$ with size $c$, the corresponding counter will be increased by $c\cdot g_k\left(  \right)$.
CSketch returns the median over $d$ counters, i.e., $\forall \ 1\le k\le d, median\{sk\left[ k,h_k(f) \right]\cdot g_k(f)\}$, as an unbiased estimate of the point query, where $sk\left[ k,h_k(f) \right]$ represents the count of flow $f$ in the $k^{th}$ array.
Counter Sum estimation Method (CSM sketch)\cite{DBLP:conf/infocom/LiCL11} splits hot items into small pieces and stores them into small counters.
Flows are hashed into \emph{l} counters, and the flow size is divided into \emph{l} roughly-equal shares, each of which is stored in one counter.
The CSM sketch randomly increments one of the hashed counters during insertions and reports the sum of all the hashed counters subtracted by the noise during queries.
Reviriego et al.\cite{DBLP:journals/tc/ReviriegoMO21} first propose protection techniques to minimize the soft errors that flips the contents of memory cells. They evaluate the effect of soft errors on CM Sketch and propose to leverage upper counter bit encoding a single parity bit per counter to detect errors.
%This algorithm sacrifices its accuracy for high speed.

\textbf{\emph{Top-$k$ detection.}} %Various methods have proposed to update the fixed-size item collection.
Randomized admission policy (RAP)\cite{DBLP:conf/infocom/Ben-BasatEFK17} and Frequent\cite{DBLP:conf/esa/DemaineLM02} update the fixed key-value table by a randomized method for top-$k$ identification and frequency estimation.
When the table is full, RAP kicks out the item \emph{c} with the minimum counter value ${c_m}$ with probability ${\frac{1}{c_m+1}}$; otherwise, the incoming item will be discarded.
Frequent decreases all counters and evicts the zero-number flows to make space for new elements.
While Space Saving\cite{DBLP:conf/icdt/MetwallyAA05} always evicts the item with the minimal count directly.
Unbiased space saving\cite{DBLP:conf/sigmod/Ting18} has more robust frequent item estimation properties
than Sample and Hold and gives unbiased estimates for any subset-sum.
CountMax\cite{DBLP:journals/ton/YuXYWH18} and MV-sketch\cite{DBLP:conf/infocom/TangHL19} apply the majority vote algorithm (\emph{MJRTY})\cite{DBLP:conf/birthday/Moore91} to track the candidate heavy flow in each bucket.
When the incoming packet is the same as the recorded, the corresponding counter will be increased; otherwise, decreased.
Furthermore, the heavy candidate item will be replaced by the new item if the counter is negative.
HeavyKeeper\cite{DBLP:journals/ton/0003ZLGUCL19} and HeavyGuardian\cite{DBLP:conf/kdd/0003GZZSL18} leverage a probabilistic method called ``\emph{exponential-weakening decay}".
When the incoming item does not match the stored items, the flow size will decay with a probability.
As a result, the elephants will be stored in the buckets, and the mouses will be decayed and replaced with high probability.
The core idea of these methods is to treat each element as the hot item and insert them into the top-$k$ data structure.
After the data structure space runs out, it gradually kicks out cold elements with a certain probability.
However, as most elements are cold ones, processing so many cold items incurs not only extra overhead but also causes significant inaccuracy for the ranking and frequency estimation of top-$k$ items\cite{DBLP:conf/bigcomp/GongTY0D0L18}.

\textbf{\emph{Various-hash design.}} Weighted Bloom filter\cite{DBLP:conf/isit/BruckGJ06} allocates $k_e$ hash functions for each element $e$ depending on the query frequency and its likelihood of being a member.
This strategy can also be gracefully integrated into the frequency estimators to maintain a set of elephant elements.
Frequency-aware Counting (FCM)\cite{DBLP:conf/icde/ThomasBAY09} addresses the problem of inaccuracy of low frequency flows.
FCM uses many hash functions per item by dynamically identifying low- and high-frequency stream items.
The low- and high-frequency items are classified by MG counter\cite{DBLP:journals/scp/MisraG82} which keeps track of counts of unique items.
This strategy is realized by setting two additional hash functions to determine the initial offset in rows and the gaps between two adjacent rows.
To make high-frequency items update fewer counters than low-frequency items, the gap of low-frequency items is smaller than the high-frequency items.
Thus the accuracy of low-frequency items will be increased.

Besides, there are various specific counter updating strategies, such as the counter updating related to the aforementioned four Section \ref{compress},\ref{virtual},\ref{hierar},\ref{variable}, sampling in Section \ref{sample}, and randomization in  Section \ref{random}. The details can be referred in the corresponding subsection.

 \subsubsection{Slot enforcement for various functionalities}

\

As shown in Fig. \ref{enforce}, because of the various measurement requirement, sketches have enforced their buckets with other cells to support different additional functions.
By augmenting timestamp, packet-header field counter, or key-$XOR$ cell, the sketch can support latency detection, heavy hitter traction, and key-enumeration functionality.

\begin{figure}
  \centering
  % Requires \usepackage{graphicx}
  \includegraphics[width=3.5in]{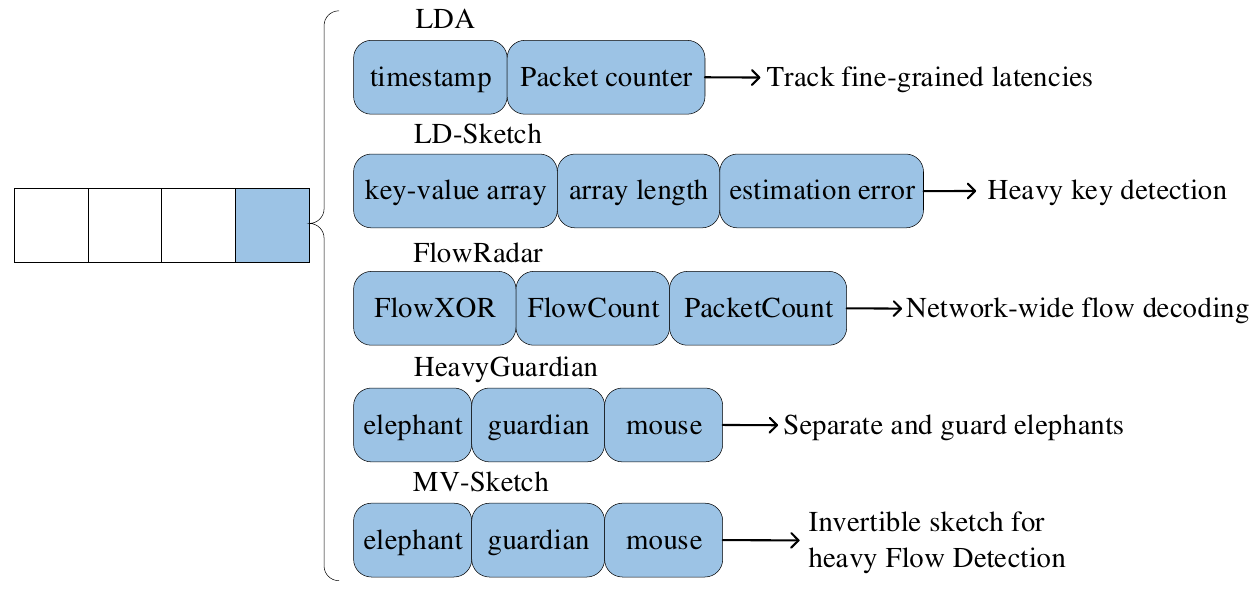}\\
  \caption{Slot enforcement strategy}\label{enforce}
\end{figure}

\textbf{\emph{Latency detection.}} Lossy Difference Aggregator (LDA)\cite{DBLP:conf/sigcomm/KompellaLSV09} is designed to measure packet loss and delay over short time scales.
LDA maintains timestamp accumulator-counter pairs in an array.
The packets are divided into groups according to the hash function. For each group, the sum of timestamps and the total count of packets are sent to the receiver.
Upon receiving such information, the receiver first applies the same hash function to divide packets into groups.
Suppose the total packet counter of the group at the receiver matches that at the sender, the receiver can calculate the average delay for such a group.
Otherwise, the receiver will discard such a group. %The average delay is calculated as follows. The sum of delays can be calculated by subtracting the sum of sending time stamps from the sum of receiving timestamps. Then the average delay can be calculated by dividing the sum of delay by the packet counter.
Therefore, LDA significantly reduces the measurement overhead by only transmitting the sum of timestamps and packet counter.
However, when packets are lost or reordered in a group, the entire group should be discarded; thus the delay for such a group cannot be estimated.

\textbf{\emph{Heavy hitter traction.}} LD-Sketch\cite{DBLP:journals/cn/HuangL15}
%comprises $r$ rows with $w$ buckets each. It further
augments each bucket $S_{i,j}$ with one additional component: $A_{i,j}$, an associative array to keep track of the hashed heavy key candidates, and three parameters: $V_{i,j}$ is incremented by the size of each incoming flow; $l_{i,j}$, the maximum length of the array; and $e_{i,j}$, the maximum estimation error for the true sums of the keys hashed to the buckets.
%In this way, when extracting information in the detection procedure, we can only focus on the associative array.
%For each incoming packet, LD-Sketch inserts the packet or evicts it, or reports the flow if its value exceeds a threshold $T$.
To keep tracking of only heavy flows candidates, LD-Sketch need to create or delete the key-value in the associative array and $l_{i,j}\mathrm{=}\lfloor \left( k+1 \right) \left( k+2 \right) -1 \rfloor$, in which $k\mathrm{=}\lfloor V_{i,j}/T \rfloor$.
%Further, LD-Sketch generates a pair of estimates for each key $x$ and each bucket ($i,j$): the lower estimate $S_{i,j}^{low}\left( x \right)$ and the upper estimate $S_{i,j}^{up}\left( x \right)$. If $x\in A_{i,j}$, $S_{i,j}^{low}\left( x \right)\mathrm{=}A_{i,j}\left[ x \right]$; otherwise $S_{i,j}^{low}\left( x \right)\mathrm{=}0$. And set $S_{i,j}^{up}\left( x \right)\mathrm{=}S_{i,j}^{low}\left( x \right)\mathrm{+}e_{i,j}$. For heavy hitter detection, at the end of each epoch, LD-Sketch first identify the bucket ($i,j$) with $V_{i,j}\ge \phi$, and then examine the keys kept in $A_{i,j}$. A key $x$ is identified as heavy hitter if $S_{i,j}^{up}\left( x \right)\ge \phi$ for all row $i$, where $1\le i\le r$, and $j\mathrm{=}f_i\left( x \right)$. For heavy changer detection, two LD-Sketch need to be maintained for two adjacent epochs. At the end of second epoch, the estimate change is given as $D_{i,j}\left( x \right) =\max \left\{ S_{i,j}^{up,1}\left( x \right) -S_{i,j}^{low,2}\left( x \right) ,S_{i,j}^{up,2}\left( x \right) -S_{i,j}^{low,1}\left( x \right) \right\}$. A key $x$ identified as heavy changer if $D_{I,j}\ge \phi$ for all row $i$, where $1\le i\le r$, and $j\mathrm{=}f_i\left( x \right)$.
However, the detection accuracy declines dramatically when the frequency of heavy flows is high\cite{DBLP:conf/lcn/ZhouZH19}.
At the beginning of detection, LD-Sketch is prone to cause frequent jitter of arrays. Moreover, the detection accuracy and memory are sensitive to the ordering of incoming keys.
The later work\cite{DBLP:journals/cn/HuangL15} proposes three enhancement heuristics to address this problem.
RL-Sketch\cite{DBLP:conf/lcn/ZhouZH19} further uses \emph{Reinforcement Learning} to evict trivial flows while leaving heavy key candidates.
MV-sketch\cite{DBLP:conf/infocom/TangHL19} augments each bucket with three components: $V_{i,j}$, the total sum of values of all flows hashed to the buckets; $K_{i,j}$ the current candidate heavy flow key in the bucket; $C_{i,j}$, which indicates if the candidate heavy flow should be kept or replaced.
Using a sufficient number of buckets, MV-sketch can significantly reduce the probability that two heavy flows are hashed into the same bucket and hence accurately track multiple heavy flows.
HeavyGuardian\cite{DBLP:conf/kdd/0003GZZSL18} enlarges each bucket to store multiple key-value pairs as heavy part and several small counters as light part.
A heavy part is used to precisely store frequencies of hot items, and a light part is used to store the frequencies of cold items approximately.
The heavy part takes \emph{Exponential Decay} strategy to decrement the count of the weakest items with a probability.
In this way, HeavyGuardian can intelligently separate and guard the information of hot items while approximately record the frequencies of cold items.

\textbf{\emph{Key-reversible functionality.}}
FlowRadar\cite{DBLP:conf/nsdi/LiMKY16} extends the Invertible Bloom Filter (IBF)\cite{DBLP:conf/sigcomm/EppsteinGUV11} design with three fields, FlowXOR, PacketCount, and FlowCount.
 It records the number of flows and number of packets that have been hashed into and $XORs$ the corresponding flow identifiers together in a cell.
 When a packet arrives, FlowRadar first checks the flow filter to see if the flow has been stored in the flow set or not.
 If the packet belongs to a new flow, all three fields will be updated. Otherwise, only PacketCount will be increased.
 When the collector receives the encoded flows, it takes the decoding strategy to derive the flow ID reversely.
 Specifically, it can decode the per-flow counters by looking for cells that include a single flow (called pure sell).
 By locating the cells, including this flow, FlowRadar removes them from flow set.
 This process ends when there are no pure cells.

By augmenting the counter with other fields, sketches can support various functionalities.
In the future, with the development of the network, there will be new-generated functionalities that require other fields in a cell/bucket.

\subsection{Sketch level optimization}
After summarizing the counter-level optimization, the base of sketch, we will get into details of the integral-sketch design.
The sketch-level optimization can be divided into multi-layer composition, multi-sketch composition and sliding window design.
The researchers have developed novel algorithms to enlarge the sketch functions skillfully by composing multiple basic units or different sketches together.

 \subsubsection{Multi-layer composition}\label{layer}

 \

Many work have proposed multi-layer design based on one basic unit to enable different measurement tasks.
We list the related work in Table. \ref{multi-layer}, which contains the basic unit and the enabled function.

 % Table generated by Excel2LaTeX from sheet 'Sheet1'
\begin{table*}[htbp]
  \centering
  \caption{Multi-layer design}
    \begin{tabular}{lllp{33em}}
    \hline
    \multirow{11}[4]{*}{Multi-layer} & Methods & Basic unit & \multicolumn{1}{l}{Enabled function} \bigstrut\\
\cline{2-4}          & Defeat\cite{DBLP:conf/imc/LiBCDGIL06} & Histogram-sketch & \multicolumn{1}{c}{Apply the subspace method\cite{DBLP:conf/sigcomm/LakhinaCD05} to detect anomalies} \bigstrut[t]\\
          & MRSCBF\cite{DBLP:journals/jsac/KumarXW06} & SCBF\cite{DBLP:journals/jsac/KumarXW06} & \multicolumn{1}{l}{Improve the estimate accuracy by different resolution} \\
          & IMP\cite{DBLP:conf/imc/ZhaoLOSWX07} & $L_p$ Sketch\cite{DBLP:conf/focs/Indyk00} &\multicolumn{1}{l} {Deduce the entropy of the intersection of two traffic} \\
          & Sequential Hashing\cite{DBLP:journals/cn/BuCCL10} & Hash table & \multicolumn{1}{l}{Enable key reversible function} \\
          & RCD sketch\cite{DBLP:conf/globecom/GuanWQ09} & Bit array & \multicolumn{1}{l}{Reverse the host with large connection degree} \\
          & SF-sketch\cite{DBLP:conf/icde/YangLYSSLCX17} & CM Sketch, Count Sketch & \multicolumn{1}{l}{Maintain a small sketch for transmission} \\
          & HashPipe\cite{DBLP:conf/sosr/SivaramanNRMR17} & match-action hash table & \multicolumn{1}{l}{Use P4Switch to HH detection} \\
          & SketchLearn\cite{DBLP:conf/sigcomm/HuangLB18} & Bit-level sketch & \multicolumn{1}{l}{Leverage multiple bit-level Gaussian distributions} \\
          & Diamond sketch\cite{DBLP:conf/infocom/0003GSWSL18} & Atom sketches & \multicolumn{1}{l}{Perform per-flow measurement on skewed traffic} \bigstrut[b]\\
    \hline
    \end{tabular}%
  \label{multi-layer}%
\end{table*}%

%Two level filtering\cite{DBLP:conf/ndss/VenkataramanSGB05} maintains two levels of filters to track the super spreader. The first level filters out \emph{SrcIP} that contact only a small number of distinct \emph{DstIP} with a small sampling probability, which is smaller than the second filter. Compare with one level filtering, two level filtering is more space-efficiency. The real super spreaders can be captured by both levels, and the \emph{SrcIP} which contact with a few \emph{DstIP} can be filtered out efficiently in the first level.

\textbf{\emph{For entropy estimation.}}
Defeat\cite{DBLP:conf/imc/LiBCDGIL06} leverages the entropy of the empirical distribution of traffic features to detect unusual traffic patterns.
It treats the abnormalities as unusual distributions of these features.
Defeat generates multiple sketches of \emph{SrcIP}, \emph{SrcPort}, \emph{DstIP}, \emph{DstPort} separately.
Moreover, multiple histogram-sketches from different switches are summed to form a global sketch, which can be utilized to detect, identify, and classify attacks.
However, the complex construction of empirical histograms makes it less appealing for online realization\cite{DBLP:journals/jsac/KallitsisSBM16}.
Intersection Measurable Property (IMP Sketch)\cite{DBLP:conf/imc/ZhaoLOSWX07} observes that the entropy of an OD flow can be estimated by a function of just two different ${L_p}$ norms ($L_{p1}, L_{p2}, p_1\ne p_2$) of OD flows.
IMP can deduce the entropy of the intersection of two streams $A$ and $B$ from the sketches of $A$ and $B$.
Specifically, every participating ingress and egress node in the network maintains a sketch of the traffic passing by it during each measurement interval.
When measuring the entropy of an OD flow between an ingress $i$ and an egress point $j$, IMP collects the sketches of ingress stream $O_i$ and egress stream $D_j$. It leverages ${L_p}$ sketch\cite{DBLP:conf/focs/Indyk00} to derive the ${L_p}$ norm of the traffic stream based on the property of the median and estimates the entropy of OD flows accordingly.

\textbf{\emph{For more accurate.}}
Diamond Sketch\cite{DBLP:conf/infocom/0003GSWSL18} composes atom sketches (counter array) into three parts: increment part, carry part, and deletion part.
The increment part records the size of the flow.
The carry part records the overflow depth of each flow.
The deletion part is used to support deletions.
It assigns an appropriate number of atom sketches to the elephants and mouses dynamically.
% where the elephant use more layers of atom sketches than the mouse.
This strategy improves the accuracy considerably while keeping a comparable speed.
Slim-Fat (SF sketch)\cite{DBLP:conf/icde/YangLYSSLCX17} takes a small-large sketch composition to relax bandwidth requirement when the local monitor transfers information to the remote controller.
The large-sketch is used to assist small-sketch during insertions and deletions to improve accuracy within a limited space. In contrast, the small sketch is transferred and used to answer queries.
Multi-Resolution Space-Code Bloom Filter (MRSCBF)\cite{DBLP:journals/jsac/KumarXW06} employs multiple SCBFs but operating at different sampling resolutions.
SCBF can approximately represent traffic flows as a multiset.
With different sampling probability,
%flow items can be represented by different SCBFs. Thus
the elements with low multiplicities will be estimated by SCBF of higher resolutions, while high multiplicities will be estimated by the lower resolution SCBF.
However, MRSCBF is very slow in terms of memory access and hash computations.
And the bitmap nature of MRSCBF determines that it is not memory efficient for counting\cite{DBLP:journals/ton/ChenCC17}.

%Rflow\cite{DBLP:conf/infocom/JangCNN17}

%HashPipe\cite{DBLP:conf/sosr/SivaramanNRMR17} implements a pipeline of hash tables that maintain elephant identifiers while evicting cold flows over time. Hashpipe divides counters into disjoint tables using emerging programmable data planes and inserting new incoming flow into the first table. HashPipe uses pipelined operation to sample multiple locations in hash tables, and evicts cold items, with updates to precisely one location and lookup per stage.

\textbf{\emph{For reversible design.}}
SketchLearn\cite{DBLP:conf/sigcomm/HuangLB18} maintains ${l\mathrm{+}1}$ levels of sketch, each level corresponds to a sketch formed by a counter matrix with \emph{r} rows and \emph{c} columns.
The level-\emph{0} sketch records the statistics of all packets, while the level-\emph{k} sketch for ${1\leq k \leq l}$ records the statistics for the $k^{th}$ bit of the flowkey. $V_{i,j}[k]$ represents the counter value of the level-$k$ sketch at $i^{th}$ row and $j^{th}$ column.
SketchLearn focuses on ${R_{i,j}\left[ k \right] =\frac{V_{i,j}\left[ k \right]}{V_{i,j}\left[ 0 \right]}}$, which denotes the ratio of the counter value ${V_{i,j}[k]}$ to the overall frequency ${V_{i,j}[0]}$ of all flows hashed to stack (${i,j}$). By leveraging the Gaussian distribution assumption, SketchLearn separates the large and small flows.
Specifically, it uses \emph{Bayes' theorem} to estimate the bit-level probabilities ${\hat{p}\left[ k \right]}$, and further obtains a template flowkey composed of zero, one and ${\ast}$.
By estimating frequencies of each \emph{k}-bit using maximum likelihood estimation, SketchLearn gets frequency estimate of candidate flows.
%Further, it associates flowkeys with bit-level probabilities to qualify the correctness of flowkeys and remove false positives.

Reversible connection degree (RCD Sketch)\cite{DBLP:conf/globecom/GuanWQ09} is a new data streaming method for locating hosts with a large connection degree.
Based on the remainder characteristics of the number theory, the in-degree/out-degree with a given host can be accurately estimated.
A connection degree sketch is denoted as $B\mathrm{=}\left( B_1,\cdots ,B_H \right)$. $B_i\left( 1\le i\le H \right)$ is a $v\times m_i$ bit arrays.
$B_i\left[ j \right] \left[ k \right] \left( 0\le j<v,0\le k<m_i \right)$ associates with a hash function $h_i:\left\{ 0,1,\cdots ,n\mathrm{-}1 \right\} \rightarrow \left\{ 0,1,\cdots ,m_i\mathrm{-}1 \right\}$, where $n$ is the size of the source space of packets.
All $B_i\left[ j \right] \left[ k \right] \left( 0\le j<v,0\le k<m_i \right)$ share a hash function $f_i:\left\{ 0,1,\cdots ,E\mathrm{-}1 \right\} \rightarrow \left\{ 0,1,\cdots ,v\mathrm{-}1 \right\}$, where $E$ is the size of combination of source and destination space, and $H$ is the number of data array.
The bits arrays are all set zero at the beginning of measurement epoch.
For an incoming packet $p_i\mathrm{=}(s_i,d_i)$, the sketch will be updated as, $\forall 1\le k\le H,\ B_k\left[ f\left( s_i||d_i \right) \right] \left[ h_k\left( s_i \right) \right]\mathrm{=}1$.
Thus, for any source $s$, the packets with source $s$ are all hashed to column $B_i\left( s \right)\mathrm{=}B_i\left[ \cdot \right] \left[ h_i\left( s \right) \right] \left( 1\le i\le H \right)$ in the sketch.
Thus we can obtain the sum of the out-degrees of source $s$.
Moreover, the host addresses which are associated with large in-degree/out-degree can be reconstructed by simple equation based on \emph{Chinese Reminder Theory} without using address information.
Sequential Hashing\cite{DBLP:journals/cn/BuCCL10} constructs reversible multi-level hash arrays for fast and accurate detection of heavy items.
The idea of Sequential Hashing is that combining the heavy sub-bits of key can efficiently discover the full heavy items' key.
The original problem is divided into nested sub-problems.
The prefixes with different lengths of flow key are extracted as the sub-keys.
It enumerates each sub-key space and combines the recovered sub-keys to form the heavy flows.
However, the update costs of both Reversible Sketch\cite{DBLP:journals/ton/SchwellerLCGGZDKM07} and Sequential Hashing\cite{DBLP:journals/cn/BuCCL10} increase with the key length. Both of them utilize the position information to recover the keys in their sketch.
The running time to recover keys in them cannot achieve the sub-linear bound\cite{DBLP:conf/infocom/LiuCG12}.

\textbf{\emph{For universal sketching.}}
Univmon\cite{DBLP:conf/sigcomm/LiuMVSB16} maintains ${log(n)}$ parallel copies of a ``${L_2}$-heavy hitter" (${L_2}$-HH) sketch instance, where \emph{n} is unique elements in traffic flow stream.
Univmon creates substreams of decreasing lengths as the ${j^{th}}$ instance is expected to have all of the hash functions agree to sample half as often as the ${(j\mathrm{-}1)^{th}}$ sketch.
Each ${L_2}$-HH instance outputs ${L_2}$ heavy hitters \emph{Q} ($Q_0$, $\cdots$, $Q_{log(n)}$) and their estimated counts.
By leveraging the results from the ${L_2}$-HH instance, Univmon can use the Recursive Sum Algorithm from the theory of universal sketching to get an unbiased estimator of \emph{G}-sum functions of interest.
Specifically, it recursively applies the result from $Q_{log(n)}$ to $Q_0$ sketch to function \emph{g} to form the interested \emph{G}-sum using these sampled streams.
However, UnivMon incurs large and variable per-packet processing overhead, resulting in a significant throughput bottleneck in high-rate packet streaming.
ActiveCM+\cite{DBLP:conf/infocom/XiaoTC20} proposes a progressive sampling techniques to solve this problem.
Specifically, it only updates the last sketch where flow $f$ is sampled instead of variably multiple sub-sketches.
Moreover, the progressive sampling also reduces the moment estimation error by more than half that UnivMon.
In this way, ActiveCM+ reduces memory footprint and improves measurement accuracy.

 \subsubsection{Multi-sketch composition}

 \

 % Table generated by Excel2LaTeX from sheet 'Sheet1'
\begin{table*}[htbp]
  \centering
  \caption{Multi-sketch design}
    \begin{tabular}{llll}
    \toprule
    \multirow{7}[4]{*}{Multi-sketch} & Methods & Combined methods & Enabled function \\
\cmidrule{2-4}          & Count-Min Heap\cite{DBLP:conf/latin/CormodeM04} & Count-min Sketch+heap & Find top-$k$ elements \\
          & DCF\cite{DBLP:journals/sigmod/Aguilar-SaboritTML06} & SBF\cite{DBLP:conf/sigmod/CohenM03}+CBF\cite{DBLP:journals/ton/FanCAB00} & {Represent the multisets compactly} \\
          & FSS\cite{DBLP:journals/isci/HomemC10} & Bitmap filter+Space Saving\cite{DBLP:conf/icdt/MetwallyAA05} & Minimize updates on the top-$k$ list \\
          & PMC\cite{DBLP:conf/infocom/LievenS10} & FM sketch\cite{DBLP:journals/jcss/FlajoletM85}+HitCounting\cite{DBLP:journals/tods/WhangVT90} & High-speed per-flow measurement \\
          & ASketch\cite{DBLP:conf/sigmod/RoyKA16} & Filter+Sketch(CM\cite{DBLP:journals/pvldb/CormodeH08}, Count sketch\cite{DBLP:conf/latin/CormodeM04}) & {Improve accuracy and overall throughput} \\
          & Bloom sketch\cite{DBLP:conf/infocom/ZhouJLZ0L18} & BF\cite{DBLP:journals/cacm/Bloom70}+multi-levels (Section \ref{hierar}) & Memory-efficient counting \\
    \bottomrule
    \end{tabular}%
  \label{Multi-sketch}%
\end{table*}%

By composing different data structures, sketch-based methods can further support various functions. We also list the related work and the corresponding basic sketch in Table. \ref{Multi-sketch}.

\textbf{\emph{For memory efficiency.}}
Dynamic count filters (DCF)\cite{DBLP:journals/sigmod/Aguilar-SaboritTML06} extends the concept of SBF\cite{DBLP:conf/sigmod/CohenM03} and improves the memory efficiency of SBF by using two filters.
The DCF is designed for speed and adaptiveness in a straightforward way.
It captures the best of SBF and CBF, the dynamic counters from SBF, and the fast memory access from CBF.
The first filter is composed of fixed size counters.
The novelty is that the size of counters in the second filter is dynamically adjusted to record the corresponding counters' overflow times in the first filter.
Unfortunately, the two filters increase the complexity of DCF, which degrades its query and update speed\cite{DBLP:conf/icde/YangLYSSLCX17}.

\textbf{\emph{For hot-cold separation.}}
Probabilistic Multiplicity Counting (PMC)\cite{DBLP:conf/infocom/LievenS10} leverages FM sketch\cite{DBLP:journals/jcss/FlajoletM85} to measure hot items and modified HitCounting\cite{DBLP:journals/tods/WhangVT90} to estimate cold items.
Specifically, the FM sketches of all estimators share their bits from a common bit pool uniformly at random so that mostly unused higher-order bits in the registers can be utilized.
PMC can record information on a passing-by packet by setting only one single bit in the packet header field.
Based on the packet header, a specific hashing mechanism determines the enabled bit position.
Moreover, PMC is designed initially to estimate flow size, but it can be easily modified for flow cardinality estimation, which other flow-size estimators do not support.
Count-Min Heap\cite{DBLP:conf/latin/CormodeM04} augments CM Sketch with a heap used to track all heavy flows and their estimated frequency.
If the incoming flow whose estimation exceeds the threshold, it is added to the heap or replaces the minimum flow when the heap is full.
The heap is kept small by checking that the estimated count for the item with the lowest count is above the threshold.
If not, the item is deleted from the heap.
At the end of each epoch, all the heavy hitters can be output by scanning the whole heap.
ActiveCM+\cite{DBLP:conf/infocom/XiaoTC20} also augments min-heap to track the top-$k$ heavy hitters within each sub-sketch.

\textbf{\emph{For filtering packets.}}
Filtered Space-Saving (FSS)\cite{DBLP:journals/isci/HomemC10} augments Space Saving with a pre-filtering approach.
A bitmap is used to filter and minimize updates on the measured list.
When a new item is incoming, the bitmap counter is first checked if there are already measured items in the bitmap; and the measured list is searched to see if this element is already there.
The update is conducted according to the check result about the bitmap and measured list.
However, FSS needs to search a certain flow key or to identify the minimum counter in the list.
It needs to traverse the whole heap in the worst case without additional memory support.
ASketch\cite{DBLP:conf/sigmod/RoyKA16} augments a pre-filtering stage to identify the elephants and leaves the mouses to sketch.
Based on the skewness of the underlying stream, Asketch improves the frequency estimation accuracy for the most frequent items by filtering them out earlier.
For the items that are stored in the sketch, Asketch reduces their collisions with the high-frequency items, as the items stored in the filter are no longer hashed into the sketch.
This reduces the possibility that a low-frequency item would appear as a high-frequency item, therefore reducing the misclassification error.
Bloom Sketch\cite{DBLP:conf/infocom/ZhouJLZ0L18} is composed of multiple layers of sketch$\mathrm{+}$BF composition.
The low layers are responsible for low-frequency items, and the high layers process the items whose multiplicity can not be counted by the lower sketches.
Moreover, the BFs record whether the high layers record the items.
It leverages BF to recognize items whose multiplicity is not successfully recorded in the low layer.
By this way, the low-frequency items are recorded in the low layer, while high-frequency items are recorded in both low and high layers.
Sketchtree\cite{DBLP:journals/tpds/WangXHZ20} comprises multiple filters, each associated with a specific task to measure both elephants' flows efficiently and aggregate mice flow.
Specifically, Sketchtree is task-oriented, and each filter is responsible for sending the related flows to the corresponding sketch and others to the next filter for other measurement tasks.

 \subsubsection{Sliding window design}\label{sliding}

 \

Traditional sketch-based methods focus on estimating flow sizes/volumes from the beginning of traffic stream (landmark window model)\cite{DBLP:conf/iwqos/ZhouZCZ17}.
In this model, beginning at a ``\emph{landmark}" time point, sketch-based methods mainly focus on the data which falls between the landmark and the current time point.
With time passing, more and more packets pass through the monitor, the sketch runs out of space and has to be reset periodically\cite{DBLP:conf/imc/GolabDDLM03}, which is the intrinsic disadvantage of this model.
For many real-time applications, the most recent elements of a stream are more significant than that arrived a long time ago.
This requirement gives rise to the sliding window sketch-model.
Based on the sketch-based methods, the sliding strategy removes the expired elements as the incoming of new elements, thereby it always maintains the most recent $W$ elements in the data stream.
A lot of effort has been invested in this area to get the most recent information.

%FCM\cite{DBLP:conf/icde/ThomasBAY09} divides traffic stream into windows of size \emph{k} and maintains a MG counters to record the frequency of items. After the ${i^{th}}$ window, the item with count less than (\emph{i} + 1) will be deleted in the list. While processing the ${(i + 1)^{th}}$ window, if a new entry is observed then its initial count is set as \emph{i}, which is the maximum number of occurrences of the item. By this design, the MG counters can classify the item and prepare for inserting items into sketch.

\emph{\textbf{First-in-first-out design.}}
This strategy inserts the new elements and evicts the stale ones.
Sliding HyperLogLog\cite{DBLP:conf/icdm/ChabchoubH10} aims to estimate the number of distinct flows over the last $w$ unit of time, which is smaller than the time window $W$.
A list of arrival timestamps and the position of the leftmost 1-bit in the binary representation of the hashed value associated with the packet are maintained to record the recent packets' information in $W$ time units.
When packets are out of the window, they are evicted from the list, and the incoming item is updated.
Although maintaining the timestamp exactly is the most accurate strategy for the recent information, it incurs a heavy space overhead.
The randomized aging strategy may be a better tradeoff between the estimation accuracy and maintenance overhead.
Sliding Window Approximate Measurement Protocol (SWAMP) \cite{DBLP:conf/infocom/AssafBEF18} stores flow fingerprints in a cyclic buffer and the frequencies are maintained in TinyTable\cite{DBLP:conf/icdcn/EinzigerF16}.
The cyclic corresponds to a measurement window $W$.
Within the insertion of a new element, its fingerprint replaces the oldest items in the buffer; furthermore, the corresponding count in TinyTable will be decreased, while the counter of arriving flow will be increased.

%LC\cite{DBLP:conf/icdm/ChabchoubH10} and PLC\cite{DBLP:journals/ccr/DimitropoulosHK08} split traffic stream into fixed-size windows and process each window sequentially to find heavy hitters. For an incoming packet, if it is already in the table, the corresponding counter is updated; otherwise, it is inserted into the table. After one measurement epoch, the data structure will evict the cold items in the record according to defined error bound and the frequency.

\textbf{\emph{Partitioned blocks design.}}
By applying partitioned blocks to record the information, the sketch can randomly perform the aging operation in the oldest block and update the operation in the newest block.
Hokusai\cite{DBLP:conf/uai/MatusevychSA12} uses a series of CM Sketch for different time intervals.
It uses large sketches for the recent intervals and small sketches for older intervals to adapt the error within a limited space.
As time passing, a large sketch can be halved into a small sketch by adding one half of the sketch to the other half.
Window Compact Space-Saving (WCSS)\cite{DBLP:conf/infocom/Ben-BasatEFK16} divides traffic stream into frames of size $W$.
%Each frame is then partitioned into $k$ equal-sized blocks.
%The window of interest is also of size $W$.
It maintains a queue for each block that overlaps with the current window.
When a block no longer overlaps with the window, the related queue will be removed.
% Furthermore, WCSS adds an empty queue for the next block.
 This strategy keeps the number of queues constant at all times.
WCSS supports point queries with constant-time
complexity under the sliding window model. %This capability is essential for frequency estimation of an arbitrary flow in real-time.
Segment Aging Counter Estimation (S-ACE) \cite{DBLP:conf/iwqos/ZhouZCZ17} uses multiple data synopses to store the elements arriving in different window segments.
This design can maintain the relative order between the window segments with their data synopses and improve the probability of deleting the correct outdated elements.
HyperSight\cite{DBLP:journals/jsac/ZhouBYGCZWZ20} also applies a partitioned Bloom filter to maintain a coarse-grained time order for testing packet behavior changes.
However, this kind of strategy keeps too many sketches which incur expensive memory overhead.

\textbf{\emph{Randomized aging design.}}
Aging Counter Estimation (ACE)\cite{DBLP:conf/iwqos/ZhouZCZ17} maintains the most recent $W$ elements and adopts the counter sharing idea\cite{DBLP:conf/infocom/LiCL11} to the sliding window model.
For ACE, an aging algorithm is proposed to eliminate one element when a new element comes.
It randomly picks a counter in the array and decreases it by one if applicable.
It is simple and efficient.
However, as $t$ becomes large, the sliding window accumulates many expired elements, introducing more noise for size estimation.
%it requires resetting the sliding window periodically. To achieve persistently accurate per-flow counting without periodical sliding window resetting,
Memento\cite{DBLP:conf/conext/Ben-BasatEKOVW18} is a family of sliding window algorithms for the HH and HHH problems in the single-device and network-wide settings.
Specifically, for each packet, Memeto performs a Full update operation (delete the stale data and add the new) with a probability or make a window update (delete outdated data).

\textbf{\emph{Time adaptive updating.}} Ada-Sketch\cite{DBLP:conf/sigmod/ShrivastavaKB16} proposes a time adaptive scheme which is inspired by the well-known digital Dolby noise reduction procedure.
Specifically, when updating, Ada-Sketch applies pre-emphasis and artificially inflates the counts of more recent items compared to the older items, i.e., multiplying the updates $c_{i}^{t}$ with $f(t)$, which is a monotonically increasing function of time $t$.
Then when querying, Ada-Sketch applies de-emphasis to estimate the frequency of item $i$ at time $t$, i.e., dividing the query results by $f(t)$.

\textbf{\emph{Scanning aging design.}}
Sliding sketch\cite{slidingsketch} maintains multiple counters in each cell and updates the most recent counter.
Unlike the above strategies, Sliding sketch adapts scanning operation to delete the out-dated by using a scanning pointer upon the sketch. The speed of the scanning pointer is determined by the length of the sliding window.
However, the process of scanning pointer will influence the accuracy of estimation a lot.

\subsection{Summary and lessons learned}

In this section, we detail the structure optimization in terms of hashing strategy, counter level optimization and sketch level optimization.
Until now, there are a lot of work developed to enforce the sketches.
From optimization techniques from flow-to-counter rules and basic counter unit to the whole sketch, the sketches can be updated with the techniques described in this section.
In the future, we believe learned hash-map\cite{DBLP:conf/sigmod/KraskaBCDP18}\cite{DBLP:journals/corr/abs-1802-00884}\cite{DBLP:conf/adcs/OosterhuisCR18} will play an important role in flow-to-counter rules design because of less space overhead and hash collision errors.
And the novel counter and sketch architecture design will be guided by traffic characteristics and the development of architecture in the network.

\section{Optimization in post-processing stage}\label{post}

After recording traffic statistics into sketches, we must perform some operations before obtaining valuable information from the sketches, such as sketch compression and merging.
Furthermore, some novel information extraction techniques are proposed to enrich the sketch functionality. In this section, we survey the optimization in the post-processing stage.

\subsection{Techniques for sketch compression and merging}

Although sketches have been designed as synopses that require only a few memory accesses for each insertion, there remains scope for improvement in terms of reducing the bandwidth bottleneck.
Remote controllers require high-speed memory to receive filled sketches from multiple measurement nodes.
Consequently, (1) the size of the filled sketch that each measurement node sends to the collector should be extremely small, and (2) the compressed or merged sketch should contain sufficient information that allows the remote collector to answer queries accurately.
The accuracy with which a sketch can provide answers to queries quantifies how close the value of the estimation from the sketch is to the actual value of the frequency \cite{DBLP:conf/icde/YangLYSSLCX17}.
Several studies have proposed merging or compressing sketches before sending them to the collector to reduce bandwidth overhead.

\subsubsection{Sketch compression}

\

Several studies have proposed compressing sketches, as shown in Fig. \ref{merge}(a). Some studies have proposed compressing sketches when adapting to the available bandwidth before sending them in order to save space for time-adaptive updating.

\begin{figure}
\centering
\subfigure[Compression operation] {\includegraphics[width=3in]{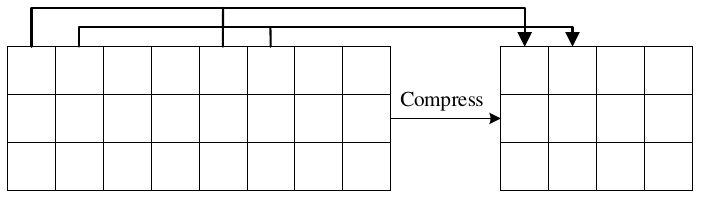}}
\subfigure[SUM merging operation] {\includegraphics[width=3in]{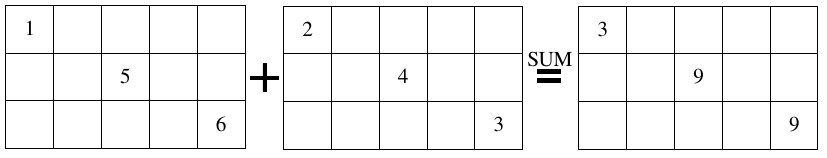}}
\caption{Sketch merging and compression}
\label{merge}
\end{figure}

\textbf{\emph{Compression for cardinality estimation.}}
Arithmetic-coding-based compression \cite{DBLP:conf/dialm/ScheuermannM07} uses arithmetic coding to compress FM sketches \cite{DBLP:journals/jcss/FlajoletM85} and HyperLogLog \cite{DBLP:conf/edbt/HeuleNH13} in a bit-wise and register-wise manner, respectively.
The compression scheme can be understood as a separation of the represented value and the significantly larger duplicate insensitivity information.
However, this approach was only designed for cardinality estimation sketches, and it does not meet the requirements for compressing CM sketches.

\textbf{\emph{Compression for transformation.}}
SF-sketch \cite{DBLP:conf/icde/YangLYSSLCX17} maintains a fat sketch and a slim sketch at a single measurement point.
The fat sketch is responsible for recording traffic flow accurately and allows the slim sketch to improve accuracy with less space overhead.
When inserting or deleting an item, SF-sketch first updates the fat sketch and then updates the slim sketch based on the estimation from the fat sketch.
With this design, the traffic flow information can be compressed into a smaller sketch for transmission.
An elastic sketch \cite{DBLP:conf/sigcomm/0003JLHGZMLU18} first groups the counters in the CM sketch and then compresses the counters in the same group into a single counter. Given a sketch $S$ of size ${zw'\times d}$ (where ${w\mathrm{=}w'\mathrm{\times}z}$ is the width, $d$ is the depth, and $z$ is an integer representing the compression rate),
an elastic sketch splits $S$ into $z$ equal divisions or sub-sketches, each of which has a size of ${w'\mathrm{\times} d}$.
By combining the same index in its division (${\left\{ S_{i}^{k}\left[ j \right] \right\} _{k\mathrm{=}1,\cdots ,z}}$) in the same group, an elastic sketch builds a compressed sketch $B$ and a set ${B_i\left[ j \right] \mathrm{=}OP_{k\mathrm{=}1}^{z}\left\{ S_{i}^{k}\left[ j \right] \right\}}$ (${1\le i\le d,\ 1\le j\le z}$), where $OP$ is the compression operation (e.g., $Max$ or $Sum$).
The authors proved that after $Sum$ compression, the error bound of the CM sketch does not change, whereas after $Max$ compression, the error bound becomes tighter.
Same as Elastic Sketch\cite{DBLP:conf/sigcomm/0003JLHGZMLU18}, Cluster-reduce\cite{DBLP:conf/kdd/YikaiZhao21} also divides counters into groups based on adjacency.
After that, the counters in the nearby groups which have similar values are rearranged into clusters.\
And each cluster are then reduced to a unique representative value.
Cluster-reduce achieves high efficiency of compression procedure, support of direct query without decompression, and
high accuracy of compressed sketches simultaneously.

\textbf{\emph{Compression for time-adaptive updating.}}
Hokusai \cite{DBLP:conf/uai/MatusevychSA12} maintains a series of CM sketches for different intervals to estimate the frequency of any item for a given time or interval.
The sketch space for recent intervals is larger than that for older intervals so as to make the sketch time-adaptive.
As time passes, recent sketches are updated with incoming traffic and older sketches are derived by compressing the most recent sketches one by one.
Specifically, a sketch is compressed by adding one half of the sketch to the other half.
Although this is a natural solution for achieving time adaptability, it still has several shortcomings, such as discontinuity, inflexibility, excessive sketches, and large overhead for shrinking \cite{DBLP:conf/sigmod/ShrivastavaKB16}.

\subsubsection{Sketch merging}

\

For collaborative measurement, distributed sketches must be merged into a large summary, as shown in Fig. \ref{merge}(b).

\textbf{\emph{Merging for CM-based sketches.}}
For merging, homogeneous sketches can be added or subtracted directly using the same hash functions and structure for all the sketches, or isomeric sketches can be used to find the corresponding counters to obtain the final result with different hash functions and sketch sizes.
SketCh REsource Allocation for Measurement (SCREAM) \cite{DBLP:conf/conext/MoshrefYGV15} allocates different sketch sizes at each switch so that each sketch may have an array of different widths and cannot be summed directly.
SCREAM sums the minimum within each sketch to formulate the merged sketch in a distributed measurement by finding the corresponding counters for each flow.
 %This approximation is always more accurate because the sum of minimums is always smaller than the minimum of sums for positive numbers.

\textbf{\emph{Merging for top-$k$ summary.}}
Mergeable summaries \cite{DBLP:journals/corr/CafaroTP14} such as MG summary \cite{DBLP:journals/scp/MisraG82} and Space Saving \cite{DBLP:conf/icdt/MetwallyAA05} are mergeable and isomorphic.
For tracking the top-$k$ flows, two top-$k$ records can be merged by subtracting the MG counters, which are lower than the ${{(k+1)}^{th}}$ counter in the merged MG summary.
The procedure can be completed with a constant number of sorts and scans of summaries of size ${O(k)}$.

Inspired by In-band Network Telemetry (INT), LightGuardian\cite{DBLP:conf/nsdi/ZhaoYLYCLZWWWZ21} divides sketch into sketchlets, which can be encoded into packets headers for further aggregation, reconstruction and analysis in the end-hosts.
By this design, LightGuardian can collect per-flow per-hop information for all the flows in the network with negligible overhead.

\subsection{Techniques for information extraction}

\subsubsection{Flow key reversible techniques}\label{reversible}

\

Traditional methods only record the volume information of flow into sketches according to the hash result of flow keys and provide a fast scheme for insertion and querying of traffic flow.
However, no scheme can efficiently recover the flow key from the sketch itself except by enumerating the flow key space.
Both direct recordings of the flow key and enumeration are inefficient for traffic measurement because of space restrictions and real-time response requirements.
Many studies have proposed solutions to this problem using bit count, sub-key count, and IBF-based or other methods.

\textbf{\emph{Direct key recording.}}
LD-sketch \cite{DBLP:journals/cn/HuangL15} maintains a two-dimensional array of buckets and associates each bucket with an associative array to track the candidate heavy flows that are hashed to the bucket.
However, updating the associative array involves high memory access overhead, which increases with the number of heavy flows.
In particular, LD-sketch occasionally expands the associative array to hold more candidate heavy flows; nevertheless, dynamic memory allocation involves high costs and is challenging to implement in hardware. MV-sketch \cite{DBLP:conf/infocom/TangHL19} also enforces each bucket with a flow key field to record the candidate heavy flow and an indicator counter to implement the majority vote algorithm (\emph{MJRTY}).
Both LD-sketch and MV-sketch detect heavy key by directly recording the candidate flow keys. Nevertheless, direct recording is not a desirable strategy for per-flow measurement.

\textbf{\emph{Bit reversibility.}}
BitCount \cite{DBLP:conf/hpsr/WangG19} maintains a counter for each bit in the IP address to count the total number of 1-bits.
It can recover the heavy hitter keys by combining the bits whose counter is larger than a given threshold.
Meanwhile, Deltoid \cite{DBLP:journals/ton/CormodeM05} comprises multiple counter groups with 1+$L$ counters each (where $L$ is the number of bits in a key), in which one counter tracks the total sum of the group and the remaining $L$ counters correspond to the bit positions of a key.
It maps each flow key to a subset of groups and increases the counters whose corresponding key bits are 1.
To recover heavy flows, Deltoid first identifies all groups whose total sums exceed the threshold.
If each such group has only one heavy flow, the heavy flow can be recovered: the bit is 1 if a counter exceeds the threshold and 0 otherwise.
Fast Sketch \cite{DBLP:conf/infocom/LiuCG12} is similar to Deltoid \cite{DBLP:journals/ton/CormodeM05}, except that it maps the quotient of a flow key to the sketch.
An incoming flow is first added to the first counter in the hashed rows.
Moreover, the quotient function determines the counters in each row to add the flow size.
After checking the rows one by one and recovering the quotient bits that contribute to the heavy change, Fast Sketch can identify the heavy changes using the inverse function according to the quotient technique.
SketchLearn \cite{DBLP:conf/sigcomm/HuangLB18} combines the sketch design of Deltoid \cite{DBLP:journals/ton/CormodeM05} with automated statistical inference.
Specifically, it maintains a data structure similar to Deltoid and provides a fundamental property of bit-level Gaussian distributions.
It proves that if there is no large flow, the bit counter values will follow a Gaussian distribution.
SketchLearn extracts large flows from multi-level sketches and leaves the residual counters for small flows to form Gaussian distributions.

\textbf{\emph{Sub-key reversibility.}}
Reversible Sketch \cite{DBLP:journals/ton/SchwellerLCGGZDKM07} finds heavy flows by pruning the enumeration space of flow keys.
It divides a flow key into smaller sub-keys that are hashed independently and concatenates the hash results to identify the hashed buckets.
To recover heavy flows, it enumerates each sub-key space and combines the recovered sub-keys to form the heavy flows.
Sequential Hashing \cite{DBLP:journals/cn/BuCCL10} follows a design similar to that of Reversible Sketch \cite{DBLP:journals/ton/SchwellerLCGGZDKM07}; however, it hashes key prefixes of different lengths into multiple smaller sketches. The update costs of both Reversible Sketch and Sequential Hashing increase with the key length.

\textbf{\emph{IBF variants.}}
FlowRadar \cite{DBLP:conf/nsdi/LiMKY16} takes an extension of IBF \cite{DBLP:conf/sigcomm/EppsteinGUV11} as the encoded flowset. Each counting table consists of three fields, namely \emph{FlowXOR}, \emph{FlowCount}, and \emph{PacketCount}, which record the $XOR$ of all flows, the number of flows, and the number of packets of all flows mapped in the bucket, respectively.
By finding the bucket that includes just one flow (called a pure cell), FlowRadar can decode the flow by leveraging hash functions to locate the other cells of this flow and remove it from all the cells (by $XORing$ with the \emph{FlowXOR} fields, subtracting the \emph{PacketCount}, and decrementing the  \emph{FlowCount}).
The decoding process will continue to find new pure cells until there is no pure cell.

Deltoid \cite{DBLP:journals/ton/CormodeM05}, Reversible Sketch \cite{DBLP:journals/ton/SchwellerLCGGZDKM07}, Sequential Hashing \cite{DBLP:journals/cn/BuCCL10}, Fast Sketch \cite{DBLP:conf/infocom/LiuCG12}, SketchLearn \cite{DBLP:conf/sigcomm/HuangLB18},  and BitCount \cite{DBLP:conf/hpsr/WangG19} are all bit-level or sub-key (multi-bit)-level sketches designed for reversibility to deduce flow keys directly.
They first iteratively check the collected sketch for bits or sub-keys to identify the candidate flow sub-key for measurement tasks.
The candidate flow key is then composed of these sub-keys and further examined with the information recorded by the entire sketch to form the final measurement result.

\subsubsection{Probability-theory-based techniques}

\

In contrast to the traditional counter estimation techniques, some methods abstract counters as \emph{random variables} and leverage probability-theory-based techniques, such as \emph{maximum likelihood estimation} and \emph{Bayesian estimation}, to obtain the estimated counter of the flow.
By formulating the probability models of the related variables, sketch methods can deduce the desired statistical properties of the flow traffic.

\textbf{\emph{Bayes' Theorem.}}
Multi-Resolution Array of Counters (MRAC) \cite{DBLP:conf/sigmetrics/KumarSXW04} first uses counting algorithms and Bayesian estimation for accurate estimation of the flow size distribution.
It increases the underlying counter for each incoming packet in a lossy data structure and infers the most likely flow size distribution from the observed counters after the collision.
Bayesian statistics are used to recover as much information from the streaming results as possible.
SketchLearn is the first approximate measurement approach to build on the characterization of the inherent statistical properties of sketches.
It abstracts the ${k^{th}}$ bit of a flow key as a random variable $k$. Further, $p[k]$ represents the probability that the ${k^{th}}$ bit is equal to 1.
The multi-level sketch provides a fundamental property, i.e., if there is no large flow, its counter values follow a Gaussian distribution.
Based on this property, SketchLearn extracts large flows from multi-level sketches and leaves the residual counters for small flows to form Gaussian distributions.

\textbf{\emph{Maximum likelihood estimation.}}
MLM sketch \cite{DBLP:conf/infocom/LiCL11} abstracts the counter in the storage vector of flow $f$ as a random variable $X$, which is the sum of $Y$, i.e., the portion contributed by the packets of flow $f$, and $Z$, i.e., the portion contributed by the packets of other flows.
Based on the probability distribution of $Y$ and $Z$, MLM sketch formulates the probability of the observed value of a counter.
Moreover, the estimated size of flow $f$ can be obtained by maximizing the likelihood function.
Persistent items Identification and Estimation (PIE) \cite{DBLP:journals/ton/DaiSLLZC18} uses the information recorded at the observation points during all the measurement periods, formulates the occurrence estimation problem as a
maximum likelihood estimation (MLE) problem, and uses
the traditional MLE approach for estimation.
It divides \emph{m} cells in the Space-Time Bloom Filter (STBF) of the measurement period $i$ into empty, singleton, and collided cells, the numbers of which are represented by ${Z_{i0}}$, $Z_{i1}$, and $Z_{iC}$, respectively.
By maximizing the likelihood function, the number of distinct items can be estimated.

\subsubsection{Machine learning techniques}

\

The statistical traffic properties change from time to time.
Owing to this uncertainty, the traditional sketch design and optimization are not suitable for various measurement tasks, which leads to a considerable burden in terms of redesigning and tuning up the sketch design.
Recently, several studies have combined machine learning (ML) techniques with a measurement framework to relieve or eliminate the binding of traffic flow characteristics and the sketch design.
By selecting a set of features from the sketch itself, where every feature is a property of the data contributing to the estimation, ML techniques can reformulate the measurement tasks with these features instead of the measured sketch and thus reduce the design complexity drastically.

\textbf{\emph{ML sketch framework.}} MLsketch \cite{DBLP:conf/sigcomm/0003WSSHJTL18} presents a generalized ML framework to reduce the dependence of the sketch accuracy on the network traffic characteristics.
It continuously retrains ML models using a small number of samples from the same traffic whose information is being stored in the sketch.
Thus, ML models can continue to adapt to any network traffic characteristic variations without requiring to manually foresee scenarios or design statistical techniques for different scenarios.
The ML model can be trained using the learning sketch and extracted features, and it reflects a mapping from the current traffic distribution to the metrics of interest.
This design is applicable to different network environments.
Furthermore, designing the features and ML model usually involves just a selection from the available features and models, which is much simpler than designing a new sketch algorithm.

\textbf{\emph{Learning assistant parameter setting.}}
To accurately evaluate the frequency and rank of the top-$k$ hot items, the thresholds that are used to determine the type of each item (cold items, potential hot items, or hot items) should be dynamically adjusted to adapt to the flow distribution.
To address this issue, SSS \cite{DBLP:conf/bigcomp/GongTY0D0L18} assumes that the data stream obeys a Zipfian distribution, employs historical data to learn the parameters of the distribution using an ML program, and sets thresholds based on the distribution function.
Specifically, it leverages the linear regression model in the ML program to learn the threshold used to identify hot items.
The intelligent SDN-based Traffic (de)Aggregation and Measurement Paradigm (iSTAMP) \cite{DBLP:conf/infocom/MalboubiWCS14} employs an intelligent sampling algorithm to select the most informative traffic flows using information collected throughout the measurement process.
The main goal is to adaptively track and measure the most rewarding/informative traffic flows, which, if measured accurately, can yield the best improvement in the overall measurement utility.
For this purpose, multi-armed bandit sequential resource allocation algorithms are used to adaptively sample the most ``rewarding" flows in order to improve the flow measurement performance.

\begin{table*}[htbp]
  \centering
  \caption{General methods}
    \begin{tabular}{lllllllll}
    \toprule
    \multirow{9}[4]{*}{General purpose} &       & Frequency & HH    & Heavy Change & Entropy & Super Spreader/DDos & Flow size distribution & Cardinality \\
\cmidrule{2-9}          & OpenSketch\cite{DBLP:conf/nsdi/YuJM13} & $\times$ & $\checkmark$ & $\checkmark$ & $\times$ & $\checkmark$ & $\checkmark$ & $\times$ \\
          & Univmon\cite{DBLP:conf/sigcomm/LiuMVSB16} & $\times$ & $\checkmark$ & $\checkmark$ & $\checkmark$ & $\checkmark$ & $\times$ & $\times$ \\
          & SketchVisor\cite{DBLP:conf/sigcomm/HuangJLLTCZ17} & $\times$ & $\checkmark$ & $\checkmark$ & $\checkmark$ & $\checkmark$ & $\checkmark$ & $\checkmark$ \\
          & SketchLearn\cite{DBLP:conf/sigcomm/HuangLB18} & $\checkmark$ & $\checkmark$ & $\checkmark$ & $\checkmark$ & $\checkmark$ & $\checkmark$ & $\checkmark$ \\
          & HeavyGuardian\cite{DBLP:conf/kdd/0003GZZSL18} & $\checkmark$ & $\checkmark$ & $\checkmark$ & $\checkmark$ & $\times$ & $\checkmark$ & $\times$ \\
          & Elastic Sketch\cite{DBLP:conf/sigcomm/0003JLHGZMLU18} & $\checkmark$ & $\checkmark$ & $\checkmark$ & $\checkmark$ & $\times$ & $\checkmark$ & $\checkmark$ \\
          & MLsketch\cite{DBLP:conf/sigcomm/0003WSSHJTL18} & $\checkmark$ & $\checkmark$ & $\times$ & $\times$ & $\times$ & $\times$ & $\checkmark$ \\
          & HyperSight\cite{DBLP:journals/jsac/ZhouBYGCZWZ20} & $\times$ & $\checkmark$ & $\checkmark$ & $\times$ & $\times$ & $\times$ & $\times$ \\
    \bottomrule
    \end{tabular}%
  \label{general}%
\end{table*}%

\textbf{\emph{Learning cardinality estimation.}}
To adapt to changes in flow size distribution, Cohen and Nezti \cite{DBLP:journals/ton/CohenN19} proposed a sampling-based adaptive cardinality estimation based on the online ML method.
The traffic flow stream was divided into batches of packets.
Each batch was sampled, the selected features and statistical properties were extracted from these samples, and the exact cardinality of each batch was calculated.
Then, these values were used to train online ML models.
Moreover, they analyzed various possible features, parameters, and online ML algorithms for their framework and proposed the most suitable combination.

\textbf{\emph{Learning frequency estimation.}}
Existing learning frequency estimation involves two strategies. On the one hand, \cite{DBLP:conf/iclr/HsuIKV19} and \cite{DBLP:journals/corr/abs-1908-05198} first learned a classifier to separately record the heavy and mouse flows in order to reduce the mutual interference collision.
On the other hand, \cite{DBLP:journals/corr/abs-2007-09261} and \cite{DBLP:conf/infocom/FuLS0C20} leveraged the concept of \emph{locality sensitivity} to cluster similar items together to reduce the approximate error. By averaging the recorded frequency, we can derive the approximate result.

\textbf{\emph{Learning membership testing.}}
Although BF has been well explored and evaluated in both academia and industry, Kraska et al. \cite{DBLP:conf/sigmod/KraskaBCDP18} suggested that ML models such as neural networks can be combined with BF to improve space utilization and query efficiency further.
They proposed a method in which ML is used to model a pre-filter stage for the backup BF.
Michael \cite{DBLP:journals/corr/abs-1802-00884} further clarified a formal model of this structure and stated what guarantees can and cannot be associated with such a structure.
The sandwiched LBF \cite{DBLP:conf/nips/Mitzenmacher18} optimizes BF by surrounding the learning model with two BF layers.
Specifically, the backup BF removes the false negatives from the learning model, while the initial BF removes the false positives upfront.
The Partitioned Learned Bloom Filter (PLBF) \cite{vaidya2020partitioned} was developed to overcome the inability of previous methods to exploit the learning model fully.
It partitions the score space into multiple regions using multiple thresholds and employs an individual backup BF for each region.
Experiments have demonstrated that PLBF can achieve near-optimal performance in many
cases. Based on the Stable Bloom Filter (SBF) \cite{DBLP:conf/sigmod/DengR06},
the Stable Learned Bloom Filter (SLBF) \cite{DBLP:journals/pvldb/LiuZSC20} replaces BFs in the sandwiched LBF \cite{DBLP:conf/nips/Mitzenmacher18} and PLBF \cite{vaidya2020partitioned} with SBF to facilitate data streaming applications.
All the above-mentioned learning sketches face a common problem in terms of updating the out-of-date classifier.

%\textcolor[rgb]{1.00,0.00,0.00}{SketchML\cite{DBLP:conf/sigmod/JiangFY018}}

\subsection{Summary and insights}
In this section, we reviewed studies related to optimization in the post-processing stage.
The linearity of CM-based sketches provides an intrinsic characteristic for sketch compression and merging.
\cite{DBLP:conf/dialm/ScheuermannM07} and \cite{DBLP:journals/corr/CafaroTP14} proposed feasible solutions for problems related to other sketches, such as cardinality registers and top-$k$ summaries.
Furthermore, we surveyed techniques for information extraction, such as flow key reversible, probability-theory-based, and ML techniques.
The information extraction techniques are directly related to the sketch structure and updating operation.
Other information extraction techniques will be proposed as novel sketch designs.

\section{Application and implementation}\label{app}

We have surveyed related studies on the sketch design from the perspective of data flow.
In the following, we survey related studies from another perspective, i.e., application and implementation of sketches.
In total, we have summarized 13 measurement tasks in terms of time and volume dimension. Furthermore, we survey the software and hardware implementation when using sketches in practice.

\begin{table*}[htbp]
  \centering
  \caption{Persistence detection}
    \begin{tabular}{lllllll}
    \toprule
    \multirow{6}[4]{*}{Persistence detection} & Methods & Enable sliding windows & Store the whole ID & Multi-structure & False positive & False negative \\
\cmidrule{2-7}          & PIE\cite{DBLP:journals/ton/DaiSLLZC18} & $\checkmark$ & $\times$ & $\checkmark$ & $\times$ & $\checkmark$ \\
          & LDF\cite{DBLP:conf/infocom/ChenJC10} & $\checkmark$ & $\times$ & $\checkmark$ & $\checkmark$ & $\checkmark$ \\
          & Lahiri et al.\cite{DBLP:journals/sadm/LahiriTC14} & $\checkmark$ & $\checkmark$ & $\times$ & $\checkmark$ & $\checkmark$ \\
          & Huang et al.\cite{DBLP:conf/infocom/HuangSCTHYY18} & $\checkmark$ & $\times$ & $\checkmark$ & $\checkmark$ & $\times$ \\
          & LTC\cite{DBLP:conf/icde/0003ZYHL19} & $\checkmark$ & $\checkmark$ & $\times$ & $\checkmark$ & $\checkmark$ \\
    \bottomrule
    \end{tabular}%
  \label{Persistence detection}%
\end{table*}%

\subsection{Measurement tasks}

Although the methods listed in Table \ref{general} are designed for supporting a comprehensive range of measurement tasks,
there are some measurement tasks that they cannot support.
To this end, many studies have designed sketches for specific or several measurement tasks.
In this subsection, we review the related measurement tasks in terms of time, volume and time-volume dimension.
%Specifically, we categorize the persistence detection and flow latency into time dimension, summarize other volume-based measurement tasks into volume dimension, including per-flow measurement, cardinality estimation, super spreader, heavy hitters, HHH, top-$k$ detection and entropy estimation and classify the change detection, persistent sketch, burst detection and item batch detection into time-volume dimension.

\subsubsection{Persistence detection}
In contrast to a frequent item, a persistent item does not necessarily occur more frequently than other items over a short period. Instead, it persists and occurs more frequently than other items only over a long period.
Persistence is typical of many stealthy types of traffic on the Internet.
For example, Xiao et al. \cite{DBLP:conf/icnp/XiaoQMC14} observed that legitimate users connect to a server intermittently, whereas attacking hosts keep sending packets to the server.
By identifying persistent items, operators can detect potentially malicious traffic behavior.
Table \ref{Persistence detection} compares existing persistence detection methods in terms of their data structure characteristics, functionality, and measurement accuracy.

\textbf{\emph{Multi-layer design.}}
Long-Duration Flows (LDFs) \cite{DBLP:conf/infocom/ChenJC10} maintain two counting Bloom filters (${B_1}$ and ${B_2}$) and one small hash table simultaneously.
At the starting time interval, LDF records all the flows that appear during this interval in ${B_1}$.
At the next time interval, LDF adds new flows, which appear in both ${B_1}$ and the current interval into ${B_2}$. Then, ${B_1}$ and ${B_2}$ are iterated by switching roles in the subsequent intervals.
When a flow is identified as LDF, it will be added to the hash table.
Lahiri et al. \cite{DBLP:journals/sadm/LahiriTC14} presented the first small-space approximation algorithm for identifying persistent items.
A hash-based filter first processes the items in each measurement period to record the flow ID and track the persistence in future periods.
Furthermore, they considered the sliding model under which only items belonging to the $W$ most recent windows are considered.
PIE \cite{DBLP:journals/ton/DaiSLLZC18} uses Raptor codes to encode the ID of each item in a flow traffic stream.
The persistent item can be retrieved and decoded from the measurement point if there are sufficient encoded bits for the ID in sufficient measurement periods.
Specifically, PIE uses STBF to record information. During each epoch, PIE maintains a new STBF and transfers it to the permanent storage region for subsequent analysis. After $T$ epochs, PIE
uses MLE to estimate the number of occurrences of any given item. However, PIE must maintain $T$ STBFs to track persistent items with high space overhead.
On-off sketch\cite{DBLP:journals/pvldb/ZhangLL0LZ020} introduces a state flag (on-off) into each counter of sketch instead of maintaining a bloom filter to record the existence of element.
At the beginning of each window, the flag is set as on, once an element is hashed into the counter, the flag turn as off and the counters are only increased when the flag bit is on. In this way, the counter is most increased once in a time window.

\textbf{\emph{k out of t persistence.}}
Although we can separate attackers from legitimates by finding flows appearing during all measurement periods, Huang et al. \cite{DBLP:conf/infocom/HuangSCTHYY18} considered the situation in which attackers might forgo sending packets during several periods.
To solve this problem, they formulated a new problem called $k$-persistent spread estimation, which measures persistent traffic elements in each flow that appear during at least $k$ out of $t$ measurement periods.
Yang et al. \cite{DBLP:conf/icde/0003ZYHL19} realized that for some applications, people care about essential items that are both persistent and frequent.
They formulated this problem and proposed LTC to solve it, including the \emph{Long-tail Replacement} technique and modified \emph{CLOCK} algorithm.

\subsubsection{Flow latency}
Delay is an essential metric for understanding and improving system performance.
When managing networks with stringent latency demands, operators must measure the latency between two observation points, such as a port in a middlebox and a network card in the end host. Moreover, as reported in \cite{DBLP:conf/sigcomm/NarayanaSNGAAJK17}, both fine-grained and coarse-grained delay measurement are of great importance in performance measurement, system diagnosis, traffic engineering, etc.

\begin{table*}[htbp]
  \centering
  \caption{Latency detection}
    \begin{tabular}{lllllll}
    \hline
    \multirow{7}[4]{*}{Flow latency} & Methods & Time stamp & Probe packet & Per-packet delay & Deal with loss and reordering  &  Abnormal delay detection \bigstrut\\
\cline{2-7}          & LDA\cite{DBLP:conf/sigcomm/KompellaLSV09} & $\times$ & $\times$ & $\times$ & $\times$ & $\times$ \bigstrut[t]\\
          & Colate\cite{DBLP:journals/ton/ShahzadL16} & $\times$ & $\times$ & $\checkmark$ & $\times$ & $\checkmark$ \\
          &  OPA\cite{DBLP:journals/ton/WangLDLL16} & $\times$ & $\times$ & $\checkmark$ & $\checkmark$ & $\checkmark$ \\
          & Marple\cite{DBLP:conf/sigcomm/NarayanaSNGAAJK17} & $\checkmark$ & $\times$ & $\checkmark$ & $\checkmark$ & $\checkmark$ \\
          & FineComb\cite{DBLP:journals/ton/LeeGKV14} & $\times$ & $\times$ & $\times$ & partially & $\times$ \\
          & RLI\cite{DBLP:conf/sigcomm/LeeDK10} & $\times$ & $\checkmark$ & $\times$ & $\checkmark$ & $\times$ \bigstrut[b]\\
    \hline
    \end{tabular}%
  \label{Latency}%
\end{table*}%

\textbf{\emph{For per-flow latency measurement.}}
MAPLE \cite{DBLP:conf/sigcomm/NarayanaSNGAAJK17} attaches a timestamp to every packet. Specifically, when measuring the packet passing through observation point $S$ to $R$, $S$ attaches a timestamp to the packets, and $R$ calculates the packet's latency from $S$ to $R$ by subtracting the timestamp from its current time.
To reduce the space required to store all the packets' latency values, MAPLE predefines a set of latency values and maps each packet to its closest value.
For each latency value, MAPLE augments a Bloom filter to test whether a given packet has been mapped to the latency value.
Reference Latency Interpolation (RLI) \cite{DBLP:conf/sigcomm/LeeDK10} uses packet probing.
Specifically, $S$ inserts probe packets with a timestamp into the flow, and $R$ calculates the latency of each probe packet similarly to MAPLE.
Suppose that the latency of two probe packets has been calculated as $l_1$ and $l_2$; to calculate the latency of the regular packets between two probe packets, $R$ uses the \emph{straight-line equation} to calculate the two probe packets' latency.
Further, Colate \cite{DBLP:journals/ton/ShahzadL16} is the first per-flow latency measurement scheme that requires no probe packets or timestamp.
It records the time information of packets at each observation point and purposely allows noise to be introduced in the recorded information to minimize storage.

\textbf{\emph{For aggregate latency.}} The aggregate latency refers to the average and deviation of the latencies experienced by all the packets that pass through two observation points. In LDA \cite{DBLP:conf/sigcomm/KompellaLSV09}, both the sender and the receiver maintain several counter vectors composed of counter pairs, the timestamp counter for accumulating packet timestamps, and the packet counter for counting the number of arriving/departing packets.
For each packet, LDA first maps the packet into the counter vector with a sampling probability.
In the long run, the packet is mapped into a counter-pair, the timestamp of the packet is added to the timestamp counter, and the packet counter is incremented by one.
LDA selects all the counter pairs with the same packet counter for both the sender and the receiver by checking the packet counters.
Then, LDA can easily calculate the total number of successfully delivered packets and the sum of their timestamps.
However, in LDA, packets belonging to one segment may be misidentified with other segments due to packet reordering.
Thus, these groups of packets cannot be used for delay measurement.

FineComb \cite{DBLP:journals/ton/LeeGKV14} proposes a data structure called stash, which maintains the information for packets near the boundary of the segments at the receiver to solve this problem.
However, for groups with lost packets, FineComb cannot calculate their per-packet delay.
Order Preserving Aggregator (OPA) \cite{DBLP:journals/ton/WangLDLL16} leverages the fact that lost and reordering packets are usually much fewer than legitimate packets to facilitate efficient ordering as well as loss of information representation and recovery.
It proposes a two-layer design to convey ordering and timestamps and efficiently derive the per-packet delay.
However, compared with LDA \cite{DBLP:conf/sigcomm/KompellaLSV09} and FineComb \cite{DBLP:journals/ton/LeeGKV14}, OPA improves the accuracy of delay measurement with additional overhead for a high data link.

As illustrated in Table \ref{Latency}, we borrow the idea from OPA \cite{DBLP:journals/ton/WangLDLL16} and compare flow latency measurement methods in terms of their functionality, performance, and the basic idea.

\subsubsection{Per-flow measurement}Providing per-flow statistics plays a fundamental role in network measurement.
As a probabilistic data structure, sketches have been extensively investigated for per-flow measurement, including flow size (the number of packets in a flow) and flow volume (the number of bytes in a flow).

Table \ref{per} summarizes related studies on per-flow measurement.
Apart from the CM sketch \cite{DBLP:conf/latin/CormodeM04} and its variants, most research focuses on the skewness characteristic of the traffic stream.
Many methods have been proposed to solve this unbalance situation in terms of the traffic size/volume. We have discussed the related work in detail in Section \ref{sketch structure}.

\begin{table*}[htbp]
  \centering
  \caption{Per-flow measurement}
    \begin{tabular}{llll}
    \toprule
    \multirow{10}[14]{*}{Per-flow measurement} & \multicolumn{2}{c}{Functionality} & Methods \\
\cmidrule{2-4}          & \multicolumn{2}{c}{\multirow{2}[2]{*}{CountMin\cite{DBLP:conf/latin/CormodeM04} and its variants}} & {Conservative update\cite{DBLP:journals/tocs/EstanV03}, CSM Sketch\cite{DBLP:conf/infocom/LiCL11}} \\
          & \multicolumn{2}{c}{} &  { Count MeanMin sketch\cite{article}, Count sketch\cite{DBLP:journals/pvldb/CormodeH08}} \\
\cmidrule{2-4}          & \multirow{7}[10]{*}{Adaptive to skewed traffic} &  Variable counters size &  {SBF\cite{DBLP:conf/sigmod/CohenM03}, DCF\cite{DBLP:journals/sigmod/Aguilar-SaboritTML06}, TinyTable\cite{DBLP:conf/icdcn/EinzigerF16}, CQF\cite{DBLP:conf/sigmod/PandeyBJP17}} \\
\cmidrule{3-4}          &       & Enlarge count range & {SAC\cite{DBLP:conf/infocom/Stanojevic07}, Cuckoo counter\cite{DBLP:conf/ancs/QiLYLL19}, SA\cite{DBLP:conf/infocom/YangXLLWBL19}} \\
\cmidrule{3-4}          &       & \multirow{3}[2]{*}{Multi-level counters} &  {Counter Braids\cite{DBLP:conf/sigmetrics/LuMPDK08} Brick\cite{DBLP:journals/ton/HuaXLZ11}, RCC\cite{DBLP:journals/ton/NyangS16}, Pyramid Sketch\cite{DBLP:journals/pvldb/YangZJCL17},} \\
          &       &       & CounterTree\cite{DBLP:journals/ton/ChenCC17}, OM sketch\cite{DBLP:conf/globecom/ZhouLJ0DL17}, ABC\cite{DBLP:conf/bigdataconf/Gong0ZYC0L17}, Cold filter\cite{DBLP:conf/sigmod/Zhou0J0YLU18}, \\
          &       &       & Diamond sketch\cite{DBLP:conf/infocom/0003GSWSL18} \\
\cmidrule{3-4}          &       & Multi-layer sketch & Bloom Sketch\cite{DBLP:conf/infocom/ZhouJLZ0L18}, ASketch\cite{DBLP:conf/sigmod/RoyKA16} \\
\cmidrule{3-4}          &       & Frequency-aware updating & FCM\cite{DBLP:conf/icde/ThomasBAY09} \\
    \bottomrule
    \end{tabular}%
  \label{per}%
\end{table*}%

\begin{table*}[htbp]
  \centering
  \caption{Cardinality estimation\cite{DBLP:conf/infocom/XiaoZC17}}
    \begin{tabular}{lllll}
    \hline
    \multirow{9}[4]{*}{Cardinality estimation} & \multicolumn{2}{c}{Methods} & Std.Err.($\delta$) & Memory units \bigstrut\\
\cline{2-5}          & \multicolumn{2}{l}{MinCount\cite{DBLP:conf/random/Bar-YossefJKST02}} & $1.00/\sqrt{m}$ & 32-bit keys \bigstrut[t]\\
          & \multicolumn{2}{l}{PCSA\cite{DBLP:journals/jcss/FlajoletM85}} & $0.78/\sqrt{m}$ & 32-bit register \\
          & \multicolumn{2}{l}{LogLog\cite{DBLP:conf/esa/DurandF03}} & $1.30/\sqrt{m}$ & 5-bit register \\
          & \multicolumn{2}{l}{HLL\cite{DBLP:conf/edbt/HeuleNH13}} & $1.04/\sqrt{m}$ & 5-bit register \\
          & \multicolumn{2}{l}{Sliding HLL\cite{DBLP:conf/icdm/ChabchoubH10}} & $1.04/\sqrt{m}$ & 5-bit register \\
          & \multicolumn{2}{l}{HLL-TailCut+\cite{DBLP:conf/infocom/XiaoZC17}} & $1.00/\sqrt{m}$ & 3-bit register \\
          & \multicolumn{2}{l}{Refined LL\cite{DBLP:journals/www/WangYWJCCL19}} & $1.00/\sqrt{m}$ & 5-bit register \\
          & \multicolumn{2}{l}{Cohen and Nezti \cite{DBLP:journals/ton/CohenN19}} & $--$  & $--$  \bigstrut[b]\\
    \hline
    \end{tabular}%
  \label{cardinality}%
\end{table*}%

\subsubsection{Cardinality estimation}\label{card} Estimating the number of distinct flows (or cardinality) is an essential issue in many applications, such as measurement and anomaly detection.

As shown in Table \ref{cardinality}, we borrow insights from the well-known cardinality estimation in \cite{DBLP:conf/infocom/XiaoZC17}, where $m$ is the total number of memory units.
Linear Counting \cite{DBLP:journals/tods/WhangVT90} maps all flows uniformly among a bit array, and each element can be mapped to a bit.
However, it cannot work efficiently under a strict condition where the memory is roughly linear to the cardinality \cite{DBLP:conf/edbt/MetwallyAA08}.
PCSA \cite{DBLP:journals/jcss/FlajoletM85}
uses \emph{stochastic averaging}, which allocates multiple registers to produce independent estimations, and returns the average.
However, to achieve high accuracy, $m$ should be large, and $m\log n$ bits is the cost of the limited memory in the monitors \cite{DBLP:conf/infocom/XiaoZC17}.
This limitation of PCSA motivates another more efficient algorithm called LogLog \cite{DBLP:conf/esa/DurandF03}.
Both LogLog \cite{DBLP:conf/esa/DurandF03} and HLL \cite{DBLP:conf/edbt/HeuleNH13} are based on the pattern ``$0R1$" in the binary representation of hashed values.
More specifically, after hashing all the packets, the position of the leftmost 1-bit is denoted by $R$.
%Then the cardinality of stream can be deduced from $R$.
However, they adopt different methods for extracting the estimation from a set of registers.
LogLog uses geometric averaging whereas HLL uses harmonic means to mitigate the impact of outliers with abnormally large estimations. Sliding HLL \cite{DBLP:conf/icdm/ChabchoubH10} improves HLL for stream processing by adding a sliding window mechanism.
%In this way, it can estimate the number of flows seen over any duration bounded by the length of the sliding window.
HLL-TailCut+ \cite{DBLP:conf/infocom/XiaoZC17} can further reduce the memory requirements of HLL by 45\% on the basis of an extended tail cutting technique that compresses the information across all registers while reducing the variance among the registers.
Its basic idea is that the memory per register may be compressed into four bits or even into three bits with no significant loss of useful information.
HLL-TailCut+ can get many registers, and a more accurate estimation result given the same space overhead.

Refined LogLog \cite{DBLP:journals/www/WangYWJCCL19} uses a fine-grained common ratio instead of 2 used in FM Sketch \cite{DBLP:journals/jcss/FlajoletM85} and LogLog \cite{DBLP:conf/esa/DurandF03} to narrow the gaps between two recordable values.
Using geometric hash functions with a smaller common ratio, Refined LogLog overcomes the restriction that the former method can only record ${2^k}$, resulting in a large gap with the actual number.
Moreover, in real-world networks, we do not have cardinality, which would enable us to choose the appropriate size of bitmaps.
A self-adaptive version of LogLog was proposed to adapt to cardinality by changing its common ratio dynamically.
%In this way, we can get rid of prior knowledge.
It achieves high accuracy when the cardinality is small and can also store large cardinality.

Cohen and Nezti \cite{DBLP:journals/ton/CohenN19} proposed a sampling-based adaptive cardinality estimation.
The traffic flow stream is divided into batches of packets.
Each batch is sampled, and the selected features are extracted from the samples.
The features are then passed to the $predict()$ operation of the online ML algorithm, which returns an estimation of the batch's cardinality.
Subsequently, the entire batch is sent for training. In the training phase, $partial{\_}fit()$ is provided with the batch's set of features and its cardinality.
The exact cardinality of a batch can be calculated using a simple hash table.

\subsubsection{Super spreader} Spreader estimation aims to identify a source IP that communicates with more than a threshold number of distinct destination IP/port pairs.

In this case, the flow is identified by the source address. The elements under measurement are the destination addresses in the headers of the packets.
An increase in the cardinality of a specific flow may signal a DDoS attack against the flow's destination address.
DDoS is a malicious attempt to disrupt regular traffic of a targeted server, service, or network by overwhelming the target or its surrounding infrastructure with a traffic flood.
In this case, a flow is identified by the destination address, and we must count the number of distinct source addresses in each flow.

\begin{table*}[htbp]
  \centering
  \caption{Super Spreader/DDos}
    \begin{tabular}{lllp{26.445em}}
    \toprule
    \multirow{7}[4]{*}{Super Spreader/DDos} & \multicolumn{2}{l}{Methods} & \multicolumn{1}{c}{Basic idea} \\
\cmidrule{2-4}          & \multicolumn{2}{l}{One/two level Filtering\cite{DBLP:conf/ndss/VenkataramanSGB05}} & \multicolumn{1}{l}{Set sampling probability to filter } \\
          & \multicolumn{2}{l}{OSM\cite{DBLP:journals/jsac/ZhaoXK06}} & \multicolumn{1}{l}{Share bitmaps} \\
          & \multicolumn{2}{l}{MultiResBitmap\cite{DBLP:journals/ton/EstanVF06}} & \multicolumn{1}{l}{One bitmap one source} \\
          & \multicolumn{2}{l}{CSE\cite{DBLP:journals/ton/YoonLCP11}} & \multicolumn{1}{l}{Bit sharing} \\
          & \multicolumn{2}{l}{Virtual Register Sharing\cite{DBLP:journals/ton/XiaoCZCLLL17}} & \multicolumn{1}{l}{Multi-bits sharing} \\
          & \multicolumn{2}{l}{RCD sketch\cite{DBLP:conf/globecom/GuanWQ09}} & \multicolumn{1}{l}{Remainder characteristics of number theory} \\
    \bottomrule
    \end{tabular}%
  \label{Super}%
\end{table*}%

Table \ref{Super} compares the studies in terms of the basic idea.
One-level filtering \cite{DBLP:conf/ndss/VenkataramanSGB05} samples distinct source-destination pairs in the packets such that each distinct pair is included in the sample with probability ${p\mathrm{=}\frac{1}{k}}$ to identify the $k$-super spreaders.
Two-level filtering \cite{DBLP:conf/ndss/VenkataramanSGB05} maintains two levels of filters.
The first level filters out the \emph{SrcIP} that contacts only a small number of distinct \emph{DstIP} with a small sampling probability, which is smaller than that of the second filter.
Compared with one-level filtering, two-level filtering is more space-efficient.
Both levels can capture the real super spreaders and the \emph{SrcIP} that contacts with a few \emph{DstIP} can be filtered out efficiently in the first level.
By embedding estimated fan-outs into binary hash strings, multiple Spread Sketch \cite{DBLP:conf/infocom/TangHL20} can be merged to provide a network-wide measurement view for recovering super spreaders and their estimated fan-outs.

MultiResBitmap \cite{DBLP:journals/ton/EstanVF06} uses bitmaps to save space instead of storing the actual source/destination addresses in each sample.
It allocates a bitmap, where each bit is set for each destination/source that the source/destination contacts.
However, such a spread estimator cannot fit in a tight space where only a few bits are available for each source \cite{DBLP:journals/ton/XiaoCZCLLL17}. OSM \cite{DBLP:journals/jsac/ZhaoXK06} allocates each source randomly to $l$ columns in a bit matrix (where columns are bitmaps) through $l$ hash functions.
%It sets one bit when storing contact.
Moreover, a column may be shared by multiple sources.
To reduce the introduced noise, OSM proposes a method to remove the noise and estimate the spread of the source.
Nevertheless, on the one hand, OSM makes $l$ memory accesses and uses $l$ bits for storing one contact; on the other hand, the noise may be too much to be removed in a compact memory space when a significant number of all the bits are set.
CSE \cite{DBLP:journals/ton/YoonLCP11} also creates a virtual bit vector for each source by taking bits uniformly at random from a common pool of available bits. %Thus, the bits are shared by all instead of sharing only bitmaps.
This sharing strategy provides an excellent property: the probability that one source's contacts cause noise in any other source is the same.
Compared with the noise in \cite{DBLP:journals/jsac/ZhaoXK06}, such uniform noise is easy to measure and remove.
Virtual Register Sharing \cite{DBLP:journals/ton/XiaoCZCLLL17} develops a framework of virtual estimators that allow sharing of the cardinality estimation solutions, including PCSA \cite{DBLP:journals/jcss/FlajoletM85}, LogLog \cite{DBLP:conf/esa/DurandF03}, and HyperLL \cite{DBLP:conf/edbt/HeuleNH13}.
Moreover, it shows that sharing at the multi-bit level is superior to sharing at the bit level. RCD \cite{DBLP:conf/globecom/GuanWQ09} locates hosts with a large connection degree by estimating the in-degree/out-degree associated with a given host through the remainder characteristics of number theory.

% Table generated by Excel2LaTeX from sheet 'Sheet1'
\begin{table*}[htbp]
  \centering
  \caption{Heavy Hitter detection}
    \begin{tabular}{lllll}
    \hline
    \multirow{13}[4]{*}{Heavy hitter} & Methods & Space & Updating time & Query time \bigstrut\\
\cline{2-5}          & CM sketch\cite{DBLP:conf/latin/CormodeM04} & $O\left( \frac{H}{\epsilon}\log \frac{1}{\delta} \right)$ & $O\left( \log \frac{1}{\delta} \right)$ & $O\left( n \right)$ \bigstrut[t]\\
          & Count-min heap\cite{DBLP:conf/latin/CormodeM04} & $O\left( \frac{1}{\epsilon}\log \frac{1}{\delta}+H\log n \right)$ & $O\left( \log \frac{H}{\delta} \right)$ & $O\left( H \right)$ \\
          &  CHHFR\cite{DBLP:journals/ieicet/MoriTPKUG07} & $---$ & $---$ & $---$ \\
          & MV-Sketch\cite{DBLP:conf/infocom/TangHL19} & $O\left( \frac{1}{\epsilon}\log \frac{1}{\delta}\log n \right)$ & $O\left( \log \frac{1}{\delta} \right)$ & $O\left( \frac{1}{\epsilon}\log ^2\frac{1}{\delta} \right)$ \\
          & LD-sketch\cite{DBLP:journals/cn/HuangL15} & $O\left( \frac{H}{\epsilon}\log \frac{1}{\delta} \right)$ & $O\left( \log \frac{1}{\delta} \right)$ & $O\left( \frac{H}{\epsilon}\log \frac{1}{\delta} \right)$ \\
          & LC\cite{DBLP:conf/vldb/MankuM02} & $O\left( \frac{1}{\epsilon}\log \left( n\right) \right)$ & $O\left(1\right)$ & $O\left( \frac{1}{\epsilon}\log \left( n\right) \right)$ \\
          & PLC\cite{DBLP:journals/ccr/DimitropoulosHK08} & $O\left( \frac{1}{\epsilon}\log \left(n \right) \right)$ & $O\left(1\right)$ & $O\left( \frac{1}{\epsilon}\log \left( n\right) \right)$ \\
          & BitCount\cite{DBLP:conf/hpsr/WangG19} & $O\left( \frac{1}{\epsilon}\cdot k \right)$ & $O\left(k\right)$ & $O\left(k\right)$ \\
          &  Sequential Hashing\cite{DBLP:journals/cn/BuCCL10} & $O\left( \frac{H}{\epsilon}\log n\log \frac{1}{\delta} \right)$ & $O\left( \log n\log \frac{1}{\delta} \right)$ & $O\left( \frac{H}{\epsilon}\log n\log \frac{1}{\delta} \right)$ \\
          & IM-SUM\cite{DBLP:conf/infocom/Ben-BasatEFK17a} & $O\left( \frac{1}{\epsilon} \right)$ & $O\left(1\right)$ & $O\left( \frac{1}{\epsilon} \right)$ \\
          & DIM-SUM\cite{DBLP:conf/infocom/Ben-BasatEFK17a} & $O\left( \frac{1}{\epsilon} \right)$ & $O\left(1\right)$ & $O\left( \frac{1}{\epsilon} \right)$ \\
          & Fast sketch\cite{DBLP:conf/infocom/LiuCG12} & $O\left( \frac{H}{\epsilon}\log \frac{1}{\delta}\log \frac{n\epsilon}{H\log \frac{1}{\delta}} \right)$ & $O\left( \log \frac{1}{\delta}\log \frac{n\epsilon}{H\log \frac{1}{\delta}} \right)$ & $O\left( \frac{H}{\epsilon}\log ^3\frac{1}{\delta}\log \frac{n\epsilon}{H\log \frac{1}{\delta}} \right)$ \bigstrut[b]\\
    \hline
    \end{tabular}%
  \label{HH}%
\end{table*}%

\subsubsection{Heavy Hitters} The flows which are identified as heavy hitters are the ones that exceed a specified volume of the total traffic.
In many applications, such as congestion and anomaly detection, identifying heavy-hitter flows is crucial \cite{DBLP:journals/ieicet/MoriTPKUG07}.

CM sketch and Count-min heap \cite{DBLP:conf/latin/CormodeM04} can detect a heavy hitter by checking whether a flow is a heavy hitter by examining if its estimated sum exceeds the threshold.
However, CM Sketch is non-invertible. Lossy Counting \cite{DBLP:conf/vldb/MankuM02} treats all incoming flows as elephant flows. If the flow's counter exists, LC increases the corresponding counter on every arrival; if the counter is not in the table, it is allocated a counter value of 1 if available.
LC keeps the table size bounded by periodically decreasing table counters and evicting items whose counter reaches 0.
However, LC requires a maximum number of $\frac{1}{\epsilon} log(n)$ table entries.
Probabilistic Lossy
Counting (PLC) \cite{DBLP:journals/ccr/DimitropoulosHK08} requires fewer table entries on average but only provides a probabilistic guarantee.
PLC makes the error bound significantly smaller than the deterministic error bound of Lossy Counting \cite{DBLP:conf/vldb/MankuM02}.
In lossy counting, the error bound reflects the potential error in the estimated frequency of an element owing to possible prior element removal(s) from the table.
%Its importance lies in its use in removing elements from the table: an element with a small error bound is more likely to be removed from table than an equal-frequency element having a significant error bound is.
PLC shows that the error of more than 90\% of the flows is significantly smaller than the deterministic error bound.
Using a probabilistic error bound instead of a deterministic error bound to remove flows in a finished epoch, PLC improves the memory consumption of the algorithm.
Furthermore, PLC reduces the false positives of lossy counting and achieves a low estimation error, albeit slightly higher than that of LC.

Caching Heavy-hitter Flows with Replacement (CHHFR) \cite{DBLP:journals/ieicet/MoriTPKUG07} leverages \emph{LRU} to track heavy hitters. It not only focuses on the update time of the traffic flows but also adds another variable $Ctr$ to find a replaced item with an underlying double-linked list of flow units to record the frequency of items.
MV-sketch \cite{DBLP:conf/infocom/TangHL19} is designed for supporting heavy flow detection with small and static memory allocation.
It tracks candidate heavy flows inside the reversible sketch data structure via the idea of majority voting.
LD-sketch \cite{DBLP:journals/cn/HuangL15} combines the traditional counter-based and sketch-based methods to reduce false positives by aggregating multiple detection results.
By augmenting an associative array in each bucket, LD-sketch can track heavy key candidates that are hashed to the bucket. BitCount \cite{DBLP:conf/hpsr/WangG19} maintains a set of counters to count the total number of 1-bits in each bit position of the binary representation of IP addresses.
However, owing to the additional property of bit counters, this method may involve inevitable false positive errors. IM-SUM \cite{DBLP:conf/infocom/Ben-BasatEFK17a} first introduces the elephant detection scheme, which focuses on the elephants in terms of volume.
It achieves constant update time consumption, i.e., an amortized time of $O(1)$, by periodically removing many small flows from the table at once instead of removing the minimum flow upon the arrival of a non-resident flow.
However, IM-SUM \cite{DBLP:conf/infocom/Ben-BasatEFK17a} performs only a small portion of the maintenance procedure to achieve a worst-case complexity of $O(1)$.

Table \ref{HH} compares various heavy hitter detection methods in terms of their space overhead, updating time, and query time.
In the table, $H$ represents the maximum number of heavy hitters in a measurement epoch, $n$ is the traffic flow key domain, $\epsilon (0\mathrm{<}\epsilon \mathrm{<}1)$ denotes the approximation parameter, and $\delta (0 \mathrm{<}\delta \mathrm{<}1)$ is the error probability.

\subsubsection{Hierarchical Heavy Hitters (HHH)}
The structure of IP addresses implies a prefix-based hierarchy.
 Network traffic characteristics can be better presented by using flow aggregates by their 5-tuple fields \cite{DBLP:conf/imc/Cho17}.
 This strategy makes aggregation of flows a powerful means for traffic measurement and a valuable component for anomaly detection to identify attacks and scans.
 Moreover, identifying pairs of source and destination prefixes or other attributes that give rise to a significant amount of global traffic refers to a novel measurement task, namely multidimensional Hierarchical Heavy Hitters (mHHH) \cite{DBLP:conf/teletraffic/TruongG09}.

\begin{table*}[htbp]
  \centering
  \caption{Hierarchical Heavy Hitters}
    \begin{tabular}{lllll}
    \hline
    \multirow{8}[4]{*}{HHH} & Methods & Space & Updating time & Enable mHHH \bigstrut\\
\cline{2-5}          &  Separator\cite{DBLP:conf/adma/LinL07} & $O\left( \frac{H^2}{\epsilon} \right)$ & $O\left( H \right)$ & $\times$ \bigstrut[t]\\
          & TCAM-based HHH\cite{DBLP:conf/nsdi/JoseY11} & $O\left( \frac{2}{T} \right)$ & $O\left( H \right)$ & $\times$ \\
          & Recursive Lattice Search\cite{DBLP:conf/imc/Cho17} & $--$  & $O\left( N\log H \right)$ & $\checkmark$ \\
          & Randomized HHH\cite{DBLP:conf/sigcomm/Ben-BasatEFLW17} & $O\left( \frac{H}{\epsilon} \right)$ & $O\left( 1 \right)$ & $\checkmark$ \\
          & Truong et al.\cite{DBLP:conf/teletraffic/TruongG09} & $O\left( \frac{H^{3/2}}{\epsilon} \right)$ & $O\left( \frac{H^{3/2}}{\epsilon} \right)$ & $\checkmark$ \\
          & Cormode et al.\cite{DBLP:journals/tkdd/CormodeKMS08} & $O\left( \frac{H\log \left( N\epsilon \right)}{\epsilon} \right)$ & $O\left( H\log \left( N\epsilon \right) \right)$ & $\checkmark$ \\
          &  Mitzenmacher et al. \cite{DBLP:conf/alenex/ThalerMS12} & $O\left( \frac{H}{\epsilon} \right)$ & $O\left( H\log \epsilon N \right)$ & $\checkmark$ \bigstrut[b]\\
    \hline
    \end{tabular}%
  \label{HHH}%
\end{table*}%

Separator \cite{DBLP:conf/adma/LinL07} consists of $H$ stream-summary structures, each corresponding to a level of the domain tree in the HHH detection problem.
It combines bottom-up and top-down processing strategies to detect HHH.
An incoming item updates the first-layer counters or higher-layer counters if there is no space available or no matched counters in the lower layer.
Randomized HHH \cite{DBLP:conf/sigcomm/Ben-BasatEFLW17} proposes a randomized constant-time algorithm for HHH, which achieves a worst-case update time of $O(1)$ for each packet.
Recursive Lattice Search \cite{DBLP:conf/imc/Cho17} revisits the commonly accepted definition of HHH and applies $Z$-ordering \cite{morton1966computer} to use a recursive partitioning algorithm.
The $Z$-order makes the ordering consistent with the ancestor-descendant relationship in the hierarchy, and it transforms the HHH problem into simple space partitioning of a quadtree.

Table \ref{HHH} compares the HHH detection methods according to their space consumption and updating time.
Specifically, $N$ represents the total number of packets, $H$ is the size of the hierarchy, and $\epsilon$ denotes the allowed relative estimation error for every single flow's frequency.
For TCAM-based HHH \cite{DBLP:conf/nsdi/JoseY11}, $T$ is a threshold.
The flows that consume at least $T$ of the link capacity in each interval can be defined as heavy hitters, and space refers to the TCAM matching rules.

\begin{table*}[htbp]
  \centering
  \caption{Top-$k$ heavy detection}
    \begin{tabular}{llll}
    \hline
    \multirow{12}[4]{*}{Top-$k$} & \multicolumn{2}{l}{Methods} & Basic idea \bigstrut\\
\cline{2-4}          & \multicolumn{2}{l}{SS\cite{DBLP:conf/icdt/MetwallyAA05}} & Evict the item with the minimal value \bigstrut[t]\\
          & \multicolumn{2}{l}{ FSS\cite{DBLP:journals/isci/HomemC10}} & Augment a sketch to filter items \\
          & \multicolumn{2}{l}{SSS\cite{DBLP:conf/bigcomp/GongTY0D0L18}} & {Insert the item according to scoring} \\
          & \multicolumn{2}{l}{CSS\cite{DBLP:conf/infocom/Ben-BasatEFK16}} & {Require no pointer and use statistical memory} \\
          & \multicolumn{2}{l}{WCSS\cite{DBLP:conf/infocom/Ben-BasatEFK16}} & {Point queries in $O(1)$ under sliding model} \\
          & \multicolumn{2}{l}{ Frequent\cite{DBLP:conf/esa/DemaineLM02}} &Evict the item by decreasing all the items \\
          & \multicolumn{2}{l}{RAP\cite{DBLP:conf/infocom/Ben-BasatEFK17}} &Randomized admission policy \\
          & \multicolumn{2}{l}{CountMax\cite{DBLP:journals/ton/YuXYWH18}} &  {Lightweight and cooperative measurement} \\
          & \multicolumn{2}{l}{HeavyKeeper\cite{DBLP:journals/ton/0003ZLGUCL19}} &Count with exponential decay \\
          & \multicolumn{2}{l}{HashPipe\cite{DBLP:conf/sosr/SivaramanNRMR17}} &Use emerging programmable (P4) \\
          & \multicolumn{2}{l}{Discrete Tensor Completion\cite{DBLP:conf/infocom/XieTWXWZ19}} &Data plane implementations \bigstrut[b]\\
    \hline
    \end{tabular}%
  \label{k}%
\end{table*}%

\subsubsection{Top-$k$ detection} Top-$k$ flows detection is a critical task in network traffic measurement, with many applications in congestion control, anomaly detection, and traffic engineering \cite{DBLP:journals/ton/0003ZLGUCL19}, including data mining, databases, network traffic measurement, and security \cite{DBLP:conf/bigcomp/GongTY0D0L18}.
 Table \ref{k} summarizes the related methods and their basic ideas.

Sketch-based methods such as CM Sketch \cite{DBLP:conf/latin/CormodeM04} and CSketch \cite{DBLP:journals/pvldb/CormodeH08} record the packet number of all flows.
By augmenting an additional heap, these methods can report the top-$k$ heavy flows. However, they also record the frequencies of mice.
Such information is useless, detrimental to finding the top-$k$ heavy flows, and requires additional memory usage \cite{DBLP:conf/bigcomp/GongTY0D0L18}.

To eliminate small flows and add suspected large flows, MG counter \cite{DBLP:journals/scp/MisraG82} performs $O(k)$ operations to update $k$ counters in a hash table; the overhead becomes significant when there are many mice to be kicked out.
Pandey et al.\cite{DBLP:conf/sigmod/Pandey0BBFJKP20} show that by using modern storage devices and building upon on recent advances in external-memory dictionaries, MG leveled external-memory reporting tables (LERTs) can process millions of stream events per second.
Space Saving \cite{DBLP:conf/icdt/MetwallyAA05} and Frequent \cite{DBLP:conf/esa/DemaineLM02} both treat each incoming packet $f$ as a suspected hot item.
They assign a high frequency to $f$ and insert it into the top-$k$ data structure. Further, they gradually evict the mice from the top-$k$ data structure with a certain probability as time passes.
However, in real traffic traces, most items are mice, and processing these mice involves additional overhead and makes the ranking and frequency estimation of the top-$k$ flows inaccurate.
FSS\ cite{DBLP:journals/isci/HomemC10} uses a filtering approach to improve Space Saving.
It uses a bitmap to filter and minimize updates on the top-$k$ data structure and estimate the error associated with each element.
SSS \cite{DBLP:conf/bigcomp/GongTY0D0L18} adds a small queue and a \emph{Scoreboard} to insert only the elephants into Stream Summary instead of all the items.
The scoreboard is responsible for recording potential hot items to indicate whether an item is an elephant item.
CSS \cite{DBLP:conf/infocom/Ben-BasatEFK16} redesigns Space Saving and achieves a reduction in space overhead of up to 85\%; it works at $O(1)$ with high probability.
Moreover, WCSS is the first algorithm that supports point queries with constant time complexity under the sliding window model. This property makes WCSS a practical choice for matching the line rate.

CountMax \cite{DBLP:journals/ton/YuXYWH18} and MV-sketch \cite{DBLP:conf/infocom/TangHL19} adopt the majority vote algorithm (\emph{MJRTY}) \cite{DBLP:conf/birthday/Moore91} to track the candidate heavy flow in each bucket.
Moreover, it has loose bounds on the estimated values of the top-$k$ flows.
Compared with SSS \cite{DBLP:conf/bigcomp/GongTY0D0L18}, CountMax consumes only 1/3-1/2 computing overhead and reduces the average estimation error by 20\%--30\% under the same memory size.
RAP \cite{DBLP:conf/infocom/Ben-BasatEFK17} is a randomized admission policy for allocating counters to non-measurement flows.
More specifically, this strategy ignores most of the tail flows and can still admit the elephants eventually.
HeavyKeeper uses the strategy from HeavyGuardian \cite{DBLP:conf/kdd/0003GZZSL18}, i.e., $count\ with\ exponential\ decay$. Specifically, HeavyKeeper only keeps track of a small number of flows.
 Mice entering the data structure will decay away to make room for the true elephants.
Discrete Tensor Completion \cite{DBLP:conf/infocom/XieTWXWZ19} solves the challenging problem of inferring the top-$k$ elephants in an efficient system with incomplete measurement data as a result of sub-sampling for scalability or missing data.

HashPipe \cite{DBLP:conf/sosr/SivaramanNRMR17} is an algorithm for tracking the $k$ heaviest flows with high accuracy within the features and constraints of programmable switches.
HashPipe is heavily inspired by Space Saving \cite{DBLP:conf/icdt/MetwallyAA05}.
It maintains both the flow key and the frequency in a pipeline of hash tables to track heavy flows by evicting lighter flows from the switch memory over time.

\subsubsection{Entropy estimation} As a critical metrics, entropy has attracted considerable attention for network measurement \cite{DBLP:conf/imc/ZhaoLOSWX07}.
Using the entropy of traffic distributions, we can perform a wide variety of network applications such as anomaly detection, clustering to reveal interesting patterns, and traffic classification \cite{DBLP:conf/sigmetrics/LallSOXZ06}.
IMP \cite{DBLP:conf/imc/ZhaoLOSWX07} is the first method to measure the entropy of the traffic between every origin-destination pair ($OD$-pair flows).
Moreover, IMP can measure the entropy of the intersection of two data streams, $A$ and $B$, from the sketches of $A$ and $B$, which are recorded from two measurement points.
Lall et al. \cite{DBLP:conf/sigmetrics/LallSOXZ06} contributed to the investigation and application of streaming algorithms to compute the entropy over network traffic streams.
They also identified the appropriate estimator functions for calculating the entropy accurately and provided proof of approximation guarantees and resource usage.
AMS-estimator \cite{DBLP:conf/soda/ChakrabartiCM07} designs an algorithm for approximating the empirical entropy of a stream of $m$ values in a single pass, using $O\left(\epsilon ^{-2}\log \left( \delta ^{-1} \right) \log m \right)$ words of space, which is near-optimal in terms of its dependence on $\epsilon$, where $\epsilon$ is the approximation parameter and $\delta$ represents the error probability.
Defeat \cite{DBLP:conf/imc/LiBCDGIL06} uses the entropy of the empirical distribution of each feature to detect unusual traffic patterns, where the traffic feature represents an entry in a packet header field.

\subsubsection{Change detection}
A sudden increase in network traffic owing to the emergence of elephant flows is essential for network provisioning, management, and security because significant patterns often imply events of interest \cite{DBLP:journals/cn/BuCCL10}.
For example, the beginning of a DDoS attack or traffic rerouting owing to link failures will cause traffic to change heavily.
A heavy changer is defined as a flow that contributes more than a threshold of the total capacity over two consecutive intervals.

To detect the significant changes in the two consecutive time intervals, Fast Sketch \cite{DBLP:conf/infocom/LiuCG12}, MV-sketch \cite{DBLP:conf/infocom/TangHL19}, LD-sketch \cite{DBLP:journals/cn/HuangL15}, Modular Hashing \cite{DBLP:journals/ton/SchwellerLCGGZDKM07}, Deltoid \cite{DBLP:journals/ton/CormodeM05}, Reversible $k$-ary sketch \cite{DBLP:conf/imc/SchwellerGPC04}, and Group testing \cite{DBLP:journals/tods/CormodeM05} all derive the difference sketch $S_d\mathrm{=}\left| S_2\mathrm{-}S_1 \right|$, where $S_1$ and $S_2$ are the sketches recorded for the two consecutive time intervals.
Because of the linearity of sketches, the difference in flows can be estimated by subtraction.
For any key whose value recorded in $S_d$ exceeds the threshold, we can conclude that this flow is a suspected heavy changer and proposed as a set of heavy change flows.
Furthermore, Deltoid \cite{DBLP:journals/ton/CormodeM05} refers to a flow as \emph{deltoid} if it has a large difference, including \emph{absolute}, \emph{relative}, or \emph{variational}, instead of only \emph{absolute} difference. MV-sketch \cite{DBLP:conf/infocom/TangHL19} and LD-sketch \cite{DBLP:journals/cn/HuangL15} both record the flow ID to achieve reversibility.
Fast Sketch \cite{DBLP:conf/infocom/LiuCG12}, Modular Hashing \cite{DBLP:journals/ton/SchwellerLCGGZDKM07}, and Reversible $k$-ary sketch \cite{DBLP:conf/imc/SchwellerGPC04} are all reversible sketch designs that adopt different strategies to identify the heavy changer.
They separately record the volume information of bit or bits. By identifying the heavy bit/bits changer, these methods can reverse the heavy change flows.

\begin{table*}[htbp]
  \centering
  \caption{Heavy change detection}
    \begin{tabular}{lllllll}
    \hline
    \multicolumn{1}{l}{\multirow{10}[4]{*}{Heavy change \newline{}}} & Methods & Memory & Updating time & Query time & False positive & False negative \bigstrut\\
\cline{2-7}          & Fast sketch\cite{DBLP:conf/infocom/LiuCG12} & $O\left( k\log \frac{n}{k} \right)$ & $O\left( \log \frac{n}{k} \right)$ & $O\left( k\log \frac{n}{k} \right)$ & $\checkmark$ & $\checkmark$ \bigstrut[t]\\
          & MV-Sketch\cite{DBLP:conf/infocom/TangHL19} & $\varTheta \left( k\log n \right)$ & $O\left( k\right)$ & $O\left( k^2 \right)$ & $\checkmark$ & $\checkmark$ \\
          & Group testing\cite{DBLP:journals/tods/CormodeM05} & $O\left( k\log \frac{n}{k} \right)$ & $O\left( \log n \right)$ & $O\left( k\log \frac{n}{k} \right)$ & $\checkmark$ & $\times$ \\
          & Defeat\cite{DBLP:conf/imc/LiBCDGIL06} & $O\left( n^{1/2} \right)$ & $O\left( 1 \right)$ & $O\left( k^2 \right)$ & $\checkmark$ & $\times$ \\
          &  Sequential Hashing\cite{DBLP:journals/cn/BuCCL10} & $\varTheta \left( k\log n \right)$ & $\varTheta \left(\log n \right)$ & $\varTheta \left(k\log n \right)$ & $\checkmark$ & $\checkmark$ \\
          & Reversible $k$-ary sketch\cite{DBLP:conf/imc/SchwellerGPC04} & $\varTheta \left( \frac{\left( \log n \right) ^{\varTheta \left( 1 \right)}}{\log\log n} \right)$ & $\varTheta \left(\log n \right)$ & $O\left( kn^{\frac{1}{\log\log n}}\log\log n \right)$ & $\checkmark$ & $\checkmark$ \\
          & Deltoid\cite{DBLP:journals/ton/CormodeM05} & $\varTheta \left( k\log n \right)$ & $\varTheta \left(\log n \right)$ & $O\left( k\log n \right)$ & $\checkmark$ & $\checkmark$ \\
          & LD-Sketch\cite{DBLP:journals/cn/HuangL15} & $\varTheta \left( k^2\log n \right)$ & $O\left( k\right)$ & $O\left( k\right)$ & $\checkmark$ & $\times$ \\
          & Modular hashing\cite{DBLP:journals/ton/SchwellerLCGGZDKM07} & $O\left( n^{\frac{1}{\log\log n}}\log\log n \right)$ & $O\left( \frac{\log n}{\log\log n} \right)$ & $O\left( kn^{\frac{1}{\log\log n}}\log\log n \right) $ & $\checkmark$ & $\checkmark$ \bigstrut[b]\\
    \hline
    \end{tabular}%
  \label{Heavy change detection}%
\end{table*}%

Interestingly, HyperSight \cite{DBLP:journals/jsac/ZhouBYGCZWZ20} exploits packet behavior changes in response to network incidents achieving both high measurement coverage and high accuracy.
By applying a series of Bloom filters, it checks whether the corresponding packet behaviors have been changed.
Moreover, HyperSight designs a packet behavior query language to express multiple measurement queries based on the packet behavior change primitive to support a wide range of network event queries.

Table \ref{Heavy change detection} summarizes existing heavy change detection methods in terms of their memory overhead, updating time complexity, and query time complexity, where $n$ represents the range of flow keys and $k$ is an upper bound on the number of anomalous keys of interest.
Moreover, we focus on the false positive and false negative errors involved in these methods.

\subsubsection{Burst detection}

Paul et al. \cite{DBLP:conf/icde/PaulP019} define burst as the acceleration over the incoming rate of an event mentioning.
Based on Persistent CM sketch (PCM) \cite{DBLP:conf/sigmod/WeiLYDW15}, they explore the efficiency and approximation quality tradeoff in the design of data summaries.
 They further present PBE to find burst events at any time for the entire history.
Zhong et al. \cite{DBLP:conf/sigmod/ZhongYLT0021} refer to a sudden increase in terms of arrival rate followed by a sudden decrease as a burst.
They propose BurstSketch first to select potential burst items using Running Track technique efficiently and then monitor the potential burst items and capture the key features of burst pattern by Snapshotting technique.
Specifically, Runnning track can be most frequent elements finding strategies, such as frequent\cite{DBLP:conf/imc/GolabDDLM03}, probabilistic decay\cite{DBLP:conf/kdd/0003GZZSL18} and probabilistic replacement\cite{DBLP:conf/infocom/Ben-BasatEFK17}.
The idea behind the Snapshotting is that a burst can be described as a sudden increase and a sudden decrease.
Compared with PBE\cite{DBLP:conf/icde/PaulP019}, BurstSketch care more about real-time burst detection in high-speed data streams.

\subsubsection{Persistent sketch} A persistent sketch, also known as a multi-version data structure in the database literature, is a data structure that preserves all its previous versions as it is updated over time\cite{DBLP:conf/sigmod/WeiLYDW15}.
By maintain such a data structure, network operators can perform historical window queries and historical queries on the whole history data instead of just a measurement period.
Wei et al. \cite{DBLP:conf/sigmod/WeiLYDW15} introduce Piecewise Linear Approximation into basic sketch structure to support historical window queries and historical queries.
They propose persistent CM Sketch and persistent AMS Sketch, which leverage piecewise-linear functions to approximate each counter of the sketch over time.
Persistent Bloom Filter\cite{DBLP:conf/sigmod/PengGLQZ18} decomposes the time into a binary representation of temporal range and maintains a BF for each to support historical window queries and historical queries.

\subsubsection{Item batch detection} Chen et al. \cite{DBLP:conf/conext/PeiqingChen21} define item batch as a consecutive sequence of identical items
that are close in time in a data stream.
They argue that item batch is a useful data stream pattern in the cache, burst detection, and APT detection.
They first propose ClockSketch to delete the stale information as much as possible while maintaining all items visited within the latest time window.

\subsection{Hardware or Software implementation}

As the Internet enters the era of big network data, traffic flow may be massive in some essential scenarios.
Modern high-speed routers forward packets at the speed of hundreds of Gigabits or even hundreds of Terabits per second \cite{DBLP:journals/ton/XiaoCZCLLL17}.
The number of traffic flows that traverse a core router may be of the order of tens of millions. Moreover, the number of bytes per flow can also be large.
Simultaneous tracking of such a large number of flows is a unique challenge for traffic measurement hardware and software.
This subsection introduces some methods that focus on the measurement implementations of hardware and software in the passive measurement area.

%\subsubsection{Implementation in hardware memory}\label{hardware}

\subsubsection{Implementation in SRAM}\label{SRAM}

To maintain high throughput, routers forward packets from incoming ports to outgoing ports via a switching fabric, bypassing the main memory and CPU.
If the measurement is implemented as an online module to process packets in real-time, one way is to implement it on network processors at the incoming/outgoing ports and use on-chip cache memory.
However, the commonly used cache on processor chips is SRAM, typically with limited hardware resources, which may have to be shared among multiple functions for routing, performance, and security purposes.
In such a context, the amount of memory allocated for measurement may be tiny.
Several methods aim to leverage SRAM to accomplish measurement tasks.
Most existing studies are devoted toward excellent counter representation to ensure that the counters occupy as little memory as possible to fit into the off-chip SRAM.
Furthermore, several studies have been conducted to break the off-chip SRAM's throughput bound to guarantee per-packet processing.

\textbf{\emph{Single counter compression.}}
Many studies have adopted a single-counter-based compressed representation approach. SAC \cite{DBLP:conf/infocom/Stanojevic07} and SA counter \cite{DBLP:conf/infocom/YangXLLWBL19} both propose a method such as floating-point number representation to enlarge the counting range or compress the counters.
SAC statically divides the available memory of $q$ bits in a unit into two parts $A$ and $mode$ of size $l$ and $k$, respectively, and represents a counter as ${A\cdot 2^{r\cdot mode}}$, while SA counter augments a flag bit to dynamically adjust the mode bits to enlarge the counting range.

\textbf{\emph{Sampling strategy.}}
Several sampling methods have been developed to reduce the processing overhead and improve the estimation accuracy of the counter estimator.
ANLS \cite{DBLP:conf/infocom/HuWTLCC08} proposes an estimator that dynamically adjusts the sampling rate for a flow depending on the number of packets that have been counted to improve the measurement accuracy for small flows.
DISCO \cite{DBLP:journals/ton/HuLZ0CCW14} extends ANLS by regulating the counter value to be a real increasing concave function of the actual flow length and supports both packet counting and byte counting with high accuracy.
Counter Estimation Decoupling for Approximate
Rates (CEDAR) \cite{DBLP:conf/infocom/TsidonHK12} decouples and shares pre-given estimation values by pointers.
By sampling the packets according to the current and following counters, CEDAR can dynamically update the pointers.
ICE-Buckets \cite{8347005} presents a closed-form representation of the optimal estimation function and introduces an independent bucket-based counter architecture to improve the accuracy further.

\textbf{\emph{Bit-sharing strategy.}}
There are several shared-bit compression approaches for reducing the wastage of high-order bits of cold items.
ABC \cite{DBLP:conf/bigdataconf/Gong0ZYC0L17} uses the space from the adjacent counters by operations such as bit borrowing and combination when a counter overflows.
Counter Tree \cite{DBLP:journals/ton/ChenCC17} shares the high-order bits of large virtual counters, which are constructed from multiple physical counters organized in a tree structure and span a path crossing multiple levels of the tree.
OM sketch \cite{DBLP:conf/globecom/ZhouLJ0DL17} shares the high-order bits by a hierarchical CSM sketch \cite{DBLP:conf/infocom/LiCL11}.
It can achieve close to one memory access and one hash computation for each insertion or query while achieving high accuracy.
Pyramid Sketch \cite{DBLP:journals/pvldb/YangZJCL17} dynamically assigns an appropriate number of bits for different items with different frequencies. It also employs a tree structure.

\textbf{\emph{Counter-sharing strategy.}}
CSM sketch \cite{DBLP:conf/infocom/LiCL11} and Counter Braids \cite{DBLP:conf/sigmetrics/LuMPDK08} are representative counter-sharing compression approaches.
CSM sketch splits each flow's size among several counters that are randomly selected from a counter pool.
This strategy is implemented by randomly selecting a counter from the flow's counter vector and increasing the counter by one.
Counter Braids establishes a hierarchical counter architecture by braiding the counter bits with random graphs to solve the central problems (counter space and flow-to-counter rule).
These approaches achieve a reasonable compression rate; however, the decompression process is relatively slow.

\textbf{\emph{Virtual estimators.}} Some virtual estimators that facilitate memory sharing for recent cardinality estimation solutions have also been proposed.
Virtual HyperLogLog \cite{DBLP:conf/globecom/ZhouZCXC16} shares a register array $C$ of $m$ HLL registers to store the packet information of all flows.
For an arbitrary flow $f$, where $f$ is the flow label, it pseudo-randomly selects $s$ registers from $C$ to logically form a virtual counter ${C_f}$ , and use ${C_f}$ to encode the packets in flow $f$.
Virtual Register Sharing \cite{DBLP:journals/ton/XiaoCZCLLL17} allocates each flow with a virtual estimator, including PCSA \cite{DBLP:journals/jcss/FlajoletM85}, LogLog \cite{DBLP:conf/esa/DurandF03}, and
HyperLogLog \cite{DBLP:conf/edbt/HeuleNH13}, and these virtual estimators share a common memory space.
It demonstrates that sharing at the multi-bit register level is superior to sharing at the bit level.

Apart from the above-mentioned methods, some other strategies have been developed to solve the throughput bound problem by exploiting the speed of SRAM.
To overcome the bottleneck, Cache-assisted Stretchable Estimator (CASE) \cite{DBLP:conf/infocom/LiWPDLL16} uses the on-chip memory as the fast cache of the off-chip SRAM.
It counts hot items in the on-chip memory and writes back to the off-chip memory according to the cache's estimation.
Because of the heavy-tailed distribution of traffic, most accesses to the counters will occur in the cache.
The throughput can be increased remarkably while the accuracy will be improved due to the non-compression counting.

\subsubsection{Implementation in DRAM}\label{DRAM}

The prevailing view is that SRAM is too expensive to implement large counter arrays, whereas DRAM is too slow to provide line speed updates.
This view is the main premise of a number of hybrid SRAM/DRAM architectural proposals \cite{DBLP:journals/micro/ShahIPM02} \cite{DBLP:conf/sigmetrics/RamabhadranV03} \cite{DBLP:conf/globecom/RoederL04} \cite{DBLP:conf/sigmetrics/ZhaoXL06} that still require substantial amounts of SRAM for large arrays.
However, some studies \cite{DBLP:journals/sigmetrics/LinX08} \cite{DBLP:conf/ancs/ZhaoWLX09}
\cite{DBLP:journals/ton/WangZLX12} \cite{DBLP:journals/tpds/WangLX13} have presented a contrarian view that modern commodity DRAM architectures, driven by aggressive performance roadmaps for consumer applications, have advanced architectural features that can be exploited to make DRAM solutions practical.

Lin and Xu \cite{DBLP:journals/sigmetrics/LinX08} first highlighted the potential of leveraging DRAM for per-flow network measurement.
They proposed two schemes, one based on the \emph{replication} of counters across memory banks and the other based on the \emph{randomized distribution} of counters across memory banks to maintain wire-speed updates to large counter arrays.
Compared with existing hybrid SRAM/DRAM counter architectures, this strategy can achieve the same update rates to counters without needing a nontrivial amount of SRAM for storing partial counts.
Further, they proposed a randomized DRAM architecture in \cite{DBLP:conf/ancs/ZhaoWLX09} \cite{DBLP:journals/ton/WangZLX12}, which harnesses the performance of modern commodity DRAM architectures by interleaving counter updates in multiple memory banks.
The architecture consists of a simple randomization scheme, a small cache, and small request queues to statistically guarantee a near-perfect load-balancing of counters to the DRAM banks.

\subsubsection{Implementation in hybrid SRAM-DRAM}\label{SRAM-DRAM}

Statistics counters are the fundamental unit of sketch-based traffic measurement.
Maintaining these counters at high line speed is a challenging research problem. Briefly, two requirements must be met to solve the per-flow measurement problem \cite{DBLP:conf/infocom/Stanojevic07}: (1) to store a large number of counters and (2) to update a large number of counters per second.
Although chip and large DRAM can easily satisfy these requirements, DRAM access times of 50-100 ns cannot accommodate many counter-updates.
Meanwhile, expensive SRAM with access times of 2-6 ns allows many more counter increments per second; however, is not economical for many counters.

A hybrid SRAM/DRAM structure can help store the massive number of counters to solve this problem.
Several methods have been developed to use DRAM to maintain statistics counters with a small, fixed amount of SRAM.
DRAM maintains \emph{N} counters of \emph{M} bits, while SRAM stores $N$ counters of ${m}$ bits, where ($m\mathrm{<}M$).
The SRAM counters will handle increments at very high speed, and once an SRAM counter reaches a value of ${2^m}$ (overflow), its value must be flushed to its corresponding DRAM counter.
Further, the access speed of DRAM (departure rate from the queue) should be higher than the average rate of counter overflow (arrival rate to the queue).
Several counter management algorithms (CMAs) have been proposed to manage updates from SRAM to DRAM while ensuring correct line-rate operations for many counters.

LCF (Largest Counter First) \cite{DBLP:journals/micro/ShahIPM02} employs a heap-based priority queue to select the counter that is the closest to overflow.
This greedy algorithm makes the best possible selection decision (without future knowledge). LCF is provably optimal in terms of the amount of SRAM required.
However, maintaining a heap in hardware involves high implementation complexity and a large amount of SRAM, i.e., around twice the amount required to store the SRAM counters.
LR($T$) (Largest Recent with threshold $T$) \cite{DBLP:conf/sigmetrics/RamabhadranV03} avoids the expensive operation of maintaining a priority queue by keeping track of only those counters that are larger than a threshold $T$ using a bitmap. A tree structure is imposed on the bitmap (resulting in a hierarchical bitmap) to allow for fast retrieval of the \emph{next counter to be flushed}.
This retrieval operation has a complexity of ${O(logN)}$, where $N$ is the number of counters in the array; however, using a large base such as 8 makes the complexity essentially a small constant.
The hardware control logic in LR($T$) is more straightforward than that in LCF and uses a much smaller amount of SRAM.

Zhao et al. \cite{DBLP:conf/sigmetrics/ZhaoXL06} removed the fundamental design objective of previous approaches that do not allow SRAM-counter overflows and achieved a theoretically minimum SRAM cost of 4-5 bits per counter when the SRAM-to-DRAM access latency ratio is between 1/10 (4 ns/40 ns) and 1/20 (3 ns/60 ns).
They augmented a small SRAM FIFO buffer between SRAM counters and DRAM counters to temporarily hold the SRAM counters' indices that have overflowed and must be made to the DRAM in the future.
A simple randomized algorithm also guarantees that SRAM counters do not overflow in bursts that are sufficiently large to fill the write-back buffer even in the worst case.
This approach consumes significantly fewer bits in SRAM and proposes a simple CMA scheme compared with a heap in LCF \cite{DBLP:journals/micro/ShahIPM02} and tree-like structure in LR($T$) \cite{DBLP:conf/sigmetrics/RamabhadranV03}.

Roeder et al. \cite{DBLP:conf/globecom/RoederL04} extended LCF \cite{DBLP:journals/micro/ShahIPM02} and LR(\emph{T}) \cite{DBLP:conf/sigmetrics/RamabhadranV03} with multilevel counter memory architecture to reduce the amount of fast memory required for nonuniform traffic patterns.
They observed that many counters in previous methods never reach a counter close to overflow; hence, using several SRAM levels to store the lower bits of every counter may be a better strategy.
Whenever a counter overflows, i.e., exceeds the maximum value, it will update the corresponding counters in the next level.
Comprehensive experiments have demonstrated that multi-level counter memory architecture can reduce the required amount of fast memory storage by as much as 28${\%}$.

Lall et al. \cite{DBLP:conf/infocom/LallOX09} propose a hybrid SRAM/DRAM algorithm for measuring all elephant and medium-sized flows with strong accuracy guarantees.
They employed a spectral Bloom filter in SRAM to filter out small flows and preferentially sample medium and large flows to a DRAM flow table.
The sketch automatically eliminates the burden of having to perform expensive updates for tens of millions of small flows typically found in network traffic.
Once the flow is identified as greater than the threshold, it creates a record for it in the flow table and samples it by SGS \cite{DBLP:conf/infocom/KumarX06} thereafter.
In all these counter array schemes \cite{DBLP:journals/micro/ShahIPM02} \cite{DBLP:conf/globecom/RoederL04}\cite{DBLP:conf/sigmetrics/ZhaoXL06}, the read operations can be performed only at DRAM speed.
To solve \emph{reading being too slow}, a few SRAM bits are augmented per counter to approximate the value range of the actual counter.

HAC \cite{DBLP:conf/infocom/Stanojevic07} uses augmented DRAM for storing information on recent history.
The idea is that the primary counter information is stored in SRAM while DRAM is used to store information only from the recent history to reduce the standard error.
For every ${\eta}$ SRAM access, there can be one DRAM access.
The counters in DRAM will be updated with probability ${\frac{1}{\eta}}$ and the counters in SRAM will be increased when the counters in DRAM overflow.
This strategy allows an estimate of full counters on a per-packet basis without DRAM access.
HAC does not have an exact counter value compared to the other counter architectures but stores only its estimate.

All these methods are designed for hybrid SRAM/DRAM architectures. LCF \cite{DBLP:journals/micro/ShahIPM02}, LR(\emph{T}) \cite{DBLP:conf/sigmetrics/RamabhadranV03}, and BCMA \cite{DBLP:conf/sigmetrics/ZhaoXL06} are CMA designs for managing updates from SRAM to DRAM for per-flow measurement.
Whereas \cite{DBLP:conf/infocom/LallOX09} leverages SRAM to filter the cold items and only measures elephant and medium-sized flows,
HAC \cite{DBLP:conf/infocom/Stanojevic07} solves the problem of \emph{reading being too slow} by switching the role of SRAM and DRAM and designs a compressed representation of counts in SRAM.

\subsubsection{Implementation in FPGA}\label{FPGA}

FPGAs can be programmed for fast pattern matching because of their ability to exploit reconfigurable hardware and parallelism.
Counter Braids \cite{DBLP:conf/sigmetrics/LuMPDK08}, HeavyKeeper \cite{DBLP:journals/ton/0003ZLGUCL19}, Reversible Sketch \cite{DBLP:journals/ton/SchwellerLCGGZDKM07}, CEDAR \cite{DBLP:conf/infocom/TsidonHK12}, and elastic sketch \cite{DBLP:conf/sigcomm/0003JLHGZMLU18} have developed their parallel versions for FPGA hardware.
Because methods related to FPGA mainly deal with program coding, we omit those details.

\subsubsection{Implementation in TCAM}\label{TCAM}

Ternary Content Addressable Memory (TCAM) is a type of memory that can perform parallel search at high speeds.
A TCAM consists of a set of entries. The top entry of the TCAM has the smallest index, whereas the bottom entry has the largest index.
Each entry is a bit vector of cells, where every cell can store one bit. A TCAM entry can be used to match the incoming packet header.
TCAM-based flow tables in commodity switches can contain statistic counters for traffic measurement.
However, owing to the cost and power consumption of TCAM, the TCAM-based flow table size is limited \cite{DBLP:journals/jcss/FlajoletM85}.

Bandi et al. \cite{DBLP:conf/sigmod/BandiMAA07} first showed how the integrated architecture of the Network Processing Unit (NPU) and TCAM could be used to develop high-speed stream summarization solutions.
They discussed the problems that occur when Lossy Counting \cite{DBLP:conf/vldb/MankuM02} and Space Saving \cite{DBLP:conf/icdt/MetwallyAA05} are adapted to the TCAM model for finding the most common elements in a data stream. Compared to their software-implemented counterparts, these techniques involve a constant and smaller TCAM operations per flow.
When performing HHH detection tasks, because the traffic keeps changing, we do not know the HHHs in advance, and there are not good rules to measure all the prefixes.
TCAM-based HHH \cite{DBLP:conf/nsdi/JoseY11} proposes three approaches to reduce the amount of TCAM required.

Considering that the TCAM-based flow table is always embedded into OpenFlow switches \cite{openflowswitch}, such as HP5406zl switches \cite{DBLP:conf/sigcomm/CurtisMTYSB11} and Broadcom Trident switches \cite{DBLP:conf/conext/StephensCFDC12}, the methods related to TCAM are implemented conducted in the context of SDN; we will introduce the related method in the following SDN subsection.

\subsubsection{Implementation in Programmable Switch}\label{P4}

P4 \cite{P4switch} is a language that specifies how switches process packets.
It provides a register abstraction that offers a form of stateful memory that can store user-defined data structures and be arranged into one-dimensional arrays of user-defined length.
Register cells can be read or updated by P4 actions and are accessible through the control plane API.
Actions are installed in match-action tables to describe specific processing logic, where each action is associated with a user-defined matching rule.
The P4 framework allows users to define the API for control-data plane communication.

Univmon \cite{DBLP:conf/sigcomm/LiuMVSB16} develops a network-wide coordination schemes and demonstrates an implementation in P4.
It is processed by sketching, storage, and a sampling table sequentially until the first level, where it does not get sampled for every packet.
The control plane then uses a simple RPC protocol to import sketching manifests and query the contents of the data plane register arrays.
SketchLearn \cite{DBLP:conf/sigcomm/HuangLB18} implements each level of counters as an array of registers.
Furthermore, the counters are updated by a stateful ALU of hash computation action in a dedicated table. A multi-thread approach is used for model learning and query runtime in the data plane to form the model and compute traffic statistics.
FCM-Sketch\cite{DBLP:conf/conext/SongKLC20} designs a lightweight implementation of CM Sketch in emerging PISA programmable switch.
By collecting data in the data-plane, FCM-Sketch enables complex measurement tasks in the control plane.
Furthermore, Elastic Sketch \cite{DBLP:conf/sigcomm/0003JLHGZMLU18}, HashPipe \cite{DBLP:conf/sosr/SivaramanNRMR17},
FlowRadar \cite{DBLP:conf/nsdi/LiMKY16},
HyperSight \cite{DBLP:journals/jsac/ZhouBYGCZWZ20},
HeavyKeeper \cite{DBLP:journals/ton/0003ZLGUCL19},
and Marple \cite{DBLP:conf/sigcomm/NarayanaSNGAAJK17} have been deployed on P4Switch and validated their performance.

\subsubsection{Implementation in Open vSwitch}\label{OVS}

Open vSwitch \cite{openvswitch} is a production quality, multi-layer virtual switch licensed under the open-source Apache 2.0 license.
It is designed to allow massive network automation through programmatic extension while supporting standard management interfaces and protocols (e.g., NetFlow, sFlow).
The architectural design of OVS \cite{DBLP:journals/usenix-login/PfaffPKJZRGWSSA15} consists of two main components: ovs-vswitchd and ovsdb-server.
The actions given by ovs-vswitchd will instruct datapath to handle packets. The DPDK version allows OVS to perform the entire packet processing
pipeline in the user space, thereby reducing the overhead of context switches and copied data between the user and kernel spaces.
The vswitchd thread has a three-tier look-up cache hierarchy. The first-level works as an Exact Match Cache (EMC) and has the highest lookup speed.
The second-level works as a Tuple Space Search and may trigger the third level managed by an OpenFlow-compliant controller.
BUS \cite{DBLP:conf/icccn/EinzigerLW17} empowers the OVS project with sampling capabilities and implements such capabilities for both physical and virtual ports.
BUS implementation redefines a forwarding action as a branch of the datapath.
Sampled packets are forwarded and unsampled packets are dropped. %Hence, the core implementation is simple and only requires a few lines of code.
DREAM \cite{DBLP:conf/sigcomm/MoshrefYGV14},
FlowRadar \cite{DBLP:conf/nsdi/LiMKY16},
CounterMap \cite{DBLP:conf/iwqos/LiuZWH17},
CO-FCS \cite{DBLP:journals/ton/XuYQLLH17},
Rflow \cite{DBLP:conf/infocom/JangCNN17},
R-HHH \cite{DBLP:conf/sigcomm/Ben-BasatEFLW17}, CountMax \cite{DBLP:journals/ton/YuXYWH18},
NSPA \cite{DBLP:conf/infocom/XuCMH19},
and NitroSketch \cite{DBLP:conf/sigcomm/LiuBEKBFS19} implemented their framework in OVS to evaluate their performance.

\subsubsection{Implementation in SDN}\label{sdn}

\begin{figure}
  \centering
  % Requires \usepackage{graphicx}
  \includegraphics[width=3in]{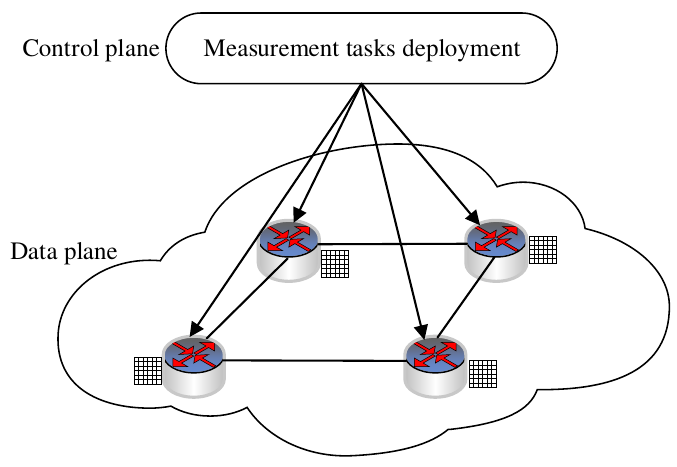}\\
  \caption{SDN measurement architecture}\label{SDN}
\end{figure}

Software network devices are gaining popularity owing to the rise of two new network architecture concepts: Software-Defined Networking (SDN) and Network Function Virtualization (NFV).
SDN facilitates network management by separating the control and data planes.
Network measurement in SDN is a lightweight task, as operators only need to install a measurement module into the controller.
Moreover, the SDN controller configures switches to measurement traffic for measurement tasks, collects statistics, produces a measurement report, and exhibits the architecture shown in Fig. \ref{SDN}.

\textbf{\emph{Measurement framework.}}
OpenSketch \cite{DBLP:conf/nsdi/YuJM13} is designed to make sketch-based measurements on network devices more general and flexible by using a customized data plane and control plane.
The OpenSketch data plane relies on hashing and wildcard rules, which are realized by the classification function of TCAM, to choose a specific flow to be collected and measured. Its control plane consists of a measurement library (CM sketch, reversible sketch, etc.) and applications. Currently, OpenFlow-based SDN is widely used in both academia and industry.
OpenFlow is the de facto standard communication interface between the control and data planes.
Furthermore, based on the packet behavior change primitives, HyperSight \cite{DBLP:journals/jsac/ZhouBYGCZWZ20} can reconcile both the measurement coverage scalabilities.
Based on the \emph{coupon collector problem}, BeauCoup \cite{DBLP:conf/sigcomm/ChenFBR20} supports multiple distinct counting queries with at most one memory access per packet.
Compared with sketch-based or sampling-based solutions, BeauCoup achieves the same accuracy while using 4$\mathrm{\times}$ fewer memory accesses.

\textbf{\emph{Collaborative measurement.}}
CO-FCS \cite{DBLP:journals/ton/XuYQLLH17} first highlights different schemes for flow statistics collection (FCS) in SDN.
It argues that both per-flow collection and per-switch collection might lead to massive costs in terms of the control plane bandwidth and long processing delay on switches in dynamic networks.
This method implements the fast and selective FCS on a switch using the wildcard-based FCS requests to avoid long-delay collection on some switches and massive traffic load on control links.
FlowCover \cite{DBLP:conf/globecom/SuWXH14} leverages the global view of the network topology and active flows to minimize the communication cost by formulating the problem as a weighted set cover, which is proved to be NP-hard.
Heuristics have been presented to obtain the polling scheme efficiently and handle flow changes practically.
OmniMon \cite{DBLP:conf/sigcomm/HuangSL0ZB20} carefully coordinates the collaboration among different types of entities in the network to execute telemetry operations. The resource constraints of each entity are satisfied without compromising accuracy.
Furthermore, from a global perspective, the controller can allocate measurement tasks to a series of switches; the related method is detailed in Section \ref{coll}.

\textbf{\emph{Resource allocation strategy.}}
DREAM \cite{DBLP:conf/sigcomm/MoshrefYGV14} is a system for a TCAM-based software-defined measurement framework.
It focuses on flow-based counters in TCAM and dynamically allocates TCAM resources to multiple measurement tasks in order to achieve their given accuracy bounds.
DREAM develops accuracy estimators for TCAM-based zoom in/out algorithms, and comprehensive evaluations have shown that DREAM is better than simple task-type agnostic schemes such as equal TCAM allocation.
By contrast, SCREAM \cite{DBLP:conf/conext/MoshrefYGV15} explores sketch resource allocation for measurement and allows dynamic resource allocation for many concurrent measurement tasks while achieving the desired accuracy for each task.

\textbf{\emph{Centralization/costs tradeoff.}}
DevoFlow \cite{DBLP:conf/sigcomm/CurtisMTYSB11} explores the \emph{centralization/costs} tradeoffs in terms of the hardware limitations in OpenFlow.
The authors argued that full control and visibility over all flows are not the best choices.
The controller should focus on significant flows and have visibility over these flows and packet samples.
In comparison, the other flows should be maintained in switches as much as possible.
DevoFlow is designed to allow aggressive use of the wild-carded OpenFlow rule in order to reduce the number of switch-controller interactions and the number of TCAM entries through new mechanisms to detect significant flows.

\textbf{\emph{Accuracy/resource tradeoff.}}
Moshref et al. \cite{DBLP:conf/sigcomm/MoshrefYG13} first qualitatively exploited the \emph{resource/accuracy} tradeoffs for measurement primitives in SDN.
They argued that different primitives need different amounts of resources.
Counters occupy precious TCAM memory; hash-based data structures need SRAM, while code fragments require CPU for processing.
These approaches differ in terms of the amount of network bandwidth that they require to communicate intermediate results with the SDN controller.
They present a qualitative understanding of these primitives, resource usage, and accuracy as a function of the spatial and temporal granularity of the measurement.
OblivSketch\cite{DBLP:conf/ndss/LaiYLY0LN21} implements the sketch-based measurement as a full-fledge service integrated with the off-the-shelf SDN framework to prevent the flow statistics leakage and abuse.
By using Intel SGX, OblivSketch employs hardware enclave for security network statistics generation and queries.

\textbf{\emph{Accuracy/computation tradeoff.}}
Alipourfard et al. \cite{DBLP:conf/sosr/AlipourfardMZ0Y18}\cite{DBLP:conf/hotnets/AlipourfardMY15} showed that saving memory through extra computation is not worthwhile in terms of achieving high performance and high accuracy in software. They argued that, in software, the key metric is not memory usage but packet processing performance (i.e., throughput and latency).
Complex design for computing the location of counters may delay the packet processing and affect the throughput.
A linear hash table and count array outperform complex data structures such as Cuckoo hashing, CM sketches, and heaps in various scenarios.

\subsection{Summary and insights}
In this section, we reviewed the corresponding research related to measurement tasks and sketch implementation.
We surveyed $10$ measurement tasks in terms of time and volume and presented a comprehensive comparison of all the measurement tasks' performance and resource overhead.
Furthermore, we reviewed the related studies on sketch hardware and software implementation. From traditional hardware to programmable switches, the functionalities for network measurement have been enhanced considerably.
Furthermore, new problems have emerged, such as coordinating multiple monitors, resource competition with other core network functions, and compatibility across different platforms. In the future, with the emergence of more
flexible and novel platform designs, sketch-based measurement technology must embrace these changes and evolve as the network itself.

\section{Open issues}\label{summary}

Based on review of existing works on the design, optimization, application, and implementation of sketches, we now outline some open issues that may be investigated in the future.

\textbf{More adaptive to traffic characteristics.}
Some methods have considered the characteristic of traffic, such as the flow size distribution.
However, as shown in Fig. \ref{distribution}, these assumptions are not always correct.
There may exist distributions of two different parameters $Zip$ in one traffic trace.
Moreover, most of these methods are designed based on the assumption of static parameters; hence, the designed sketch may not be accurate when applied to the dynamic network.
Elastic sketch \cite{DBLP:conf/sigcomm/0003JLHGZMLU18} is designed to be adaptive to the flow size distribution.
However, it only focuses on the heavy flow part and proposes a dynamical adjustment of the heavy part.
In the future, it will be necessary to design a more flexible sketch design that is more adaptive to various situations of skewed traffic traces.

\textbf{More adaptive to the evolution of the network.}
With the increasing deployment of the IPv6 protocol, the network has emerged as double-stack wherein IPv4 and IPv6 are implemented.
IPv6 was invented and put into practice to resolve the problem of IP address shortage by offering $2^{128}$ IP addresses.
Despite the consistently increasing use of IPv6, IPv4 still dominates the network.
%There are many reasons for such a slow transition, including considerations of investment and compatibility, adoption of emerging technologies such as NAT and DHCP, and some social reasons.
Due to the investment and compatibility, adoption of emerging technologies such as NAT and DHCP, the double-stack circumstance is expected to last for a long time.
D-Sketch\cite{DBLP:journals/icl/LiLGF21} proposes that IPv4 and IPv6 flows are inequivalent in terms of both flow cardinality and flow size.
%This situation will introduce further inference between the measurement results of traffic flows.
Besides, the bit lengths of IPv4 and IPv6 are different.
Therefore, invertible sketch design will not be applicable.
It is necessary to design a new strategy to make the sketch more adaptive to double-stack networks.
Moreover, with the development of SDN\cite{Software53:online}, NFV\cite{DBLP:conf/conext/YenWSVGR20} and SDN-NV\cite{DBLP:conf/infocom/YangJKMY20}\cite{DBLP:conf/sosr/NamSS18}, the new technologies have emerged in new areas, such as 5G\cite{Preface--11:online} and SD-WAN.
The novel sketch-based measurement design is expected in the corresponding areas.

\textbf{More adaptive to SDN and programmable switch.}
SDN ensures the separation of the control unit from the underlying routers and switches and introduces programmability into the network.
How to coordinate the network management and measurement task together under SDN is crucial for the network's performance.
Many network functions are moving from hardware to software to achieve better programmability at a lower cost.
Network policy is defined in the control plane; the control plane enforces the policy while the data plane executes it by forwarding data accordingly \cite{DBLP:journals/pieee/KreutzRVRAU15}.
Meanwhile, the control plane is responsible for managing the management and monitoring logic and then allocating this logic into data plane implementation,
whereas the data plane corresponds to the networking devices responsible for (efficiently) forwarding data. The control plane represents the protocols and measurement logic used to populate the data plane elements' forwarding tables.
Moreover, TCAM-based flow tables are also used to provide measurement statistics information;
however, owing to the limited TCAM entries and high power consumption, the counters can only provide aggregate statistics of matching flows.
To achieve fine-grained measurement, researchers have also proposed performing the measurement task in SRAM. However, the SRAM size in the programmable switch is limited, and the space resources must also be shared with all the online network functions for routing, management, performance, and security purposes.
How to keep up with the online measurement and maintain the network function stability are also important.

\textbf{Enforce with ML.}
There are two aspects in which we can make the sketch more efficient and accurate. First, as reported in \cite{DBLP:conf/sigmod/KraskaBCDP18}, hash maps typically use a hash function to deterministically map keys to positions inside an array, which is essential to point query in the sketch data structure.
By contrast, the idea of learning a model to replace the hash function in order to reduce the conflict may be a better choice.
The core idea is to avoid too many distinct keys from being mapped into the same position inside the hash map.
Second, during the information extraction phase, the traditional sketch algorithm uses the basic counter unit to obtain statistical information by tracking the entire data structure.
According to \cite{DBLP:conf/sigcomm/0003WSSHJTL18}, combining sketches with ML extraction technology can make measurement tasks more accurate.
Moreover, in the future, researchers may focus on adopting ML to solve some tasks, such as frequency estimation and anomaly detection.

\section{Conclusion}
Sketches have been widely employed in traffic measurement because of their high space efficiency.
As stated in this survey, sketches can act as basic data synopses to record and retrieve statistical traffic information.
This survey reviews existing sketch variants from two main aspects, i.e., data processing pipeline and deployment.
From the aspect of data processing, we survey the preparation for sketch-based network measurement, design and optimization for the sketch data structure itself, and optimization in the post-processing stage.
This perspective provides us with a complete view from processing the raw packet traffic to statistical information useful for network management.
Furthermore, we survey existing sketch applications and implements them for network measurement.
In addition, we comprehensively compare sketch-based network measurement tasks, including time and volume. We also discuss the corresponding implementation in hardware, software, and SDN.
Finally, we summarize the open issues for future work in the field of sketch-based network measurement.
We anticipate developing further applications and redesigning sketches in the next generation of networks, communication systems, and beyond.

\bibliography{IEEEabrv,survey}

\bibliographystyle{IEEEtran}

\begin{IEEEbiography}[{\includegraphics[width=1in,height=1.25in,clip,keepaspectratio]{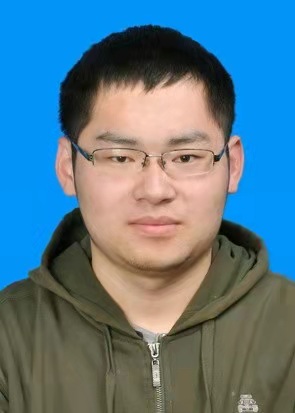}}]{Shangsen Li} received the B.S. degree in automation,
in 2019, from the Northeastern University, Shenyang, China. He is currently working toward the M.S. degree with the College of
Systems Engineering in National university of defense technology. His research interests include network measurement, SDN and sketch data structure.
\end{IEEEbiography}

\begin{IEEEbiography}[{\includegraphics[width=1in,height=1.25in,clip,keepaspectratio]{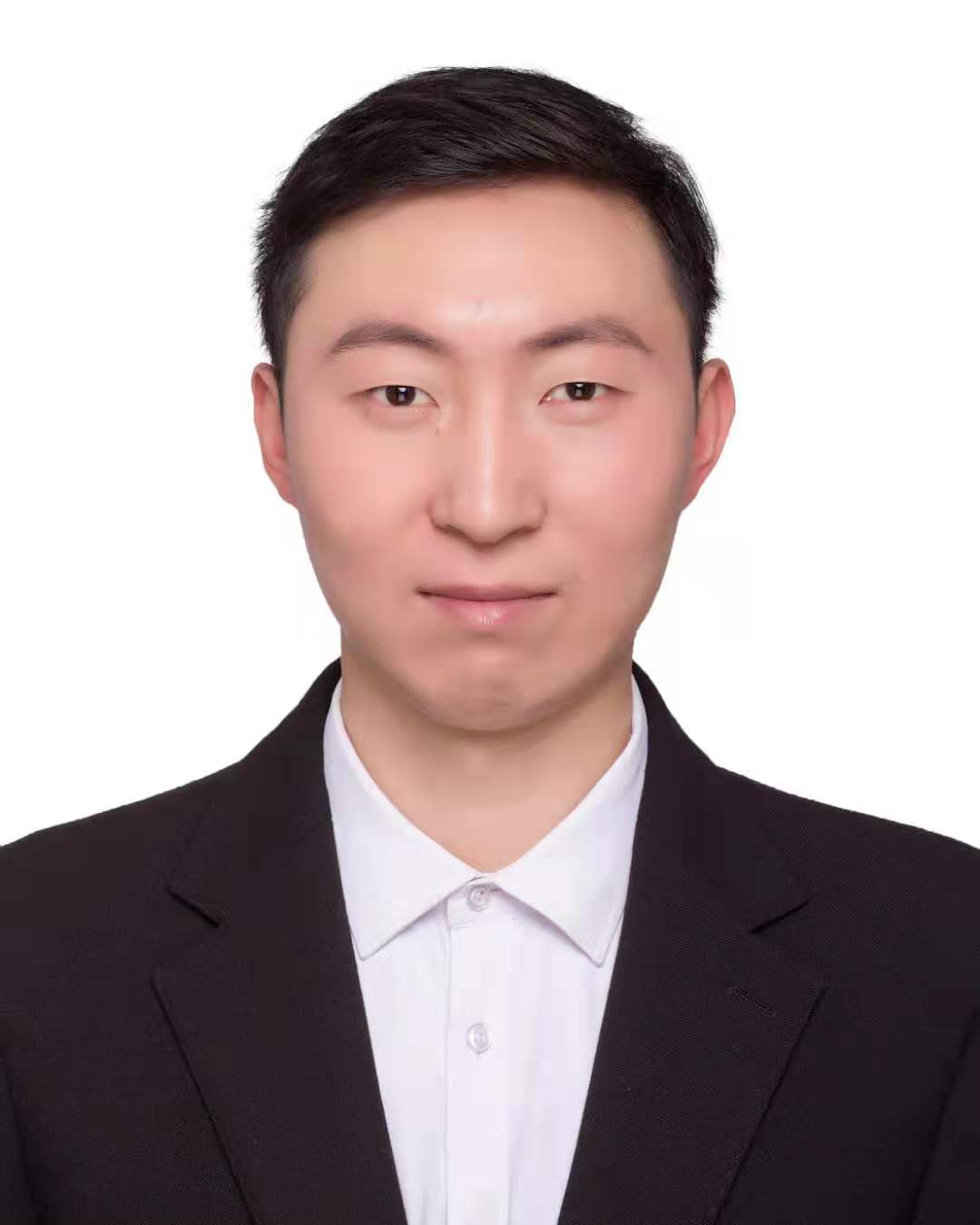}}]{Lailong Luo} received the BS, MS, and PhD
degrees from the College of Systems Engineering, National University of Defence Technology, Changsha, China, in 2013, 2015, and 2019, respectively. He is currently a lecturer with the School of Systems, National University of Defense Technology, Changsha, China. His research interests include data structure and distributed networking systems.
\end{IEEEbiography}

\begin{IEEEbiography}[{\includegraphics[width=1in,height=1.25in,clip,keepaspectratio]{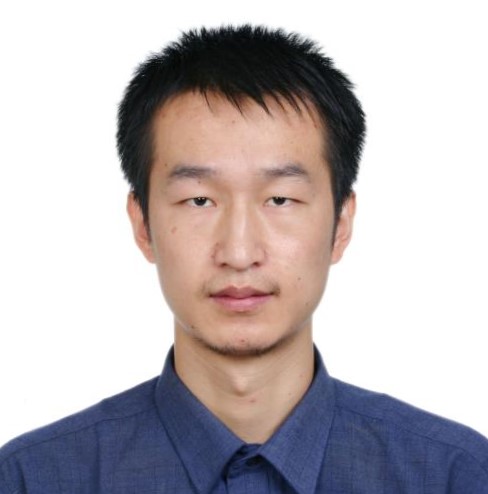}}]{Deke Guo }(Senior Member, IEEE) received the
BS degree in industry engineering from the Beijing University of Aeronautics and Astronautics, Beijing, China, in 2001 and the PhD degree in management science and engineering from the National University of Defense Technology, Changsha, China, in 2008. He is currently a professor with the College of Systems Engineering, National University of Defense Technology. His research interests include distributed systems, software-defined networking, data center networking,
wireless and mobile systems, and interconnection networks. He is a member of ACM.
\end{IEEEbiography}

\begin{IEEEbiography}[{\includegraphics[width=1in,height=1.25in,clip,keepaspectratio]{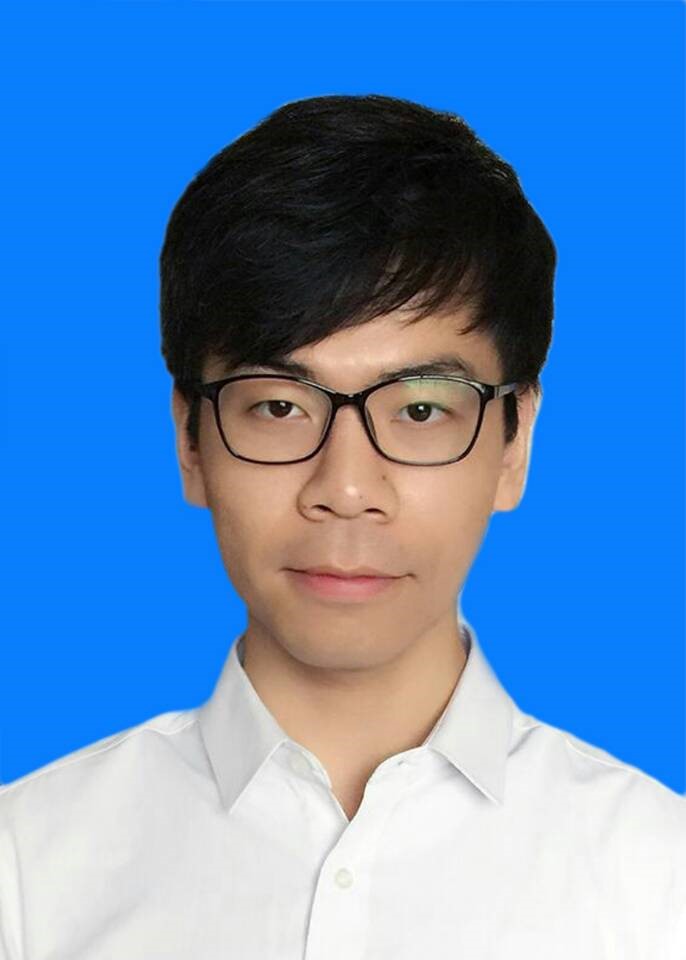}}]{Qianzhen Zhang} received his BSc and MSc degrees in Computer Science from Zhengzhou University and Guangxi University in 2014 and 2018, respectively. Currently, he
is a PhD student at College of Systems Engineering at National University of Defense Technology, China. His research interests include Continuous Subgraph Matching, Graph Data Analytics and Knowledge Graph.
\end{IEEEbiography}

\begin{IEEEbiography}[{\includegraphics[width=1in,height=1.25in,clip,keepaspectratio]{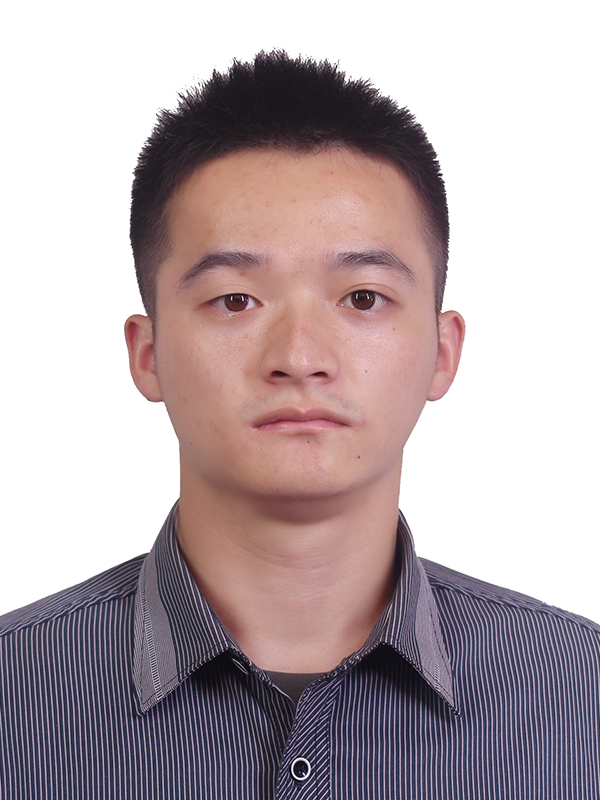}}]{Pengtao Fu} received the bachelor degree from
the National University of Defense Technology,
China, in 2020, where he is currently pursuing
the master degree with the Science and
Technology on Information Systems Engineering
Laboratory. His research interests include data
structure and network measurement.
\end{IEEEbiography}

\end{document}